\documentclass[prd,aps,floatfix,showpacs,nofootinbib]{revtex4}

\usepackage{graphicx}
\usepackage{longtable}

\newcommand{\bra}{\left\langle}
\newcommand{\ket}{\right\rangle}

\def\simgt{\,\rlap{\lower 3.5pt\hbox{$\mathchar\sim$}}\raise 1pt\hbox {$>$}\,}
\def\simlt{\,\rlap{\lower 3.5pt\hbox{$\mathchar\sim$}}\raise 1pt\hbox {$<$}\,}

\begin{document}

\draft

\title{
{\normalsize \hfill {\sf UTHEP-486}} \\
\vspace*{-2pt}
{\normalsize \hfill {\sf UTCCS-P-1}} \\
\vspace*{-2pt}
{\normalsize \hfill {\sf April, 2004}} \\
Light hadron spectroscopy in two-flavor QCD with small sea quark masses
}

\author{CP-PACS Collaboration:
  Y.~Namekawa\rlap,$^{a}$%
  \footnote{Present address:
      Department of Physics,
      Nagoya University,
      Nagoya 464-8602, Japan
 }
  S.~Aoki\rlap,$^{a}$
  M.~Fukugita\rlap,$^{b}$
  K-I.~Ishikawa\rlap,$^{a,c}$%
  \footnote{Present address:
      Department of Physics,
      Hiroshima University, 
      Higashi-Hiroshima, Hiroshima 739-8526, Japan
 }
  N.~Ishizuka\rlap,$^{a,c}$
  Y.~Iwasaki\rlap,$^{a}$
  K.~Kanaya\rlap,$^{a}$ 
  T.~Kaneko\rlap,$^{d}$ 
  Y.~Kuramashi\rlap,$^{d}$%
  \footnote{Present address:
      Center for Computational Physics,
      University of Tsukuba, Tsukuba, Ibaraki 305-8577, Japan
 }
  V.I.~Lesk\rlap,$^{c}$%
  \footnote{Present address:
      Department of Biological Sciences,
      Imperial College, London SW7 2AZ, U.K.
 }
  M.~Okawa\rlap,$^{e}$ 
  A.~Ukawa$^{a,c}$
  T.~Umeda\rlap,$^{c}$%
  \footnote{Present address:
      Yukawa Institute for Theoretical Physics,
      Kyoto University, Kyoto 606-8502, Japan
 }
  and
  T.~Yoshi\'e$^{a,c}$
 }

\affiliation{
$^{a}$Institute of Physics,
      University of Tsukuba, Tsukuba, Ibaraki 305-8571, Japan \\
$^{b}$Institute for Cosmic Ray Research,
      University of Tokyo, Kashiwa 277-8582, Japan \\
$^{c}$Center for Computational Sciences,
      University of Tsukuba, Tsukuba, Ibaraki 305-8577, Japan \\
$^{d}$High Energy Accelerator Research Organization
      (KEK), Tsukuba, Ibaraki 305-0801, Japan \\
$^{e}$Department of Physics,
      Hiroshima University, 
      Higashi-Hiroshima, Hiroshima 739-8526, Japan 
}

\date{\today}


\begin{abstract}
   We extend the study of the light hadron spectrum
   and the quark mass in two-flavor QCD
   to smaller sea quark mass,
   corresponding to $m_{PS}/m_{V}=0.60$--0.35.
   Numerical simulations are carried out using the RG-improved
   gauge action and the meanfield-improved clover quark action
   at $\beta=1.8$ ($a = 0.2$~fm from $\rho$ meson mass).
   We observe that the light hadron spectrum for small sea
   quark mass
   does not follow the expectation from chiral extrapolations
   with quadratic functions made from the region of
   $m_{PS}/m_{V}=0.80$--0.55.
   Whereas fits with either polynomial or continuum
   chiral perturbation theory (ChPT) fails, the Wilson ChPT
   (WChPT) that includes $a^2$ effects associated with explicit
   chiral symmetry breaking successfully fits the whole data:
   In particular, WChPT correctly predicts the light quark mass
   spectrum from simulations for medium heavy quark mass, such as
   $m_{PS}/m_V \simgt 0.5$.
   Reanalyzing the previous data 
   with the use of WChPT,
   we find the mean up and down quark mass being smaller
   than the previous result from quadratic chiral extrapolation
   by approximately 10\%,
   $m_{ud}^{\overline{\rm MS}}(\mu=2~\mbox{GeV}) = 3.11(17)$~[MeV]
   in the continuum limit.
\end{abstract}


\maketitle


\section{Introduction}
\label{section:introduction}

Recent years have witnessed steady progress in the lattice 
QCD calculation of 
the light hadron spectrum~\cite{plenary_lattice}. 
In the quenched approximation
ignoring quark vacuum polarization effects,
well-controlled
chiral and continuum extrapolations 
enabled a calculation of hadron masses 
with an accuracy of 0.5--3\%~\cite{quench.CP-PACS}.
At the same time the study established a systematic deviation of the 
quenched light hadron spectrum from experiment
by approximately 10\%.
We then have made an attempt of full QCD calculation 
that allows chiral and continuum extrapolations within 
a consistent set of simulations~\cite{Spectrum.Nf2.CP-PACS}.
The deviations from experiment in the 
light hadron spectrum are significantly 
reduced and the light quark mass decreases by about 25\%
with the inclusion of dynamical $u$ and $d$ quarks. 
With currently available computer power and simulation algorithms,
however, the sea quark mass that can be explored is  
far from the physical value and a long chiral
extrapolation is involved to get to the physical $u$ and $d$ 
quark mass.

An attempt has been made to push down the simulation 
to a small quark mass
corresponding to $m_{PS}/m_{V} \approx 0.3$ in 
full QCD with the Kogut-Susskind(staggered)-type 
quark action~\cite{Spectrum.Nf2.MILC}.
The staggered action, however, poses a problem of flavor mixing, 
which would modify the hadron spectrum and its quark mass dependence 
near the chiral limit.
The staggered action also suffers from 
ambiguities in hadron
operators and has a potential problem of non-locality.
The Wilson-type quark actions have the advantage of simplicity:
they are local and respect flavor symmetry,
but a larger computational cost limits
the simulations to relatively large quark masses
corresponding to
$m_{PS}/m_{V} \simgt 0.6$~\cite{Spectrum.Nf2.SESAM,
Spectrum.Nf2.TchiL,Spectrum.Nf2.SESAM-TchiL,
Spectrum.Nf2.UKQCD.csw176,Spectrum.Nf2.UKQCD,
Spectrum.Nf2.CP-PACS,
Spectrum.Nf2.QCDSF,Spectrum.Nf2.JLQCD}.
An important problem is to examine whether chiral extrapolations
from such a quark mass range lead to results viable
in the chiral limit.

Chiral extrapolations are usually made
with polynomials in the quark mass.
The problem is that they are not consistent with the logarithmic 
singularity expected in the chiral limit.  
In reality, the physical quarks are not exactly massless
and hence the polynomial extrapolation should in principle 
work.
However, increasingly higher orders are needed should one wish to 
increase the accuracy of the extrapolation.  
It is compelling to estimate the systematic errors
due to higher order contributions 
when the data are extrapolated using a low-order polynomial.

An alternative choice for chiral extrapolations is to 
incorporate chiral perturbation theory 
(ChPT)~\cite{ChPT.Gasser}.
The present lattice data, however, are not quite consistent with 
the ChPT predictions.  
The high-statistics JLQCD simulation of
two-flavor full QCD, using the plaquette gauge action
and the $O(a)$-improved Wilson quark action at $\beta=5.2$
($a=0.0887(11)$~fm; the spatial size $L \simeq 1.06$--1.77~fm), 
shows no signature for the logarithmic singularity
in the pion mass and pion decay 
constant~\cite{Spectrum.Nf2.JLQCD}.
%
%
A possible reason for the failure to find the chiral logarithm is
that sea quark masses, 
corresponding to $m_{PS}/m_V = 0.8$--0.6, 
are too large.  Higher order corrections of ChPT may have to be included
to describe the data, as suggested from 
a partially quenched analysis, which shows that 
$m_{PS}/m_V = 0.4$--0.3 is required for the convergence of one-loop 
formula~\cite{PQChPT.Sharpe,PQChPT.Durr}.
%
%
%
Another possibility is explicit chiral symmetry breaking
of the Wilson quark actions that may invalidate the
ChPT formulae. Modifications due to finite lattice spacings
may be needed for an analysis of data obtained on a coarse lattice.

Recently studies were made to adapt 
ChPT to the Wilson-type fermion at finite lattice spacings (WChPT)
~\cite{WChPT.Sharpe,WChPT.Rupak,WChPT.Oliver,WChPT.Sinya},
with subtle differences in the order counting, and hence the resulting
formulae for observables, among the authors. 
The work~\cite{WChPT.Rupak} assumes the $O(a)$ chiral symmetry
breaking effects being smaller than
those from the quark mass,
and only the effects linear in lattice spacing are retained
in the chiral Lagrangian.
This contrasts to the authors of Refs.~\cite{WChPT.Oliver,WChPT.Sinya}
who include the $O(a^2)$ effects 
in the chiral Lagrangian,
however, with different order countings.
In Ref.~\cite{WChPT.Oliver} the $O(a)$ terms
are treated as being comparable to the quark mass term
while the $O(a^2)$ terms are assumed to be subleading:
in this case, $O(a)$ effects are essentially absorbed into
the redefinition of the quark mass in the one-loop formulae
and the $O(a^2)$ terms provide additional counter terms.
In Ref.~\cite{WChPT.Sinya}, on the other hand,
the terms of $O(a^2)$ are kept at the leading order,
because the existence 
of parity-broken phase and vanishing of pion mass depend on them in a 
critical way~\cite{WChPT.Sharpe}.
The coefficients 
of chiral logarithm terms receive $O(a)$ contributions,
and hence 
the logarithmic chiral behavior is modified at a finite lattice spacing. 
Similar attempts to include the $O(a^2)$ flavor mixing
for the staggered-type quark action 
were made in Refs.~\cite{SChPT.Lee,SChPT.Bernard,SChPT.Aubin}.

%
%
The qq+q collaboration~\cite{qq+q} applied
the one-loop ChPT and WChPT with the prescription of 
Refs.~\cite{WChPT.Rupak,WChPT.Oliver}
to their data obtained at $m_{PS}/m_{V}=0.9$--0.5.
Their simulations were made at coarse lattices of 
$a = 0.19$~fm $(\beta=5.1)$ 
and $0.28$~fm $(\beta=4.68)$ 
using the plaquette gauge action and
the unimproved Wilson quark action ($L \approx 3$~fm).
They reported that their data are described by these formulae.
However, their sea quark masses are not quite small, and, 
since large scaling violation is suspected
with unimproved actions at 
coarse lattice spacings and lattice artifacts are suggested
at strong couplings~\cite{lattice_artifact.MILC},
it should be demonstrated at weaker couplings 
in order that the discretization effects are actually under control.
The UKQCD collaboration reported a result at 
$m_{PS}/m_{V}=0.44(2)$ obtained with the actions and
the lattice spacing the same as those of JLQCD, 
with $L \approx 1.6$~fm~\cite{Spectrum.Nf2.UKQCD_light}.
They indicated the pion decay constant to bend slightly 
downward at this quark mass, but further work is required for 
quantitative comparison with the ChPT predictions.

%
%
In this paper, we follow up on our previous two-flavor full 
QCD work~\cite{Spectrum.Nf2.CP-PACS} with an RG-improved gauge action 
and tadpole-improved $O(a)$-improved Wilson-clover quark action 
at $m_{PS}/m_V = 0.80$--0.55 and attempt to lower the quark mass
to give  $m_{PS}/m_V$ down to 0.35.
Since the computational costs grows rapidly toward the chiral limit, 
roughly proportional to $(m_{PS}/m_{V})^{-6}$~\cite{cost.Ukawa},
we concentrate our effort on the coarsest lattice of $a \approx  0.2$~fm 
at $\beta=1.8$, while using improved actions.

Generation of configurations below $m_{PS}/m_V\approx 0.5$ demands 
technical improvements. The BiCGStab 
algorithm sometimes fails to converge, which we overcome by an 
improvement called BiCGStab(DS-$L$)~\cite{BiCGStab_DSL,BiCGStab_IDSL.Itoh}. 
Another problem is the emergence of instabilities 
in the HMC molecular dynamics evolution~\cite{spike.Jansen,spike.UKQCD}.  
This seems to be caused by very small eigenvalues of the 
Dirac operator, leading to the change of the molecular dynamics 
orbit from 
elliptic to hyperbolic.
The only resolution at present is to reduce the time step size. 
In this manner, we generated
4000 trajectories at $m_{PS}/m_V\approx 0.6$, 0.5 and 0.4
and 1400 trajectories at the smallest quark mass of 
$m_{PS}/m_V\approx 0.35$
on a $12^3 \times 24$ lattice with $L \approx 2.4$~fm.
To examine the finite-size effect,
we also generated 2000 trajectories at $m_{PS}/m_V\approx 0.6$ and 0.5
on a $16^3 \times 24$ lattice with $L \approx 3.2$~fm.

We calculate the light hadron spectrum and the quark mass on
these configurations, and examine the validity of the quadratic 
chiral extrapolations by comparing the extrapolations
made in the previous work with our new data at smaller quark masses.
It turns out that the new data are increasingly lower than 
the extrapolation toward a smaller sea quark mass. 
We then examine how our data compare with the WChPT formulae,  
and whether WChPT fits using only the previous data at large 
quark masses predict correctly the new small quark mass data.   
This serves as a test to verify the viability of WChPT  
and of chiral extrapolations.

Computing for the present work was made on the VPP5000/80 at the 
Information Processing Center of University of Tsukuba.  
We used 4 or 8 nodes, each node having the 
peak speed of 9.6~Gflops.
The present simulation costed
0.119~Tflops$\cdot$years of computing time measured in terms 
of the peak speed.

This paper is organized as follows.
We describe configuration generations 
in Sec.~\ref{section:simulation}.
The method of measurement of hadron masses, decay constants,
quark masses and the static quark potential
is explained in Sec.~\ref{section:measurement}.
The finite-size effects on hadron masses are also discussed
in the same section.
Sec.~\ref{section:chiral_extrapolation_polynomials}
discusses chiral extrapolations with conventional polynomials,
and those based on ChPT are
presented in Sec.~\ref{section:chiral_extrapolation_ChPT}.
Our conclusion is given in Sec.~\ref{section:conclusion}.
Preliminary results of these calculations were reported
in Ref.~\cite{Spectrum.Nf2.CPPACS.lat02.Namekawa}.


\section{Simulation}
\label{section:simulation}

%
%
For the gauge part
we employ the RG improved action defined by 
\begin{equation}
S_g = \frac{\beta}{6}
      \left\{ c_0 \sum_{x,\mu\nu} W_{\mu\nu}^{1\times 1}(x) 
             +c_1 \sum_{x,\mu\nu} W_{\mu\nu}^{1\times 2}(x)
      \right\}.
\label{equaton:RGaction} 
\end{equation}
The coefficients $c_0 = 3.648$ of the $1\times 1$ Wilson loop
and $c_1 = -0.331$ of the $1\times2$ Wilson loop 
are determined by an approximate renormalization group 
analysis~\cite{RGaction}.  They satisfy the normalization condition 
$c_0+8c_1=1$, and $\beta=6/g^2$.
%
%
For the quark part we use the clover quark action~\cite{SWaction}
defined by  
\begin{eqnarray}
S_q & = & \sum_{x,y}\overline{q}_x D_{x,y}q_y,\\
D_{x,y} & = & \delta_{xy}
- \kappa \sum_\mu \left\{(1-\gamma_\mu)U_{x,\mu} \delta_{x+\hat\mu,y}
      + (1+\gamma_\mu)U_{x,\mu}^{\dag} \delta_{x,y+\hat\mu} \right\}
- \, \delta_{xy} c_{\rm SW} \kappa \sum_{\mu < \nu}
         \sigma_{\mu\nu} F_{\mu\nu},
\label{equation:clover}
\end{eqnarray}
where $\kappa$ is the hopping parameter, $F_{\mu\nu}$ the standard 
clover-shaped lattice discretization of the field strength
and $\sigma_{\mu\nu} = (i/2)[\gamma_{\mu},\gamma_{\nu}]$.
For the clover coefficient 
we adopt a meanfield improved value
$c_{\rm SW} = u_0^{-3}$~\cite{tadpole_improvement} 
where
\begin{equation}
 u_0 = \left ( W^{1\times 1} \right )^{1/4} 
        = \left (1-0.8412 \beta^{-1} \right )^{1/4},
\label{equation:u0} 
\end{equation}
using the plaquette $W^{1\times 1}$ calculated
in one-loop perturbation theory~\cite{RGaction}.
This choice is based on our observation
that the one-loop calculation
reproduces the measured values well~\cite{Comparative.Nf2.CP-PACS}.

Our simulation is performed at a single value of $\beta=1.8$ using 
two lattice sizes $12^3\times 24$ and $16^3\times 24$ to study  
finite size effects. The lattice spacing fixed from $m_{\rho}$
at the physical sea quark mass is 0.2~fm.
We adopt four values of the sea quark mass 
corresponding to the hopping parameter
$\kappa_{sea}=0.14585$, $0.14660$, $0.14705$ and $0.14720$.
This choice covers $m_{PS}/m_{V}=0.60$--0.35, extending 
the four values $\kappa_{sea}=0.1409$, 0.1430, 0.1445, and 0.1464
corresponding 
to $m_{PS}/m_V=0.80$--0.55 studied in Ref.~\cite{Spectrum.Nf2.CP-PACS}.
The simulation parameters are summarized 
in Table~\ref{table:simulation_parameters}, where we also list the 
number of nodes (PE's) employed and the CPU time per trajectory.
%
%
Gauge configurations are generated using 
the Hybrid Monte Carlo (HMC) algorithm~\cite{HMC.Duane,HMC.Gottlieb}.
The trajectory length in each HMC step 
is fixed to unity.
We use the leap-frog integration scheme 
for the molecular dynamics equation.

\begin{figure}[h]
    \includegraphics[width=75mm]
   {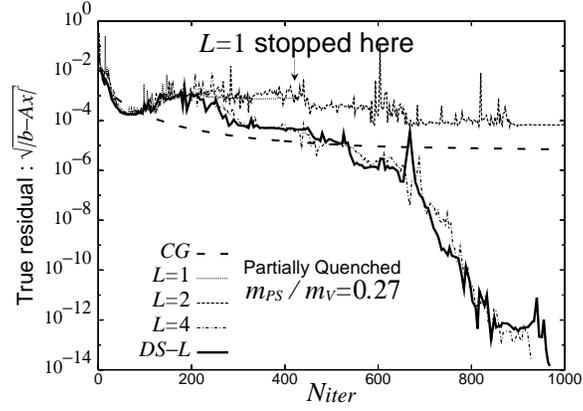}
   \caption{
      Comparison of convergence with various algorithms
      for inversions of the quark matrix
      at $\kappa_{sea}=0.14585$ ($m_{PS}/m_V=0.60$)
      and $\kappa_{valence}=0.14850$ ($m_{PS}/m_V=0.27$)
      on $12^3 \times 24$ lattice.
      Conventional BiCGStab corresponds to $L=1$.
      We define an iteration $N_{iter}$ as a dimension of
      the Krylov subspace to which approximate solutions
      belong~\cite{BiCGStab_L.Sleijpen}.
      The number of matrix-vector products
      to obtain an approximate solution
      is $2 \times N_{iter}$.
   }
   \label{figure:comparison_of_convergence-beta_1.80-kappa_0.14585}
\end{figure}

\begin{figure}[h]
  \includegraphics[width=75mm]
   {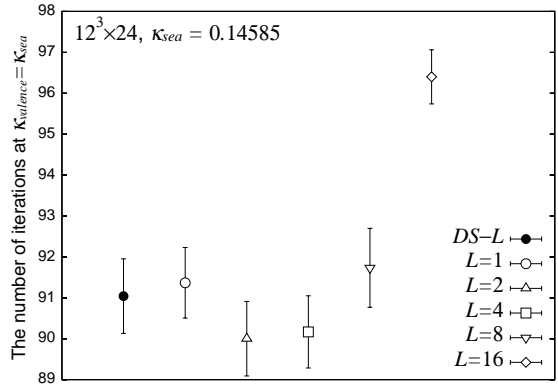}
   \caption{
      The number of iterations required with BiCGStab($L$)
      and BiCGStab(DS-$L$)
      for inversions of the quark matrix
      at $\kappa_{sea}=\kappa_{valence}=0.14585$ ($m_{PS}/m_V=0.60$)
      on $12^3 \times 24$ lattice.
      Conventional BiCGStab corresponds to $L=1$.
  }
   \label{figure:comparison_of_iterations-all-RC-beta_1.80-kappa_0.14585}
\end{figure}

%
The even/odd preconditioned BiCGStab~\cite{BiCGStab}
is one of
the most optimized  algorithms for the Wilson quark matrix inversion
to solve the equation $D_{xy}G_{y}=B_{x}$.
However, BiCGStab sometimes
fails to converge at small sea quark masses.
While the CG algorithm is guaranteed to converge,
it is time-consuming.
We find that the BiCGStab($L$) algorithm~\cite{BiCGStab_L.Sleijpen}, 
which is an extension of BiCGStab to $L$-th order minimal residual 
polynomials, is more stable~\cite{BiCGStab_IDSL.Itoh}.
Figure~\ref{figure:comparison_of_convergence-beta_1.80-kappa_0.14585}
illustrates for a very light valence quark mass corresponding to 
$m_{PS}/m_V=0.27$ that the BiCGStab($L$), while not convergent for 
$L=1$ and 2, succeeds to find the solution for $L=4$.
In practice, however, too large $L$ also 
frequently introduces another instability
from possible loss of conjugacy among the $L$ vectors. 
The optimum value of $L$ depends on simulation parameters. 
To avoid a tuning of $L$ at each simulation point, 
we employ the BiCGStab(DS-$L$) algorithm~\cite{BiCGStab_DSL}.
This is a modified BiCGStab($L$) in which a candidate of the optimum $L$ 
is dynamically selected. 
We find that BiCGStab(DS-$L$) is much more robust than
the original BiCGStab at small quark masses.
We also find that, at large quark masses
where the conventional BiCGStab converges, 
the computer time required for BiCGStab(DS-$L$) is comparable.
See Fig.~\ref{figure:comparison_of_iterations-all-RC-beta_1.80-kappa_0.14585}.
Therefore, we adopt BiCGStab(DS-$L$) at all values of our sea quark masses.

\begin{figure}[h]
   \includegraphics[width=75mm]
   {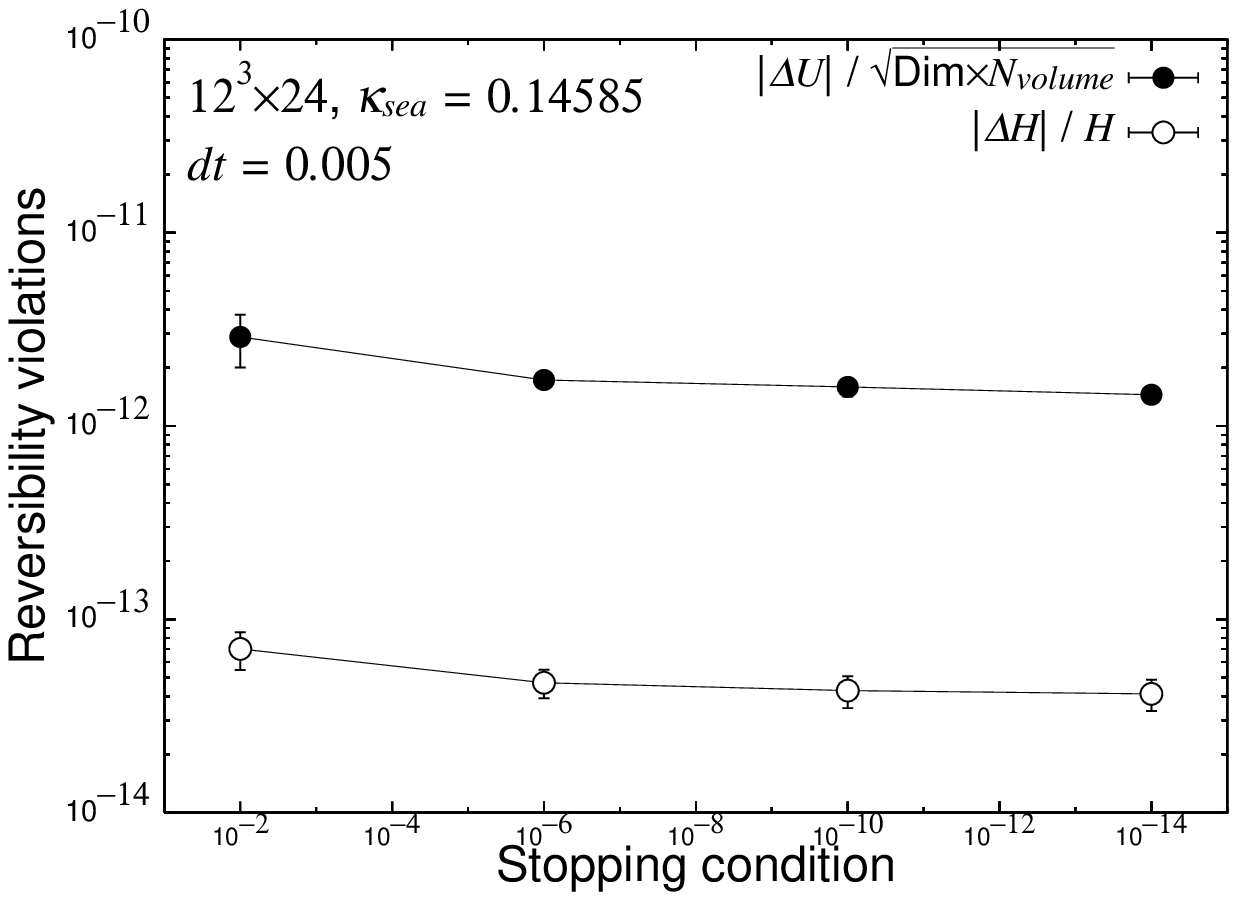}
   \includegraphics[width=75mm]
   {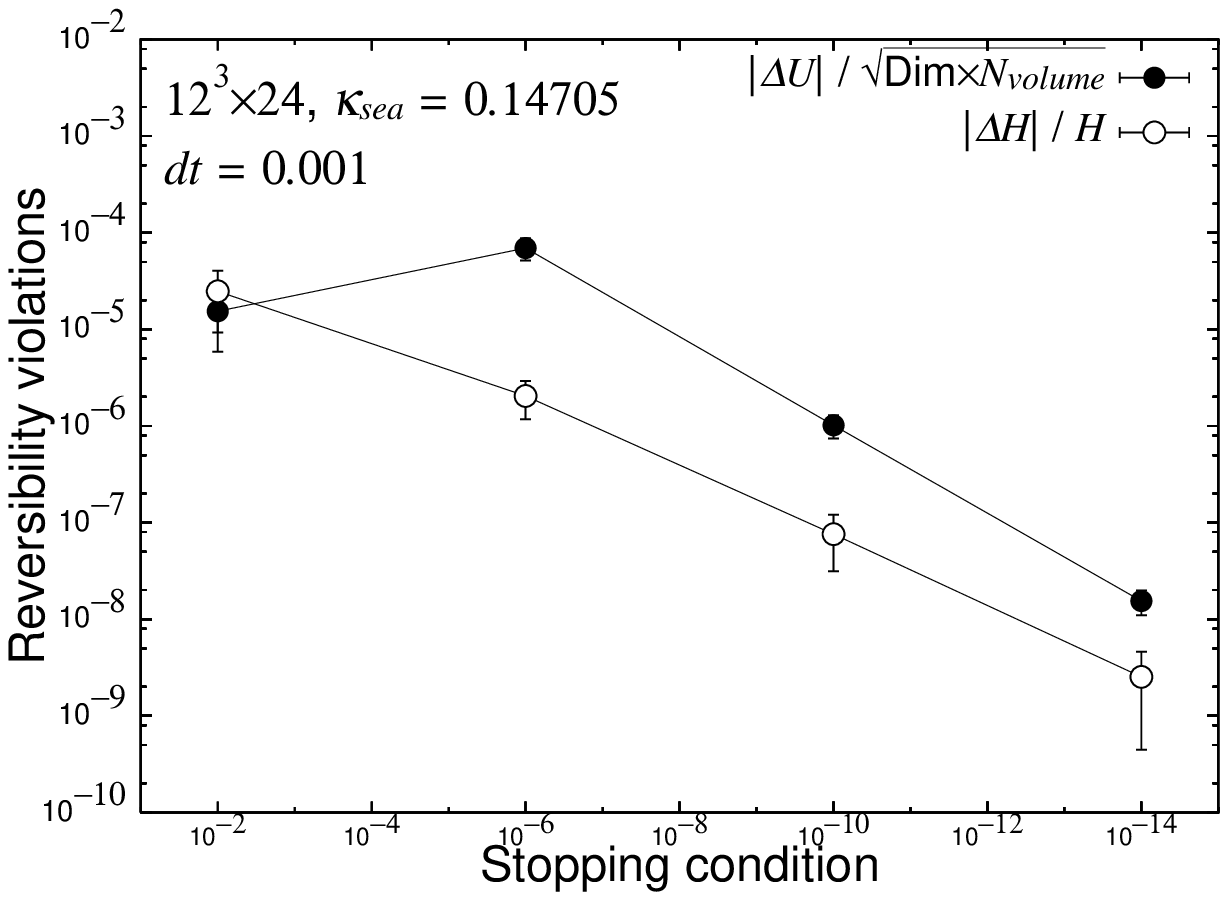}
   \caption{
      Reversibility violation
      at large sea quark mass of $\kappa_{sea}=0.14585$ ($m_{PS}/m_V=0.60$)
      (left panel)
      and small sea quark mass of $\kappa_{sea}=0.14705$ ($m_{PS}/m_V=0.40$)
      (right panel)
      on $12^3 \times 24$ lattice.
   }
   \label{figure:reversibility_violation}
\end{figure}

%
%
We employ the stopping condition $|| DG - B || < \Delta$ 
in HMC.
The value of $\Delta$ in the evaluation of the fermionic force
is chosen so that the reversibility over 
unit length is satisfied to a relative precision of
order $10^{-8}$ or smaller for the Hamiltonian,
\begin{equation}
 |\Delta H| = |H_{reversed} - H_{0}|,
\end{equation}
where $H_{reversed}$ is the value of the Hamiltonian
obtained by integrating to $t=1$ and integrating back to $t=0$.
We also check the reversibility violation
in the link variable,
\begin{equation}
 |\Delta U| = \sqrt{ \sum_{n,\mu,a,b} U_{\mu,a,b}^{reversed}(n)
                                    - U_{\mu,a,b}^{0}(n) },
\end{equation}
where the sum is taken over all sites $n$, colors $a,b$
and the link directions $\mu$.
We illustrate our check in Fig.~\ref{figure:reversibility_violation},
where results at $\kappa_{sea} = 0.14585$ and $\kappa_{sea} = 0.14705$
on 20 thermalized configurations separated by 100 trajectories
are shown.
When the sea quark mass is large ($\kappa_{sea} = 0.14585, m_{PS}/m_V=0.6$),
the violation does not show any clear dependence
on the stopping condition.
For small sea quark mass  
($\kappa_{sea} = 0.14705$, $m_{PS}/m_V=0.4$), however,
it depends on the stopping condition significantly.
We must be careful with the choice of the stopping condition
at small sea quark mass.
We use a stricter stopping condition
in the calculation of the Hamiltonian 
in the Metropolis accept/reject test.
Table~\ref{table:simulation_parameters} shows 
our choice of $\Delta$ 
together with the average number, $N_{inv}$, of the 
BiCGStab(DS-$L$) iterations
in the quark matrix inversion for
the force calculation.

\begin{figure}[h]
   \includegraphics[width=75mm]
   {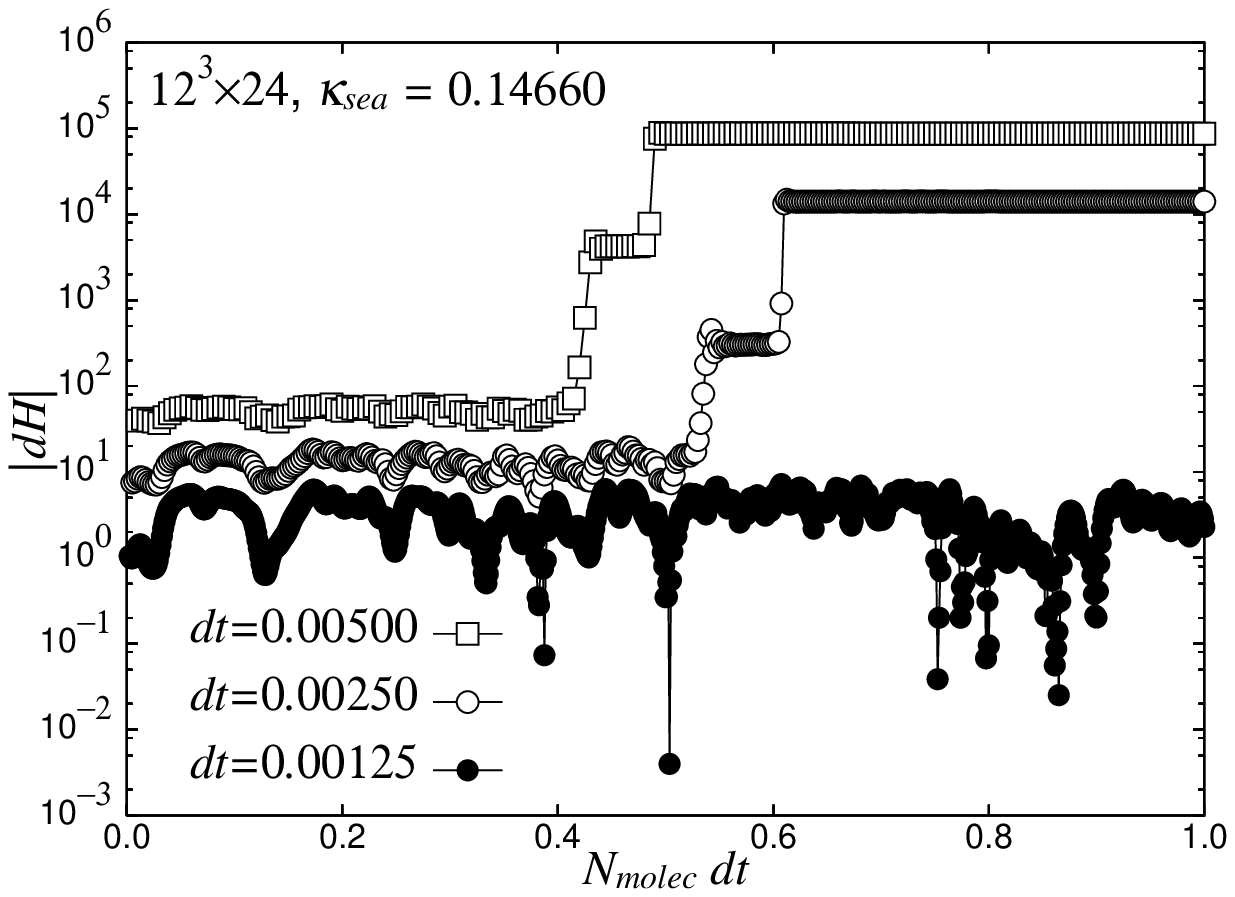}
   \includegraphics[width=75mm]
   {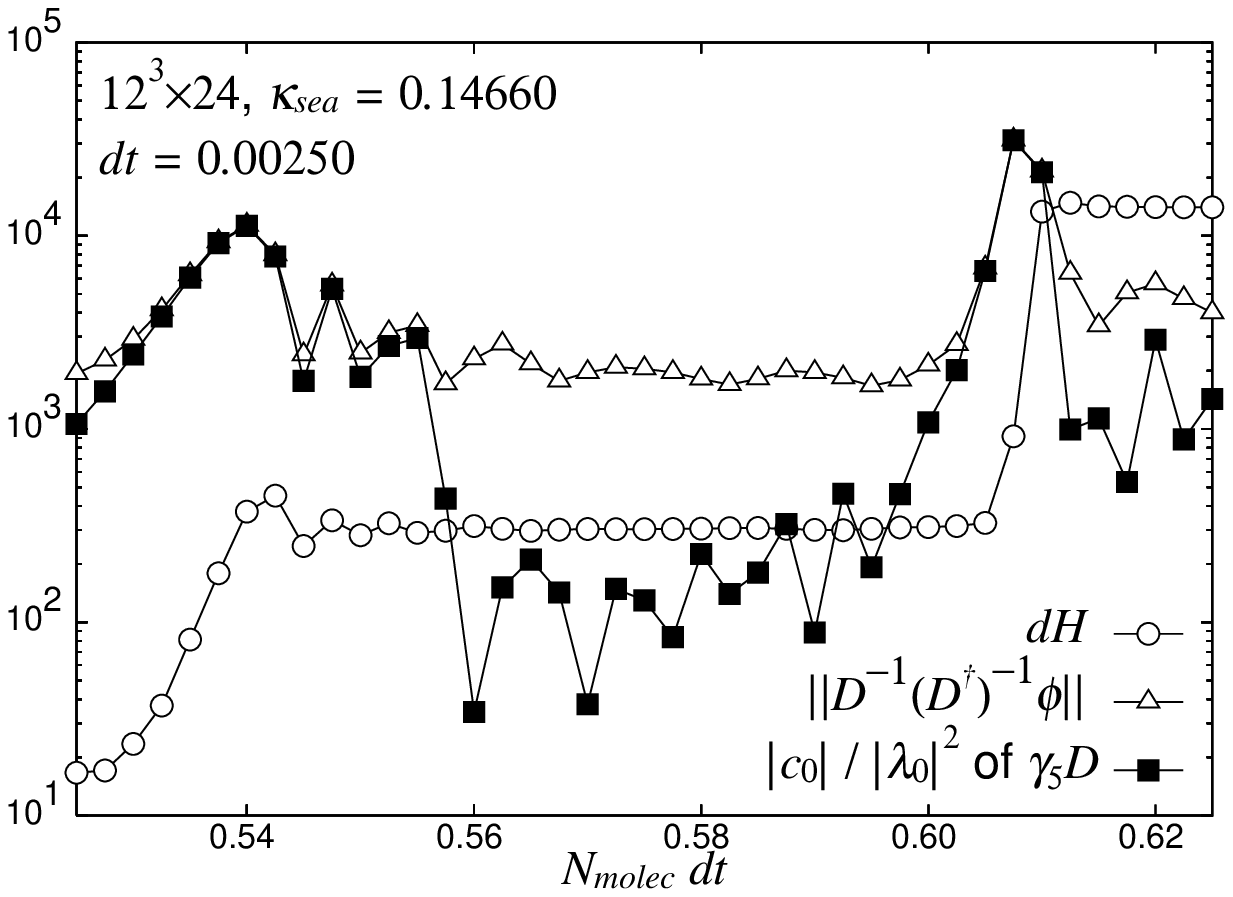}
   \caption{
      Effect of the molecular dynamics step size $dt$
      on the appearance of spikes in $dH \equiv H_{trial} - H_{0}$
      at $\kappa_{sea}=0.14660$ ($m_{PS}/m_{V} = 0.50$)
      on $12^3 \times 24$ lattice (left panel).
      The right panel is an enlargement around the spikes
      in the case of $dt = 0.0025$.
      $||D^{-1} (D^{\dagger})^{-1} \phi||$
      with the Wilson-clover operator $D$ and the pseudofermion field $\phi$
      as well as
      the corresponding contribution 
      with the smallest eigenvalue $\lambda_0$
      and its overlap $c_0 \equiv (x_0,\phi)$ for $\gamma_5 D$
      are also plotted,
      where $x_i$ is an eigenfunction of $\gamma_5 D$
      such that $\phi = \sum_{i} c_i x_i$.
   }
   \label{figure:dt_dependence_of_spike}
\end{figure}

In the course of configuration generation by the HMC algorithm,
we sometimes encountered extremely large values of
$dH \equiv H_{trial} - H_{0}$,
the difference of the trial and starting Hamiltonians.  
Similar experiences have been reported by other 
groups~\cite{spike.Jansen,spike.UKQCD}.
Empirically this phenomenon occurs more frequently for smaller sea quark 
masses at a fixed step size, and can be suppressed by decreasing the 
step size.
A typical example is shown in Fig.~\ref{figure:dt_dependence_of_spike}.
In our runs we employ a step size $dt$ small enough 
for this purpose.  As a consequence our runs have a 
rather high acceptance 80--90\%.
It is possible that this phenomenon is connected to the appearance of very 
small eigenvalues of the Wilson-clover operator toward small quark masses. 
In the right panel of Fig.~\ref{figure:dt_dependence_of_spike},
we show the norm $\vert\vert D^{-1}(D^\dagger)^{-1}\phi\vert\vert$ 
(triangles) and the contribution  of the smallest eigenvalue 
of $\gamma_5 D$ to the norm (filled squares). 
We observe that the jump of $dH$ (open circles) is associated with 
a peak of the norm, and that the peak is saturated by the contribution of 
the smallest eigenvalue. 
We suspect that such small eigenvalues cause some modes of the 
HMC molecular dynamics evolution to change its character from elliptic 
to hyperbolic, leading to divergence of the Hamiltonian.
We defer a further study of this problem to future publications.

We accumulate 4000 HMC trajectories
at $\kappa_{sea}=0.14585$, $0.14660$ and $0.14705$
and 1400 trajectories at $\kappa_{sea}=0.14720$
on the $12^3 \times 24$ lattice.
We also accumulate 2000 trajectories
at $\kappa_{sea}=0.14585$ and $0.14660$
on the $16^3 \times 24$ lattice.
Measurements of light hadron masses and the static quark 
potential are carried out at every 5 trajectories.

\section{Measurement}
\label{section:measurement}

\subsection{Hadron masses}
\label{section:measurement:hadron}

The meson operators are defined by 
\begin{equation}
   M(x) = \bar{q}^{(f)}(x) \Gamma q^{(g)}(x), \hspace{5mm}
   \Gamma = I, \gamma_5, \gamma_{\mu}, 
            \gamma_5 \gamma_{\mu},
    \label{equation:meson}
\end{equation}
where $f$ and $g$ are flavor indices and 
$x$ is the coordinates on the lattice.
%
%
The octet baryon operator is defined as
\begin{equation}
   O^{fgh}(x)
 = \epsilon^{abc}
   \left( q^{(f)a}(x) {}^{T}C \gamma_5 q^{(g)b}(x) \right) 
   q^{(h)c}(x),
   \label{equation:octet_baryon}
\end{equation}
where $a, b, c$ are color indices and 
$C = \gamma_4 \gamma_2$ is the charge conjugation matrix.
Decuplet baryon correlators are calculated 
using an operator defined by 
\begin{equation}
   D^{fgh}_{\mu}(x)
 = \epsilon^{abc}
   \left( q^{(f)a}(x) {}^{T}C \gamma_{\mu} q^{(g)b}(x) \right) 
   q^{(h)c}(x).
   \label{equation:decuplet_baryon}
\end{equation}

For each configuration
quark propagators are calculated
with a point and a smeared source.
For the smeared source,
we fix the gauge configuration to the Coulomb gauge
and use an exponential smearing function
$\psi(r) = A \exp(-Br)$ for $r>0$ with $\psi(0)=1$.
We chose $A=1.25$ and $B=0.50$
as in our previous study~\cite{Spectrum.Nf2.CP-PACS}.
%
%
In order to reduce the statistical fluctuation of hadron correlators,
we repeat the measurement for two choices of 
the location of the hadron source,
$t_{src} = 1$ and $N_t/2 +1(=13)$
and take the average over the two~\cite{Spectrum.Nf2.JLQCD}:
\begin{equation}
   \frac{1}{2} 
       \left( \left\langle H(t_{src}+t)
                           H(t_{src})^{\dagger} 
              \right\rangle_{t_{src}=1}
             +\left\langle H(t_{src}+t)
                           H(t_{src})^{\dagger} 
              \right\rangle_{t_{src}=N_t/2 +1}
       \right).
   \label{equation:source_shift}
\end{equation}
This procedure reduces
the statistical error of hadron correlators typically by 30 to 40\%,
which suggests that the statistics are increased effectively
by a factor of 1.7 to 2.
For further reduction of the statistical fluctuation,
we take the average over three polarization states
for vector mesons, two spin states for octet baryons
and four spin states for decuplet baryons.

\begin{figure}[h]
   \includegraphics[width=75mm]
   {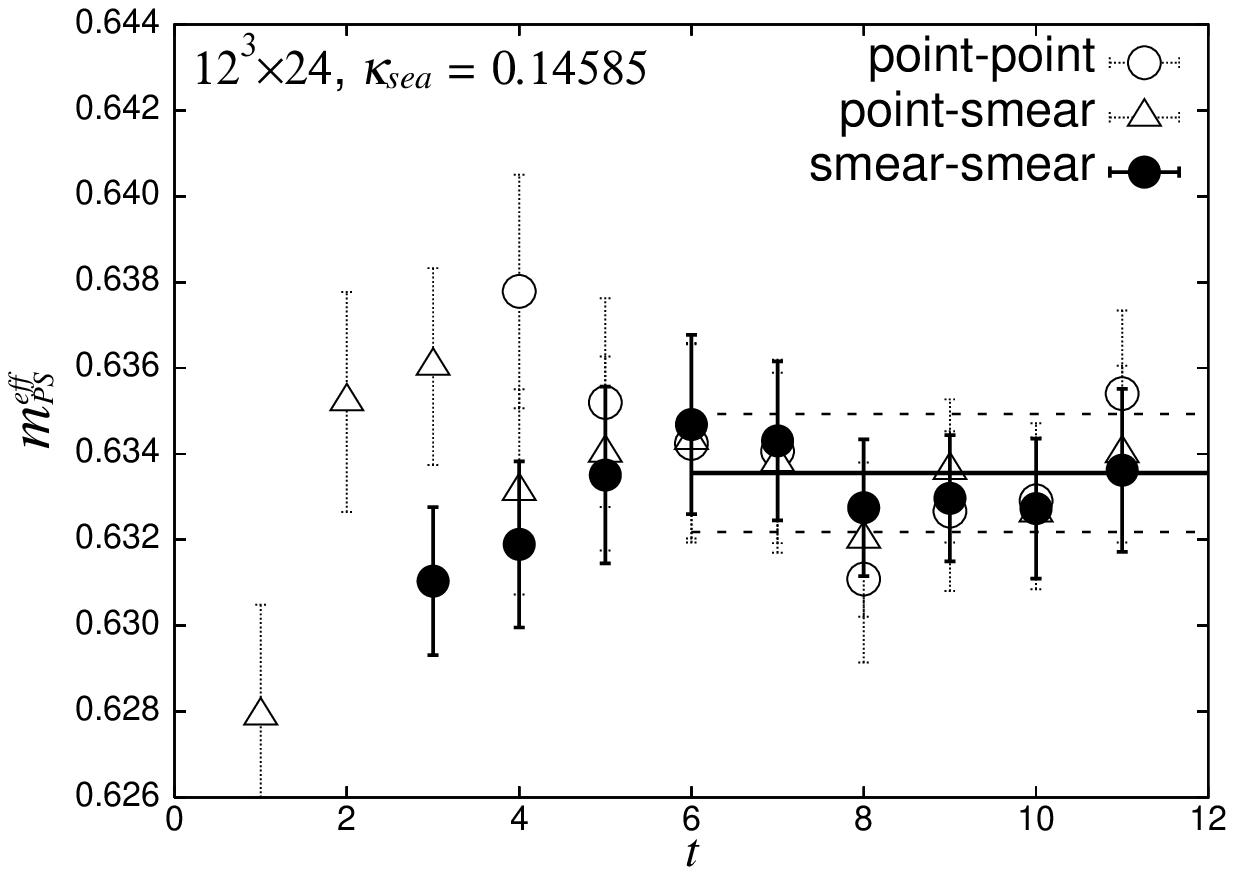}
   \includegraphics[width=75mm]
   {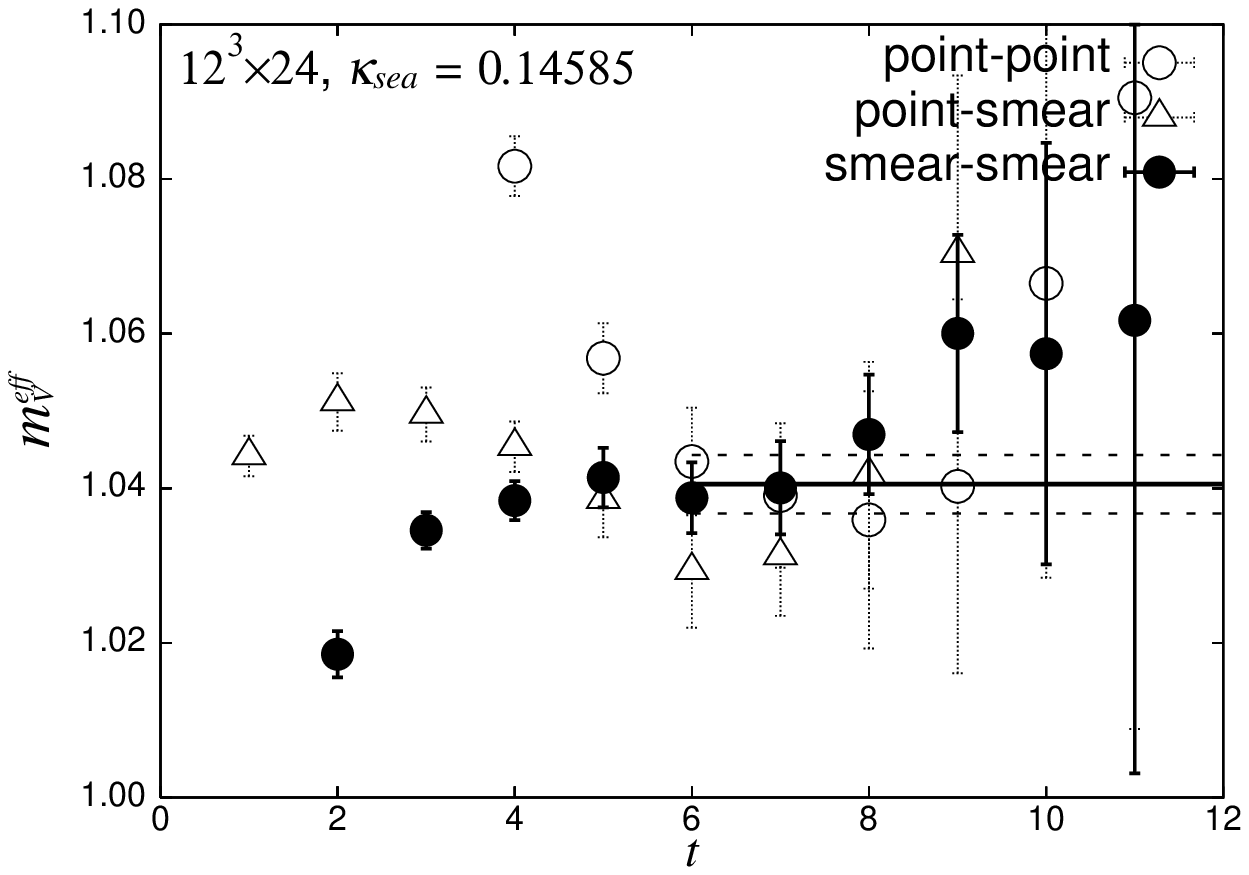}
   \caption{
      Effective masses of pseudoscalar (left panel)
      and vector meson (right panel)
      at $\kappa_{sea}=0.14585$ ($m_{PS}/m_{V} = 0.60$)
      on $12^3 \times 24$ lattice.
   }
   \label{figure:m_eff_PS_V-12x24_kappa_sea_0.14585}
\end{figure}

\begin{figure}[h]
   \includegraphics[width=75mm]
   {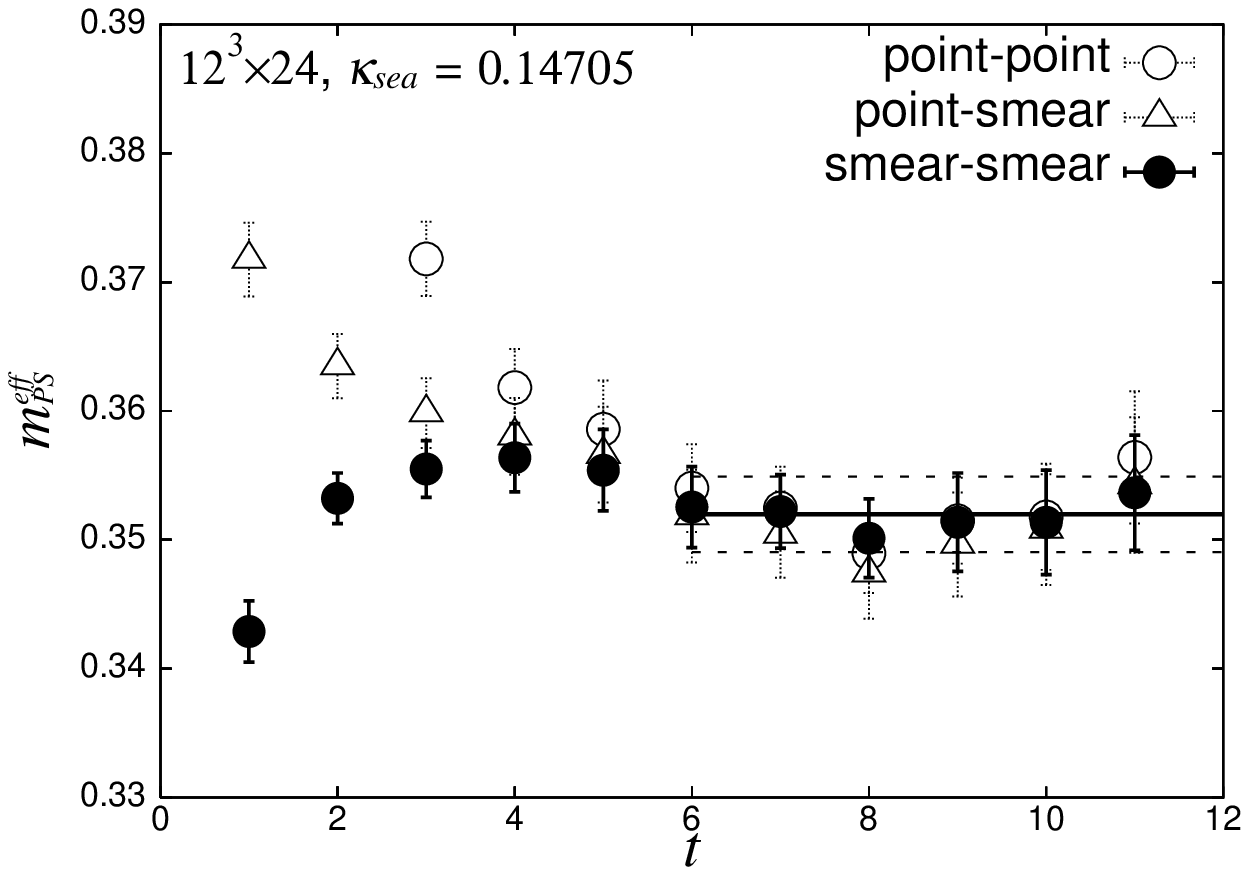}
   \includegraphics[width=75mm]
   {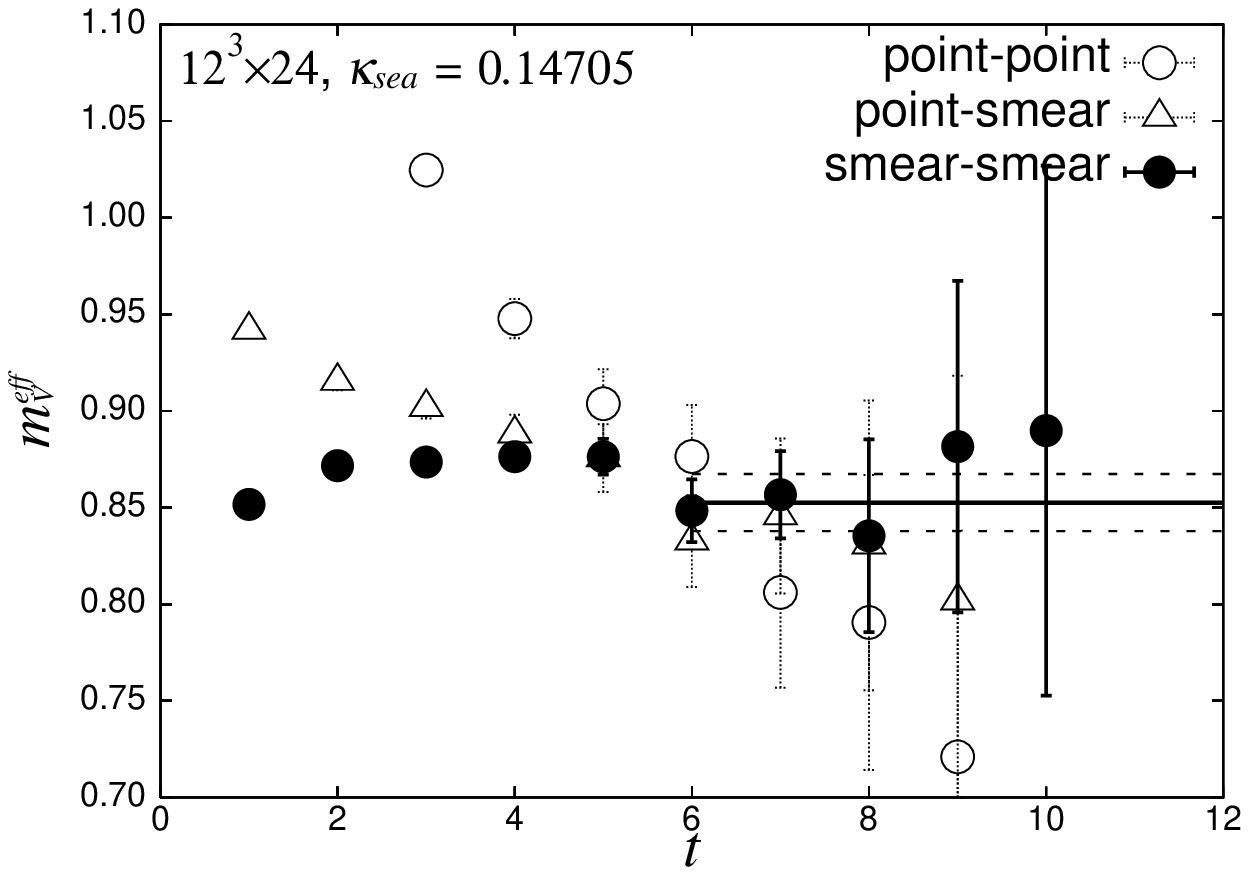}
   \caption{
      Effective masses of pseudoscalar (left panel)
      and vector meson (right panel)
      at $\kappa_{sea}=0.14705$ ($m_{PS}/m_{V} = 0.40$)
      on $12^3 \times 24$ lattice.
   }
   \label{figure:m_eff_PS_V-12x24_kappa_sea_0.14705}
\end{figure}

Figures~\ref{figure:m_eff_PS_V-12x24_kappa_sea_0.14585}
and \ref{figure:m_eff_PS_V-12x24_kappa_sea_0.14705}
illustrate the quality
of effective mass plots.
For mesons, an acceptable plateau of the effective mass is obtained 
from hadron correlators with the point sink 
and the doubly smeared source.
Signals are much worse for baryons.

We carry out $\chi^2$ fits to hadron correlators,
taking account of correlations among different time slices.
A single hyperbolic cosine form is assumed for mesons,
and a single exponential form for baryons.
We set the lower cut of the fitting range as $t_{min}=6$ for mesons
and $t_{min}=5$ for baryons,
which is determined by inspecting stability of the resulting mass.
The upper cut ($t_{max}$) dependence of the fit  
is small and, therefore, we fix $t_{max}$ to $N_t/2$
for all hadrons.
Our choice of fit ranges and the detailed
results of hadron masses are given in tables of
Appendix~\ref{section:appendix_m_hadron}.
Statistical errors of hadron masses are estimated 
with the jack-knife procedure.
We adopt the bin size of 100 trajectories from an analysis of
the bin size dependence of errors as discussed 
below in Sec.~\ref{section:autocorrelation}.

\subsection{Quark masses}
\label{section:quark_mass}

We calculate the mean up and down quark mass through
both vector
and axial-vector Ward identities. 
The two types of quark masses,
denoted by $m_{VWI}$ and $m_{AWI}$ respectively,
differ at finite lattice spacings 
because of explicit violation of chiral symmetry
by the Wilson term.

%
%
A bare VWI quark mass is defined by
\begin{equation}
 m_{quark}^{VWI} = \frac{1}{2}
                   \left( \frac{1}{\kappa} - \frac{1}{\kappa_c} \right).
 \label{equation:m_quark_VWI}
\end{equation}
The critical hopping parameter $\kappa_c$ is determined
by chiral extrapolations as discussed
in Sec.~\ref{section:chiral_extrapolation_polynomials}
and \ref{section:chiral_extrapolation_ChPT}.
%
%
A bare AWI quark mass is calculated
using the fourth component of the improved axial-vector current
\begin{equation}
   A_4^{\rm imp} = A_4 + c_A \partial_4 P,
   \label{equation:A4_imp}
\end{equation}
where $P$ is the pseudoscalar meson operator,
Eq.~(\ref{equation:meson}) with $\Gamma = \gamma_5$,
and $\partial_4$ is the symmetric lattice derivative.
Then, $m_{quark}^{AWI}$ is obtained through
\begin{equation}
   m_{quark}^{AWI} =
   \frac{m_{PS} C_A^s}{2 C_P^s}.
   \label{equation:m_quark_AWI}
\end{equation}
The amplitudes $C_A^s$ and $C_P^s$ are calculated as follows.
We determine the pseudoscalar meson mass $m_{PS}$ and $C_P^s$ by
\begin{equation}
 \langle P^l(t) P^s(0)^{\dagger} \rangle
 = C_P^s \left [ \exp(-m_{PS} t) + \exp(-m_{PS} (L_t-t) ) \right ],
\end{equation}
where the superscripts $l$ and $s$ distinguish local and smeared operators.
Keeping $m_{PS}$ fixed, we extract $C_A^s$ from
\begin{equation}
  \langle A_4^{{\rm imp},l}(t) P^s(0)^{\dagger} \rangle
  = C_A^s \left [ \exp(-m_{PS} t) - \exp(-m_{PS} (L_t-t) ) \right ].
\end{equation}

The renormalized quark masses
in the $\overline{\mbox{MS}}$ scheme at 2~GeV
are obtained as follows.
%
%
The VWI up and down quark mass
\begin{equation}
 m_{ud}^{VWI} = \frac{1}{2}
                \left( \frac{1}{\kappa_{ud}} - \frac{1}{\kappa_{c}}
                \right),
\end{equation}
with $\kappa_{ud}$ the hopping parameter at the physical point,
is renormalized using one-loop renormalization constants
and improvement coefficients at $\mu = 1/a$:
\begin{equation}
 m_{ud}^{VWI,\overline{\rm MS}}(\mu = 1/a)
 = Z_m \left( 1 + b_m \frac{m_{ud}^{VWI}}{u_0} \right)
             \frac{m_{ud}^{VWI}}{u_0}.
 \label{equaton:renormalized_m_quark_VWI}
\end{equation}
%
%
Similarly the renormalized AWI quark mass is obtained by
\begin{equation}
 m_{ud}^{AWI,\overline{\rm MS}}(\mu = 1/a) 
 = \frac{Z_A \left( 1 + b_A \frac{m_{ud}^{VWI}}{u_0} \right)}
                  {Z_P \left( 1 + b_P \frac{m_{ud}^{VWI}}{u_0} \right)}
             m_{ud}^{AWI},
 \label{equaton:renormalized_m_quark_AWI}
\end{equation}
where $m_{ud}^{AWI}$ is the value of $m_{quark}^{AWI}$ 
extrapolated to $\kappa_{ud}$.
The determination of $\kappa_{ud}$ is discussed
in Sec.~\ref{section:chiral_extrapolation_polynomials}
and \ref{section:chiral_extrapolation_ChPT}.
%
%
Since non-perturbative values for the renormalization coefficient
$Z_A$ and the improvement 
parameters $c_A$, $b_A$ etc.\
are not available for our combination of actions in two-flavor QCD,
we adopt one-loop perturbative values calculated 
in Refs.~\cite{Z_factors.Sinya1,Z_factors.Sinya2}
improved with the tadpole procedure using $u_0$ given
in Eq.(\ref{equation:u0}).
%
%
The $\overline{\mbox{MS}}$ quark masses at $\mu = 1/a$ 
are evolved to $\mu = 2$~GeV
using the four-loop
beta function~\cite{Run_factor.Chetrykin,Run_factor.Vermaseren}.

\subsection{Decay constants}
\label{section:decay_constant}

The pseudoscalar meson decay constant is calculated by
\begin{equation}
f_{PS} = 2 \kappa u_0 Z_A \left ( 1+ b_A \frac{m_{quark}^{VWI}}{u_0} \right ) 
         \frac{C_A^s}{C_P^s}\sqrt{\frac{C_P^l}{m_{PS}}},
 \label{equation:f_PS}
\end{equation}
where $C_P^l$ is determined by
\begin{equation}
 \langle P^l(t) P^l(0)^{\dagger} \rangle
 =  C_P^l \left [ \exp(-m_{PS} t) + \exp(-m_{PS} (L_t-t) ) \right ],
\end{equation}
keeping $m_{PS}$ fixed to the value
from $\langle P^l(t) P^s(0)^{\dagger} \rangle$.

The vector meson decay constant $f_{V}$ is defined as
\begin{equation}
\langle 0 | V_i | {V} \rangle = \epsilon_i f_{V} m_{V},
\end{equation}
where $\epsilon_i$ is a polarization vector.
%
%
The procedure to obtain the vector meson decay constant
is parallel to that for $f_{PS}$.
The vector meson correlator with a smeared source is fitted with
\begin{equation}
 \langle V^l(t) V^s(0)^{\dagger} \rangle
 = C_V^s \left [ \exp(-m_{V} t) + \exp(-m_{V} (L_t-t) ) \right ],
\end{equation}
which determines $m_{V}$ and $C_V^s$.
Using $m_{V}$ as an input we fit the correlator 
\begin{equation}
 \langle V^l(t) V^l(0)^{\dagger} \rangle
 = C_V^l \left [ \exp(-m_{V} t) + \exp(-m_{V} (L_t-t) ) \right ],
\end{equation}
where the amplitude $C_V^l$ is the only fit parameter.
A renormalized vector meson decay constant is then obtained through 
\begin{equation}
 f_{V} = 2 \kappa u_0 Z_V \left ( 1+ b_V \frac{m_{VWI}}{u_0} \right )
 \sqrt{ \frac{C_V^l}{m_{V}} },
 \label{equation:f_V}
\end{equation}
where we also use one-loop perturbative values
for $Z_V$ and $b_V$~\cite{Z_factors.Sinya1,Z_factors.Sinya2}.
We do not include the improvement term
$c_V \tilde{\partial}_\nu T_{n\mu\nu}$
because the corresponding correlator is not measured.

\subsection{Static quark potential}
\label{section:potential}

\begin{figure}[h]
   \includegraphics[width=75mm]
   {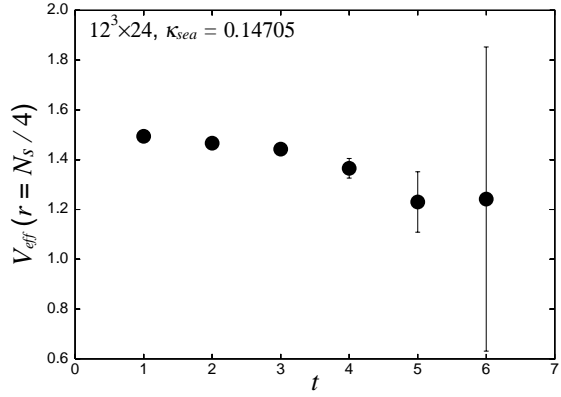}
   \caption
   {
      Effective potential energies 
      $V_{eff}(r=N_s/4,t)$
      at $\kappa_{sea}=0.14705$ ($m_{PS}/m_{V} = 0.40$)
      on $12^3 \times 24$ lattice.
   }
   \label{figure:m_eff_potential}
\end{figure}

\begin{figure}[h]
   \includegraphics[width=75mm]
   {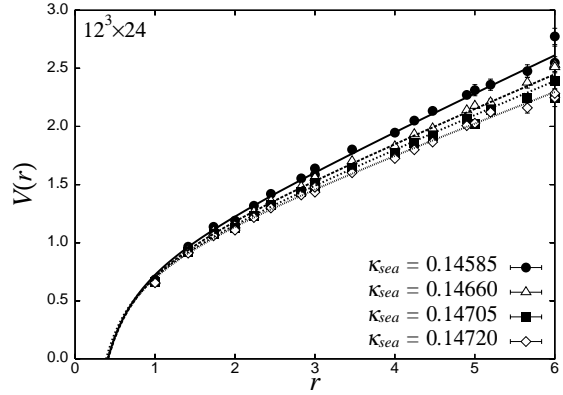}
   \caption
   { Static quark potentials
     at $\kappa_{sea}=0.14585,0.14660,0.14705$ and $0.14720$
     correspond to $m_{PS}/m_V=0.60,0.50,0.40$ and $0.35$
     on $12^3 \times 24$ lattice.
   }
   \label{figure:r_V_eff-12x24}
\end{figure}

We calculate the static quark potential $V(r)$
from the temporal Wilson loops $W(r,t)$
\begin{equation}
   W(r,t) = C(r) \exp \left( - V(r)t \right).
   \label{equation:potential}
\end{equation}
%
%
We apply the smearing procedure of Ref.~\cite{Potential.Bali}.
The number of smearing steps is fixed to 
its optimum value $N_{opt}=2$ at which 
the overlap to the ground state $C(r)$ takes the largest value.
%
%
Let us define an effective potential
\begin{equation}
   V_{eff}(r,t) = \log \left[ W(r,t)/W(r,t+1) \right].
   \label{equation:effective_potential}
\end{equation}
Examples of $V_{eff}$ are plotted in Fig.~\ref{figure:m_eff_potential}, 
from which we take the lower cut of $t_{min}=2$.
%
%
As shown in Fig.~\ref{figure:r_V_eff-12x24},
we do not observe any clear indication of the string breaking.
Therefore, we carry out a correlated fit to $V(r)=V_{eff}(r,t_{min})$
with 
\begin{equation}
 V(r) = V_0 - \frac{\alpha}{r} + \sigma r .
 \label{equation:string_tension}
\end{equation}
Here we do not include the lattice correction to the Coulomb term 
calculated perturbatively 
from one lattice-gluon exchange diagram~\cite{deltaV},
since rotational symmetry is well restored for our RG-improved action.
%
%
The Sommer scale $r_0$ is defined through~\cite{r_0} 
\begin{equation}
   r_0^2 \left. \frac{d V(r)}{d r} \right|_{r=r_0}
   =
   1.65.
   \label{equation:r_0_definition}
\end{equation}
We determine $r_0$ from 
the parameterization of the potential $V(r)$:
\begin{equation}
   r_{0} = \sqrt{\frac{1.65-\alpha}{\sigma}}.
   \label{equation:r_0}
\end{equation}

%
%
The lower cut of the fit range in Eq.~(\ref{equation:string_tension})
is determined as $r_{min} = \sqrt{2}$
from inspection of the $r_{min}$ dependence of $r_0$.
With $r_{min} < \sqrt{2}$, $\chi^2/{dof}$ takes
an unacceptably large value,
while $\alpha$ becomes ill-determined with $r_{min}>\sqrt{3}$.
On the other hand, the $r_{max}$ dependence of $r_0$ is mild.
Therefore, we fix $r_{max}$ to $N_s/2$.
%
%
We estimate the systematic error of the fit as follows.
The fit of Eq.~(\ref{equation:string_tension}) is repeated 
with other choices of the range:
$t_{min}=3$ or $r_{min} = \sqrt{3}$.
The variations in the resulting parameters and $r_0$ 
are taken as systematic errors.
%
%
The parameters in Eq.~(\ref{equation:string_tension}) 
and $r_0$
are presented in Table~\ref{table:potential}.

\subsection{Autocorrelation} 
\label{section:autocorrelation}

\begin{figure}[h]
   \includegraphics[width=75mm]
   {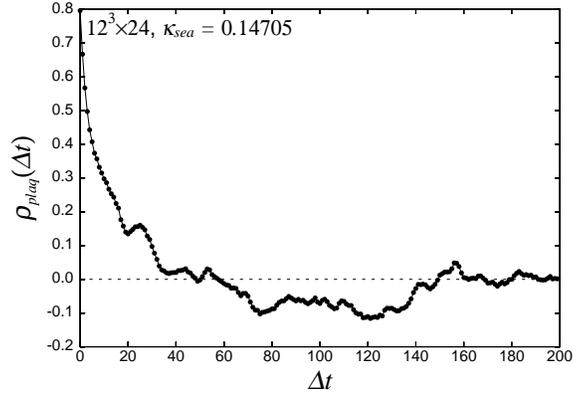}
   \caption
   {  Autocorrelation function of plaquette
      at $\kappa_{sea}=0.14705$ ($m_{PS}/m_{V} = 0.40$)
      on $12^3 \times 24$ lattice.
   }
   \label{figure:autocorrelation_function-plaq}
\end{figure}

\begin{figure}[h]
\begin{center}
   \includegraphics[width=75mm]
   {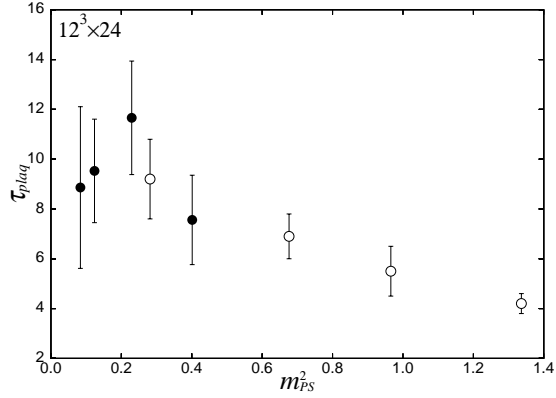}
   \caption
   {  Sea quark mass dependence of
      the cumulative autocorrelation time
      of plaquette
      on $12^3 \times 24$ lattice.
      Open symbols are the results obtained
      in our previous study \protect\cite{Spectrum.Nf2.CP-PACS}.
   }
   \label{figure:m_PS2_autocorrelation_time_plaq}
\end{center}
\end{figure}

The autocorrelation in our data is studied 
by the cumulative autocorrelation time 
\begin{equation}
   \tau^{cum}_{\mathcal{O}}(\Delta t_{max}) 
  =\frac{1}{2}  
   +\sum_{\Delta t=1}^{\Delta t_{max}}
    \rho_{\mathcal{O}}(\Delta t),
   \label{equation:autocorrelation_cumulative}
\end{equation}
where $\rho_{\mathcal{O}}(t)$ is the autocorrelation function
\begin{equation}
   \rho_{\mathcal{O}}(\Delta t) 
   = 
   \frac{\Gamma_{\mathcal{O}}(\Delta t)}
         {\Gamma_{\mathcal{O}}(0)}, \\
  ~~~ \Gamma_{\mathcal{O}}(\Delta t) 
   = 
   \langle \left( \mathcal{O} (t) - \langle \mathcal{O} \rangle
           \right)
           \left( \mathcal{O} (t+\Delta t) - \langle \mathcal{O} \rangle
           \right)
   \rangle .
   \label{equation:autocorrelation_rho}
\end{equation}
A conventional choice for $\Delta t_{max}$ is the first point 
where $\rho_{\mathcal{O}}$  vanishes
because $\rho_{\mathcal{O}}$ should be positive when the statistics are 
sufficiently high.
We take $\Delta t_{max}=50$ from the plaquette 
shown in Fig.~\ref{figure:autocorrelation_function-plaq}.
In Table~\ref{table:autocorrelation},
we give $\tau^{cum}_{\cal O}$ for
(i) the plaquette which is measured at every trajectory,
(ii) the pseudoscalar meson propagators at $t=N_t/4$, and
(iii) the temporal Wilson loop with $(r,t)=(2,2)$.
Fig.~\ref{figure:m_PS2_autocorrelation_time_plaq} shows 
the autocorrelation time for the plaquette.
Combining the previous (open circles) and the new (filled circles) data, 
we observe a trend of increase for smaller quark masses. 
A sharp rise expected toward the chiral limit, however, is not seen.  
Our statistics may not be sufficient 
to estimate autocorrelation times reliably near the chiral limit.

\begin{figure}[h]
   \includegraphics[width=75mm]
   {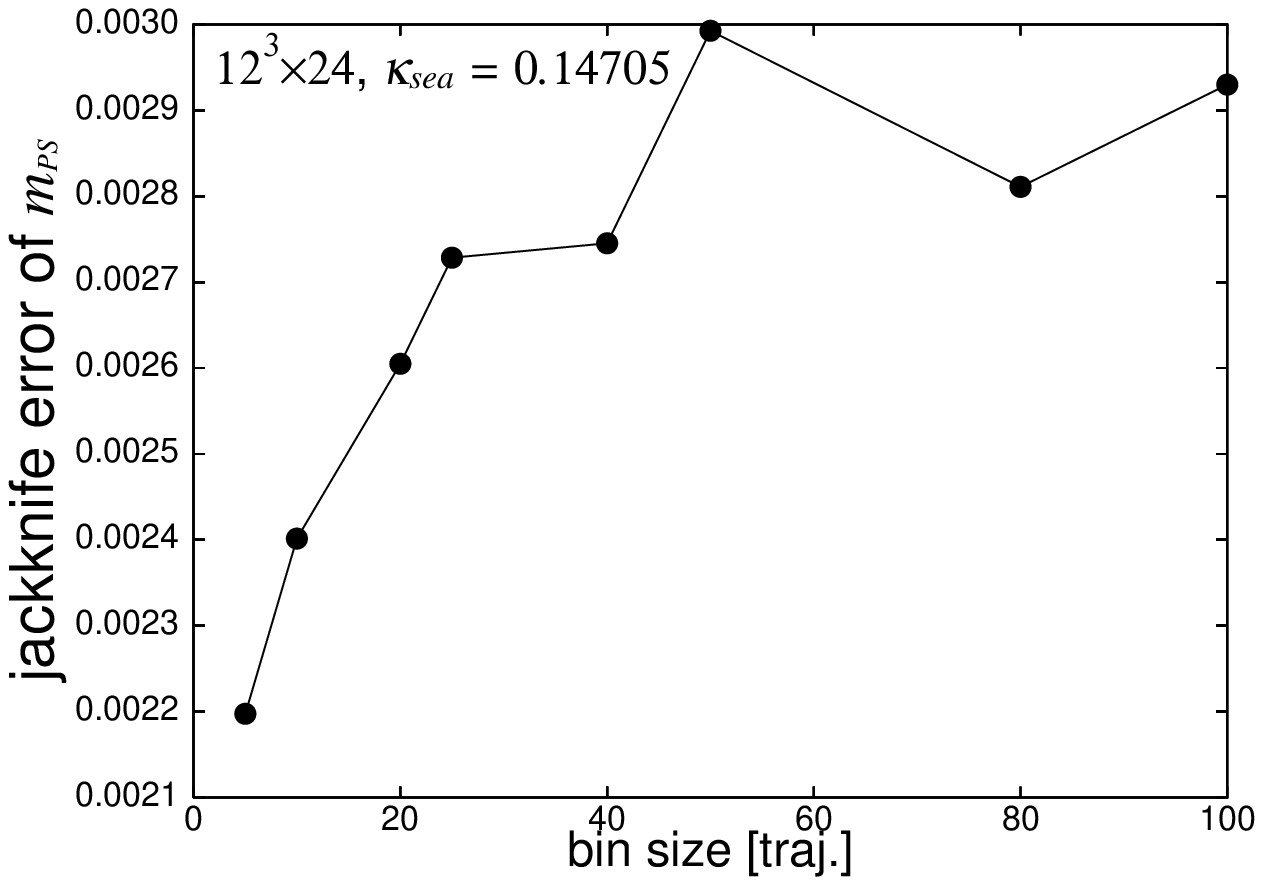}
   \includegraphics[width=75mm]
   {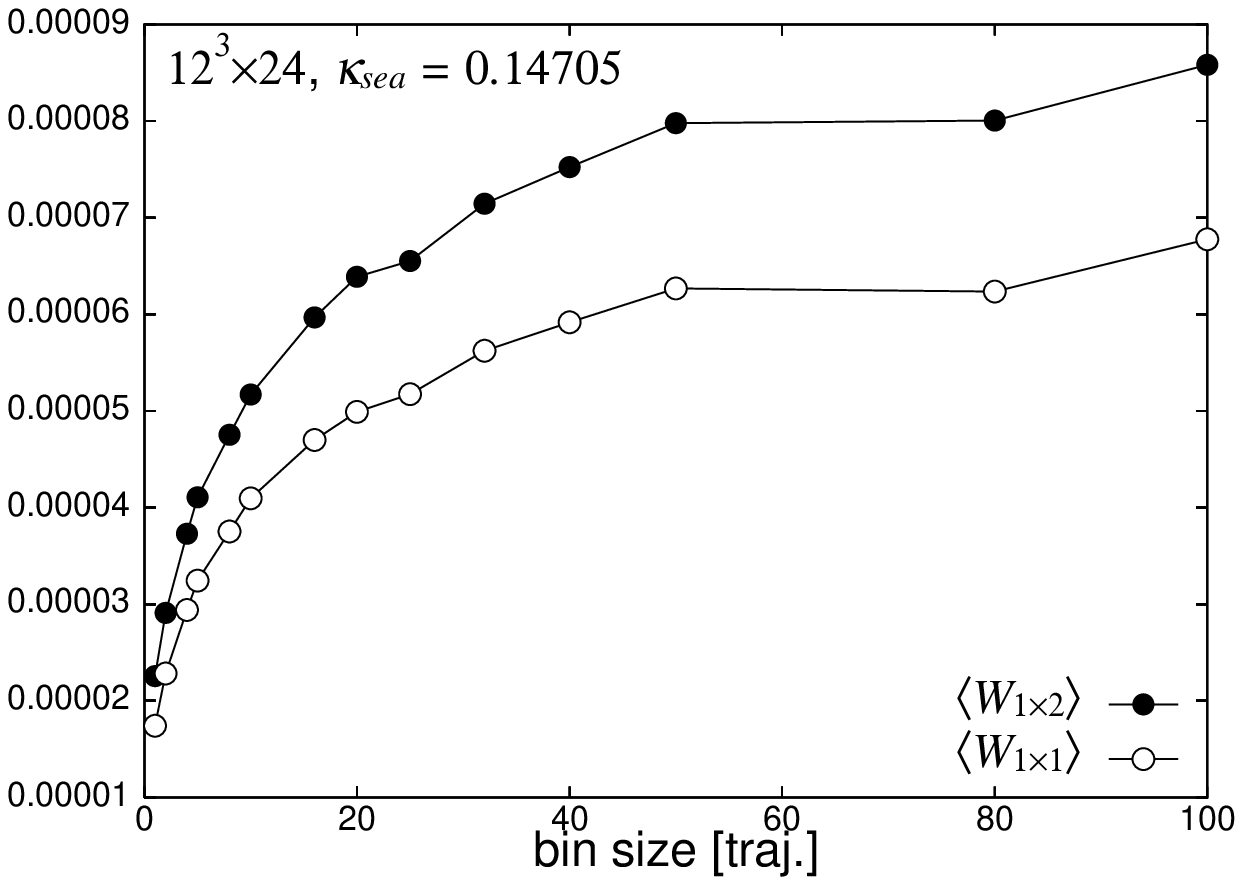}
   \caption
   {  Bin size dependence of jack-knife error of
      pseudoscalar meson mass (left panel),
      and plaquette and rectangular loop (right panel)
      at $\kappa_{sea}=0.14705$ ($m_{PS}/m_{V} = 0.40$)
      on $12^3 \times 24$ lattice.
   }
   \label{figure:bin_size_dependence-m_PS_wloop}
\end{figure}

The bin size dependence of the jack-knife errors of 
hadron masses and Wilson loops is exhibited 
in Figs.~\ref{figure:bin_size_dependence-m_PS_wloop}.
The jack-knife errors reach plateaus at 
bin size of 50--100 trajectories.
The situation is similar on $16^3 \times 24$.
Therefore, we take the bin size of 100 trajectories 
in the error analysis.

\subsection{Finite-size effects}
\label{section:finite_size_effects}

\begin{figure}[h]
\begin{center}
   \includegraphics[width=75mm]
   {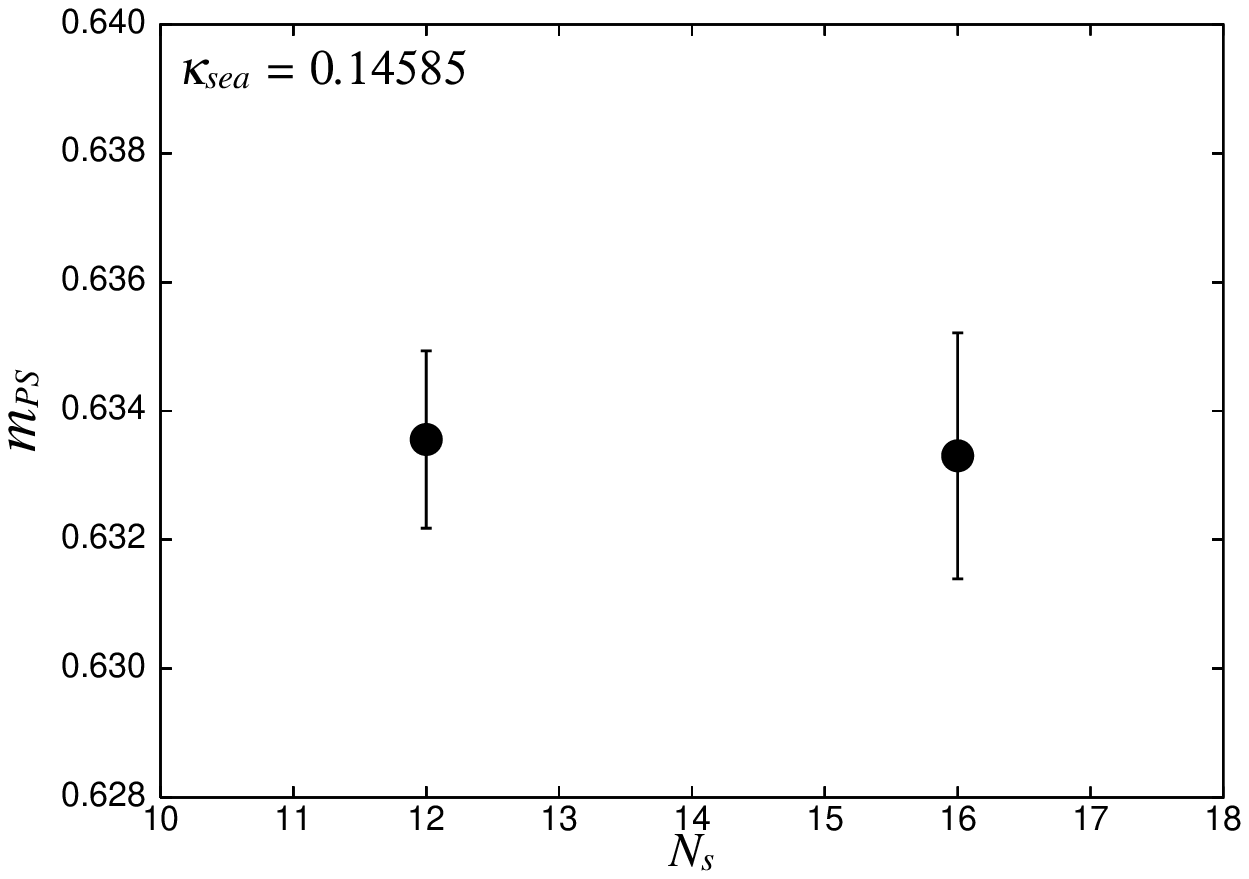}
   \includegraphics[width=75mm]
   {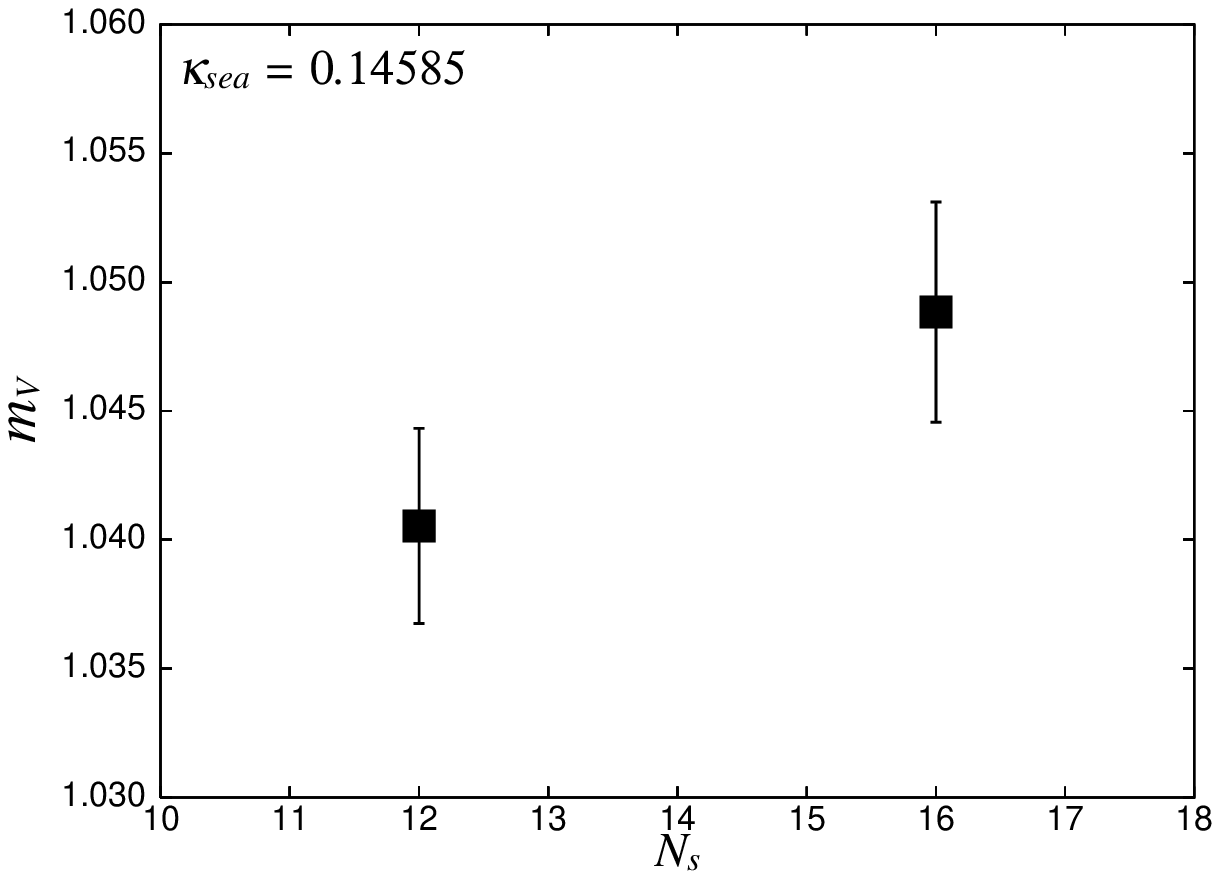}
   \includegraphics[width=75mm]
   {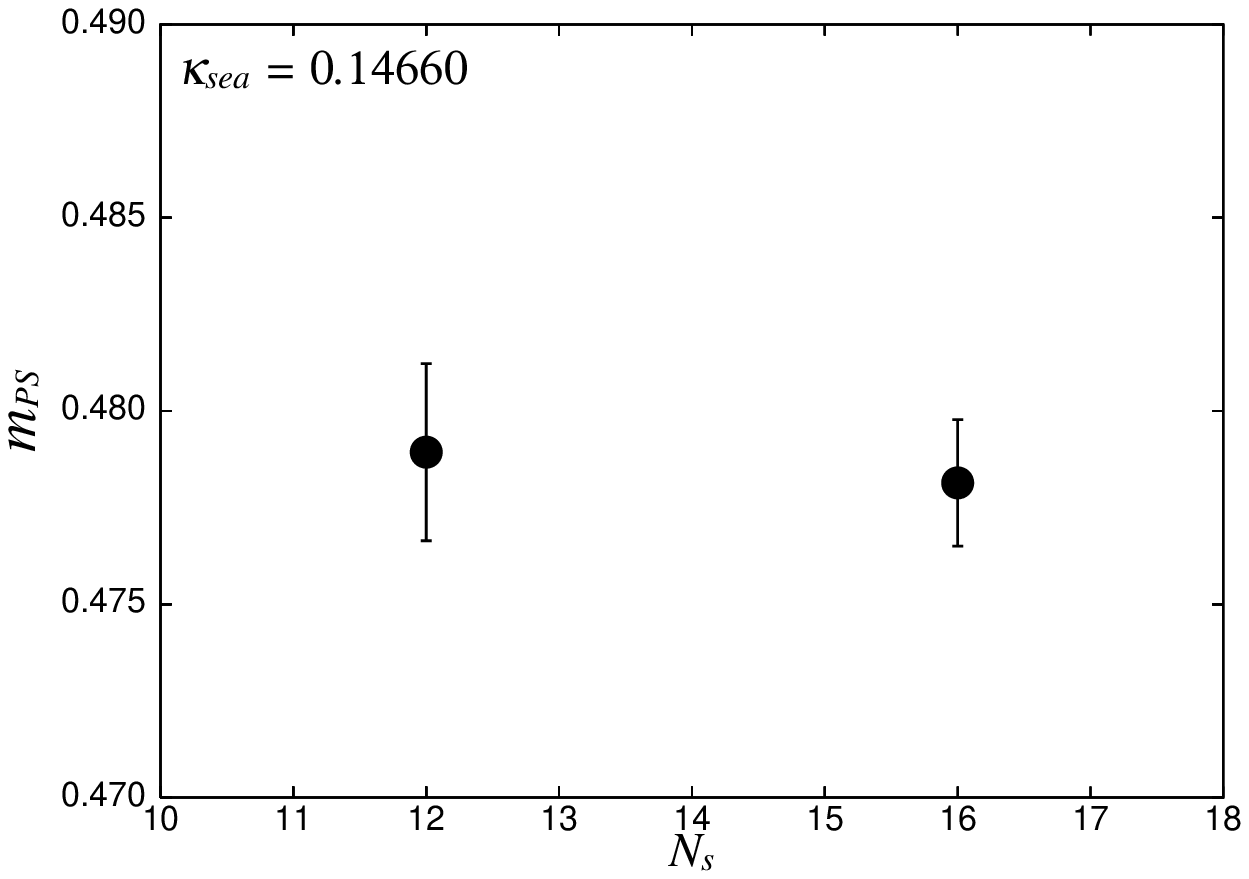}
   \includegraphics[width=75mm]
   {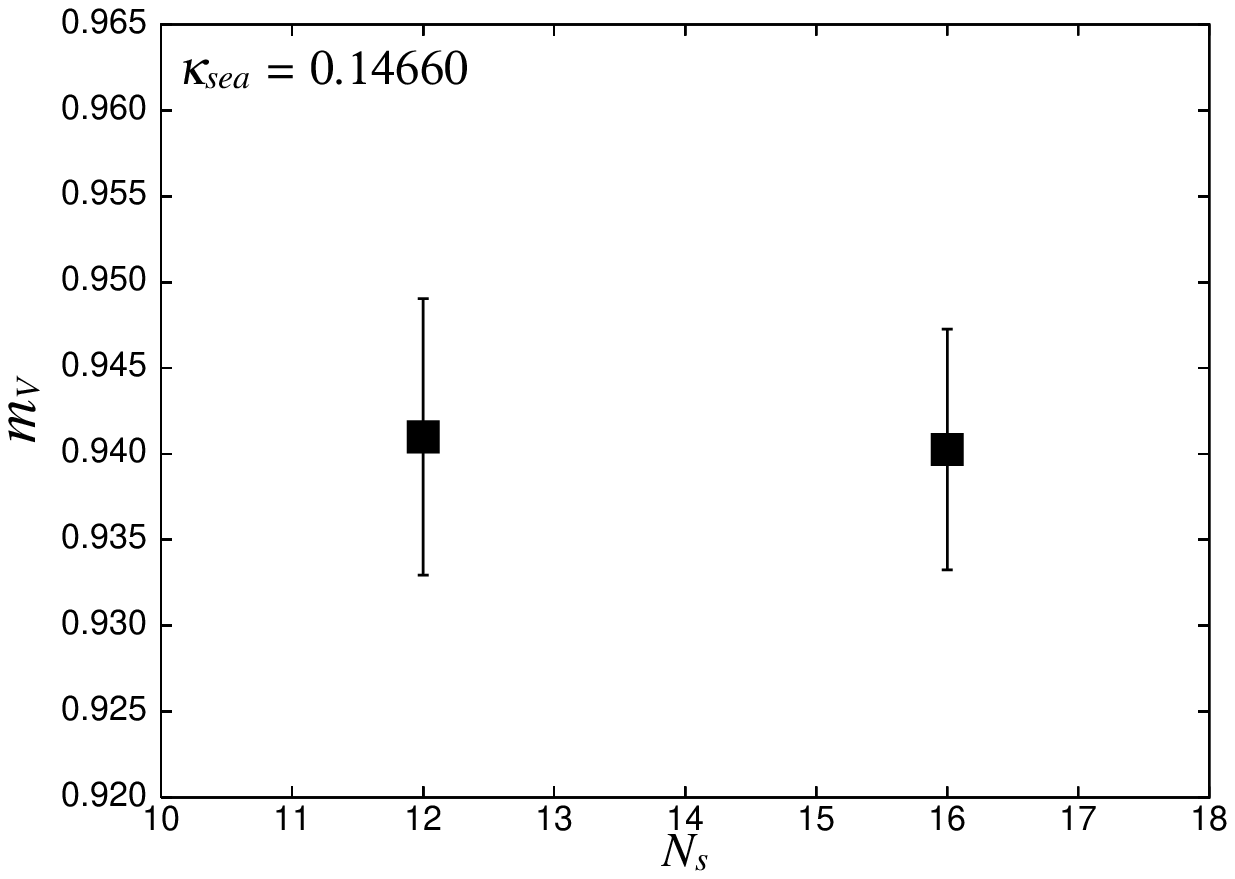}
   \caption
   {  Volume dependence of 
      pseudoscalar (left panel) and vector meson masses (right panel)
      at  $\kappa_{sea}=0.14585$ ($m_{PS}/m_{V} = 0.60$)
      and $\kappa_{sea}=0.14660$ ($m_{PS}/m_{V} = 0.50$).
   }
   \label{figure:FSE_meson}
\end{center}
\end{figure}

\begin{figure}[h]
\begin{center}
   \includegraphics[width=75mm]
   {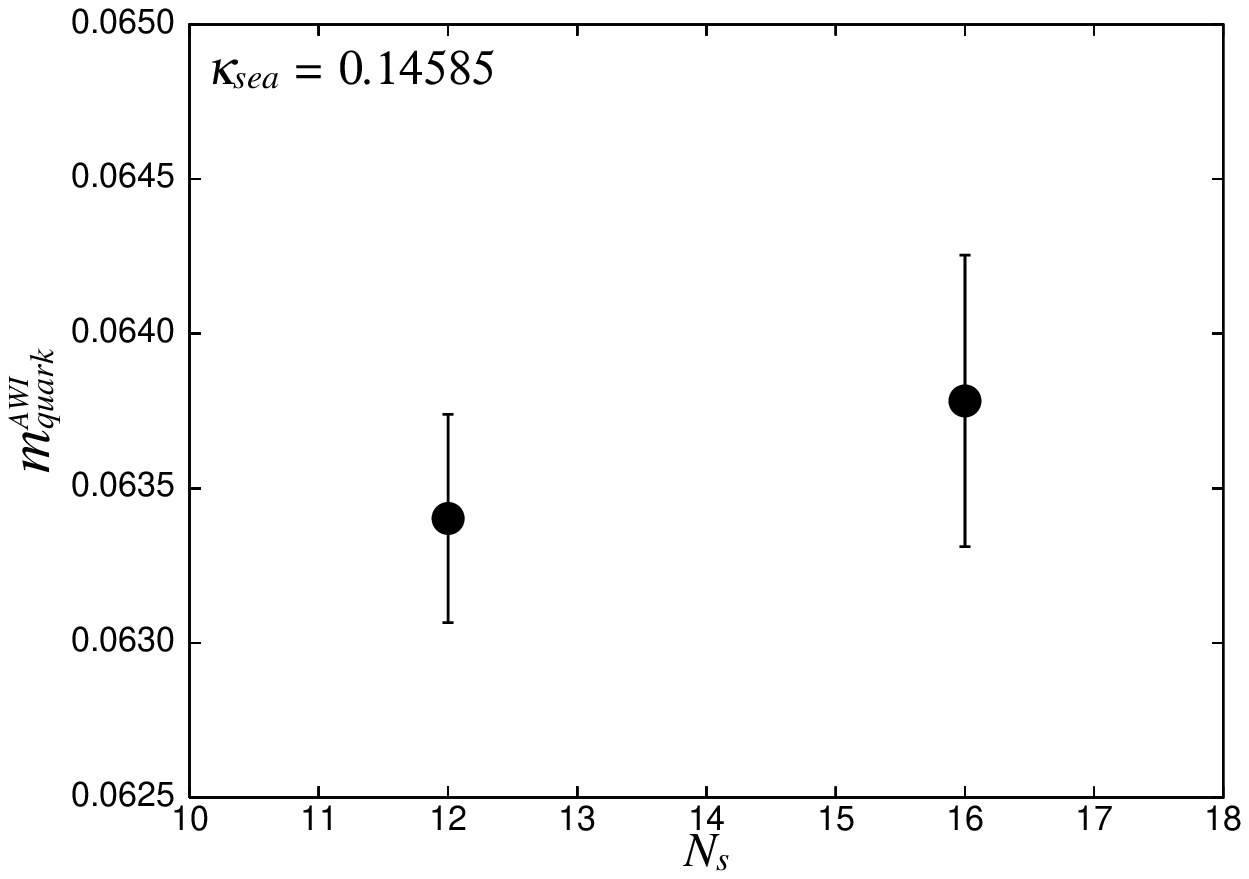}
   \includegraphics[width=75mm]
   {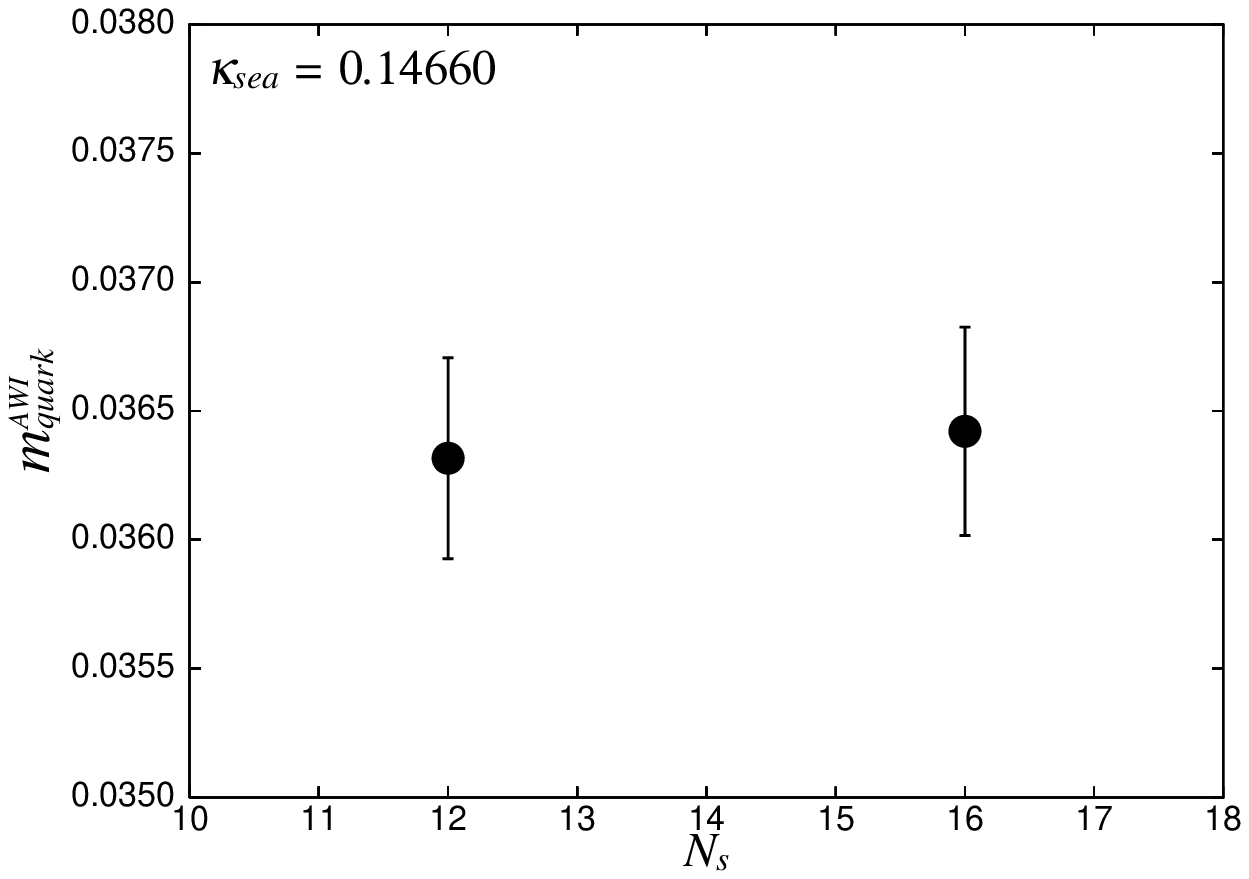}
   \caption
   {  Volume dependence of AWI quark masses
      at  $\kappa_{sea}=0.14585$ ($m_{PS}/m_{V} = 0.60$)
      and $\kappa_{sea}=0.14660$ ($m_{PS}/m_{V} = 0.50$).
   }
   \label{figure:FSE_m_quark_AWI}
\end{center}
\end{figure}

\begin{figure}[h]
   \includegraphics[width=75mm]
   {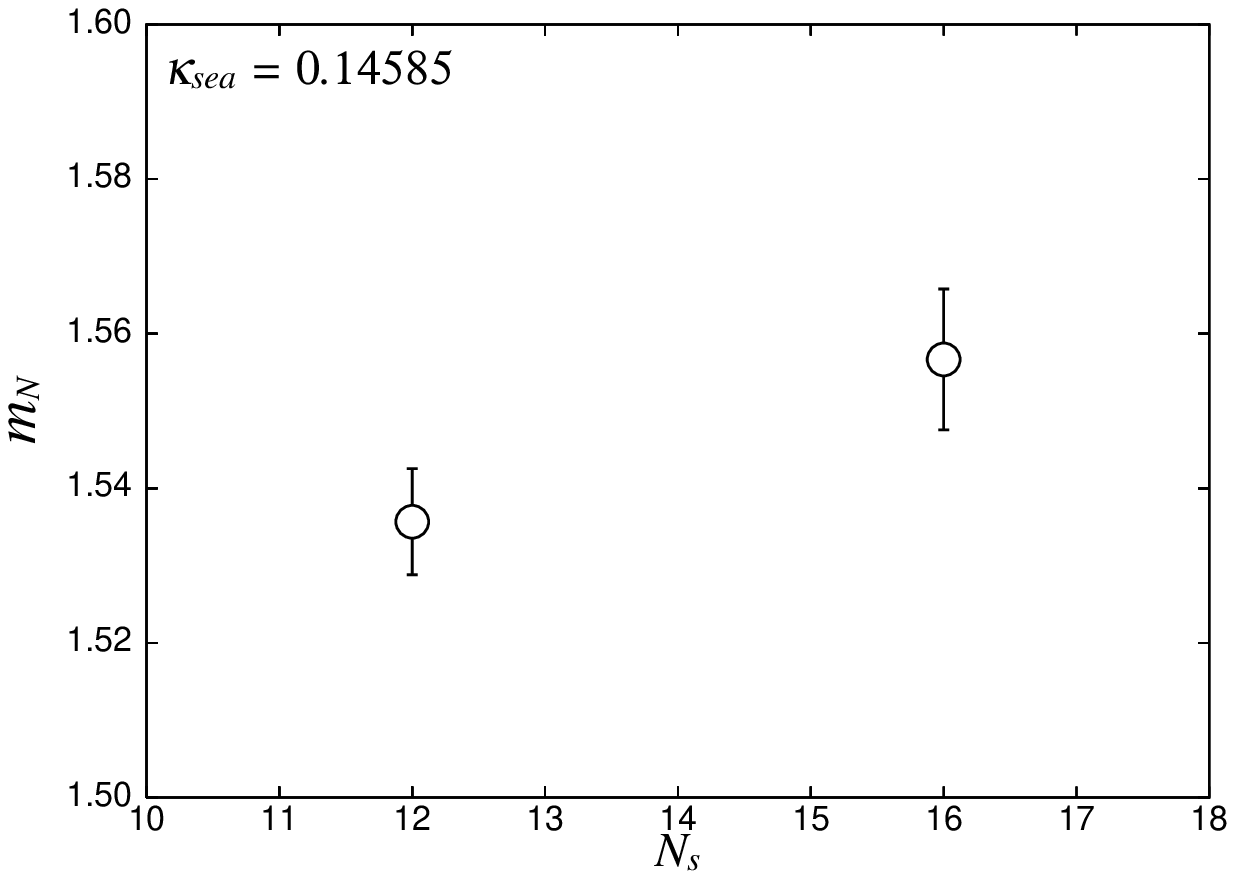}
   \includegraphics[width=75mm]
   {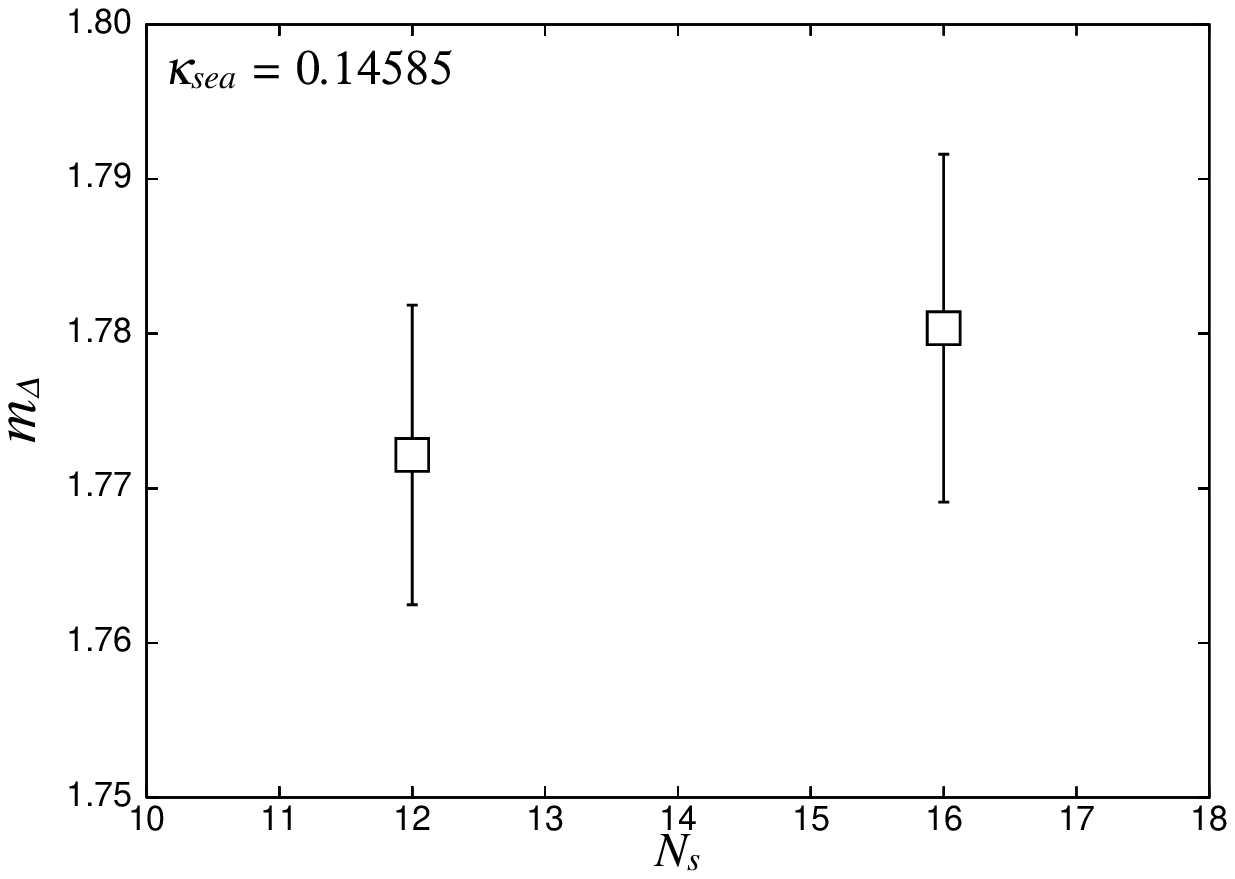}
   \includegraphics[width=75mm]
   {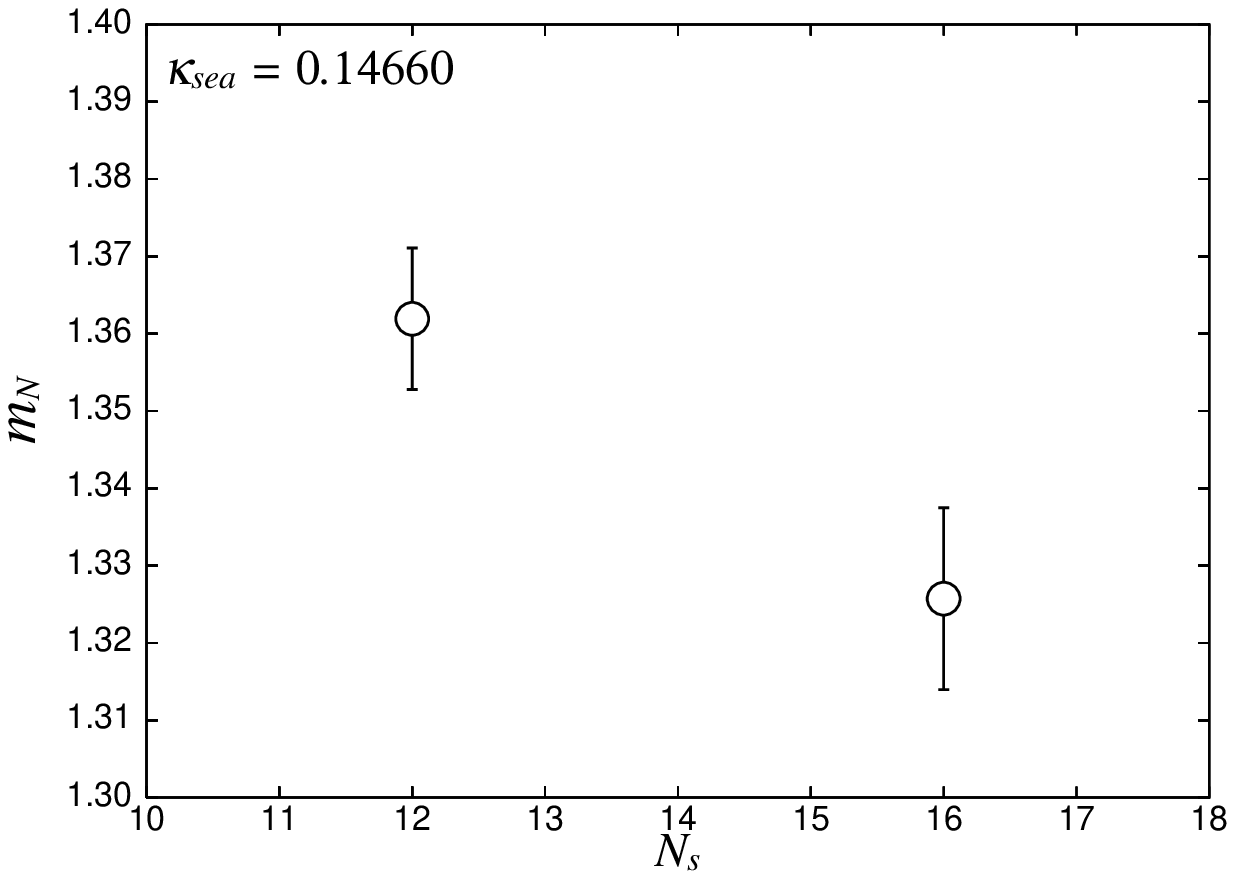}
   \includegraphics[width=75mm]
   {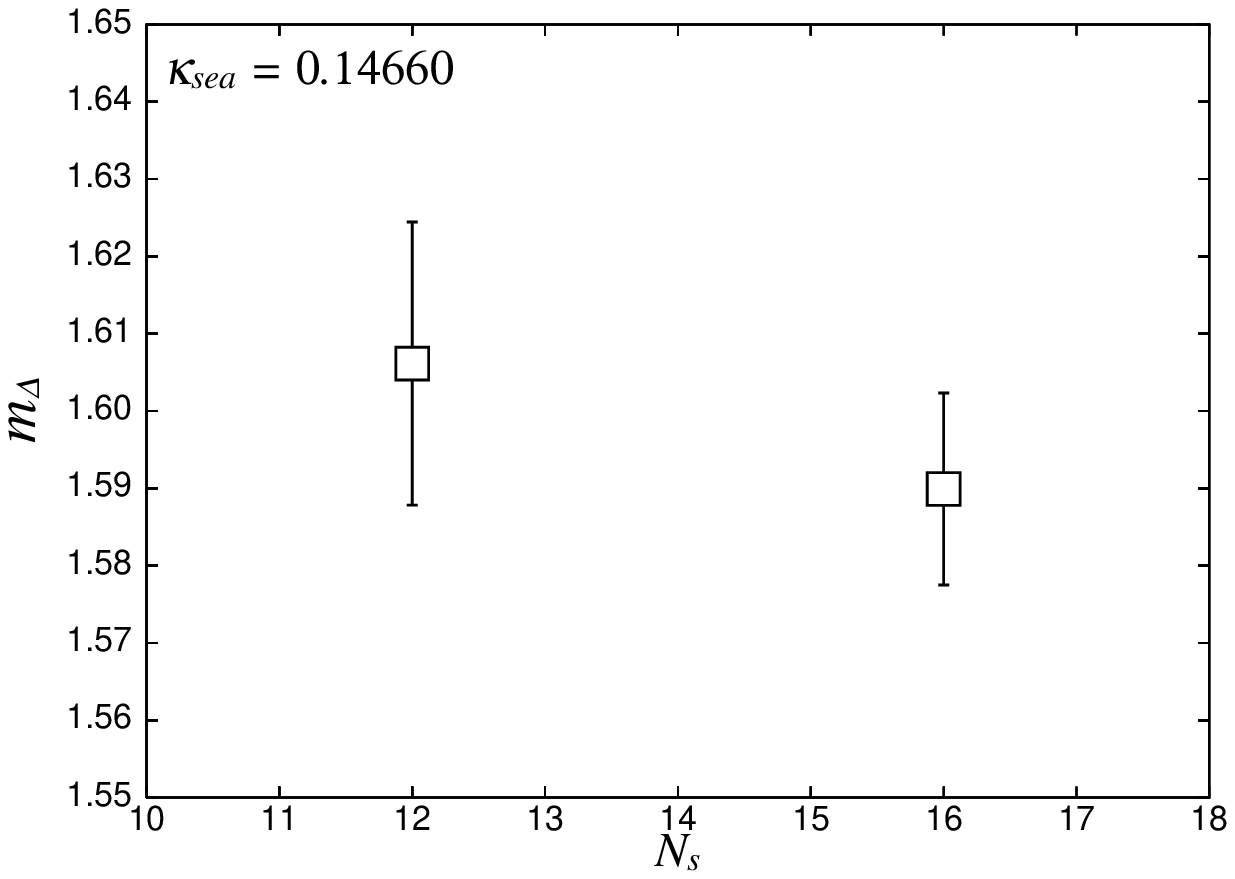}
   \caption
   {  Volume dependence of 
      octet (left panel) and decuplet baryon masses (right panel)
      at  $\kappa_{sea}=0.14585$ ($m_{PS}/m_{V} = 0.60$)
      and $\kappa_{sea}=0.14660$ ($m_{PS}/m_{V} = 0.50$).
   }
   \label{figure:FSE_baryon}
\end{figure}

\begin{figure}[h]
   \includegraphics[width=75mm]
   {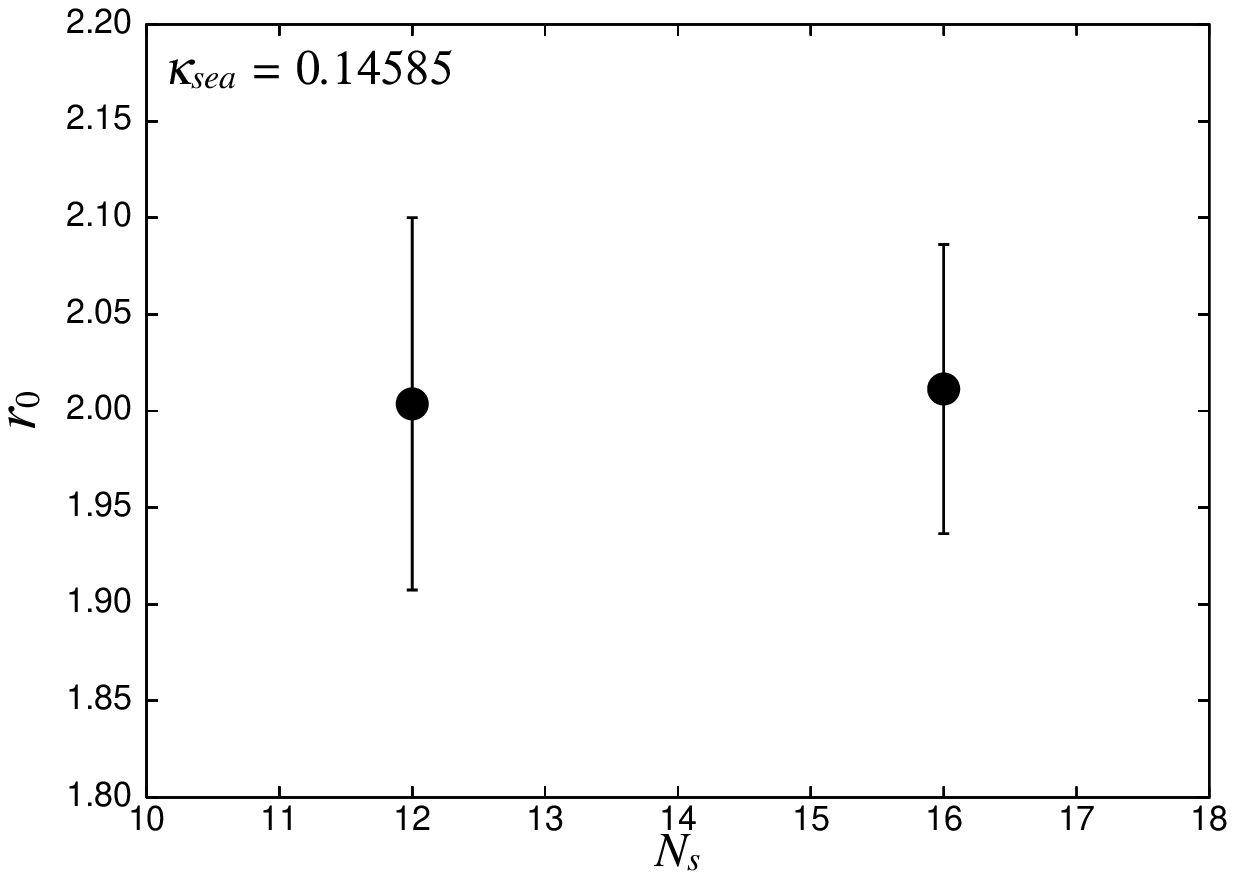}
   \includegraphics[width=75mm]
   {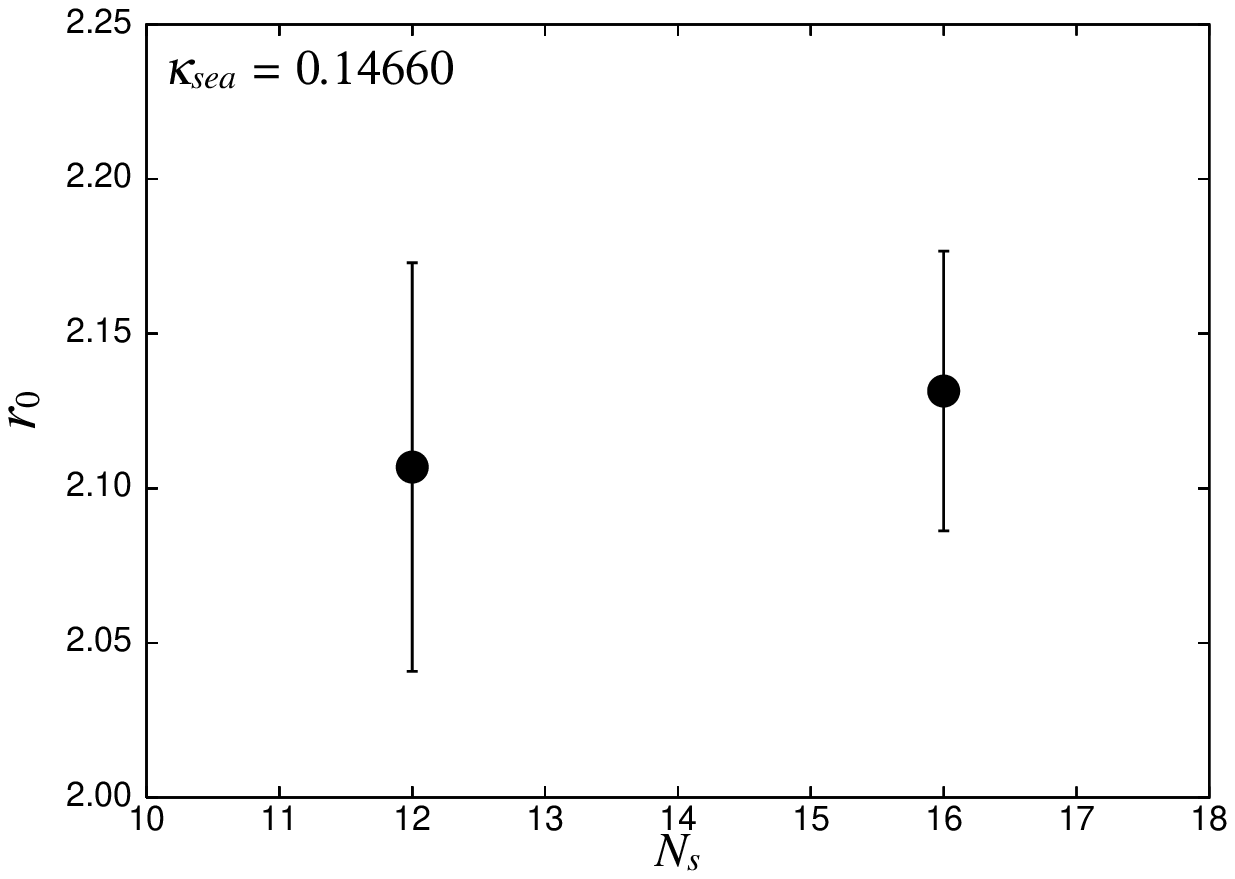}
   \caption
   {  Volume dependence of Sommer scales
      at  $\kappa_{sea}=0.14585$ ($m_{PS}/m_{V} = 0.60$)
      and $\kappa_{sea}=0.14660$ ($m_{PS}/m_{V} = 0.50$).
   }
   \label{figure:FSE_r_0}
\end{figure}

In Figs.~\ref{figure:FSE_meson} and \ref{figure:FSE_m_quark_AWI},
we present meson and AWI quark masses
as a function of the spatial volume.
The results obtained on $12^3 \times 24$ and $16^3 \times 24$ lattices
are mutually consistent within errors.
%
%
For baryons, there may be some indication in our data 
at $m_{PS}/m_{V}=0.50$ ($\kappa_{sea}=0.14660$) that 
the light baryon masses $m_{N}$ and $m_{\Delta}$
decrease by 1--3\% (0.8--3.1$\sigma$)
as shown in Fig.~\ref{figure:FSE_baryon}.
The effect is only around 2$\sigma$, and 
higher statistics are needed to confirm if the difference can be 
attributed to finite-size effects.
%
%
Finite-size effects in $r_0$ are expected to be much smaller
than those in hadron masses.
Our results in Fig.~\ref{figure:FSE_r_0} confirm this.
In the following analysis, we use data obtained on the 
$12^3 \times 24$ lattice.

\section{Chiral extrapolation with polynomials}
\label{section:chiral_extrapolation_polynomials}

Extrapolation of the lattice simulation data
to physical values requires some 
parameterization of the data
as functions of the quark mass.
In this section, we employ polynomials in quark masses.
We work with the two data sets, the one obtained in the previous
work that covers $m_{PS}/m_{V}=0.80$--0.55 
(the large quark mass data set),
and the other obtained in the present work that covers 
$m_{PS}/m_{V}=0.60$--0.35 (the small quark mass data set), and with the
combined data set of the two. For the large mass data set we borrow 
the fit from the previous work. 

%
We fit hadron masses in lattice units
rather than those normalized by $r_0$. 
With our choice of the improved actions, 
$r_0$ exhibits only a mild sea quark mass dependence
as shown below in 
Sec.~\ref{section:decay_constants_baryon_masses_Sommer_scale}, 
and hence
introducing $r_0$ does not change convergence of chiral extrapolations. 
From practical side, $r_0$ suffers from a large systematic error 
on coarse lattices with $a = 0.2$~fm.
Hence fits become less constraining 
if hadron masses are normalized by $r_0$.

  \subsection{Pseudoscalar meson mass and AWI quark mass}

\begin{figure}[h]
\begin{center}
   \includegraphics[width=75mm]
   {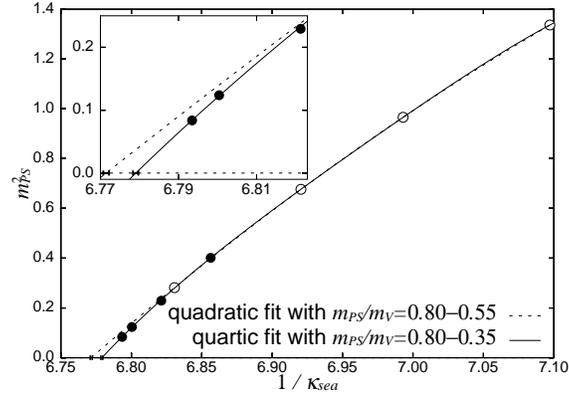}
   \caption
   {  Chiral extrapolation of pseudoscalar meson mass.
      Open symbols show the results obtained
      in the previous calculation \protect\cite{Spectrum.Nf2.CP-PACS}
      and filled symbols are our new results.
      Lines are polynomial fits as explained in the figure.
   }
   \label{figure:kappa_inv-m_PS2}
\end{center}
\end{figure}

\begin{figure}[h]
\begin{center}
   \includegraphics[width=75mm]
   {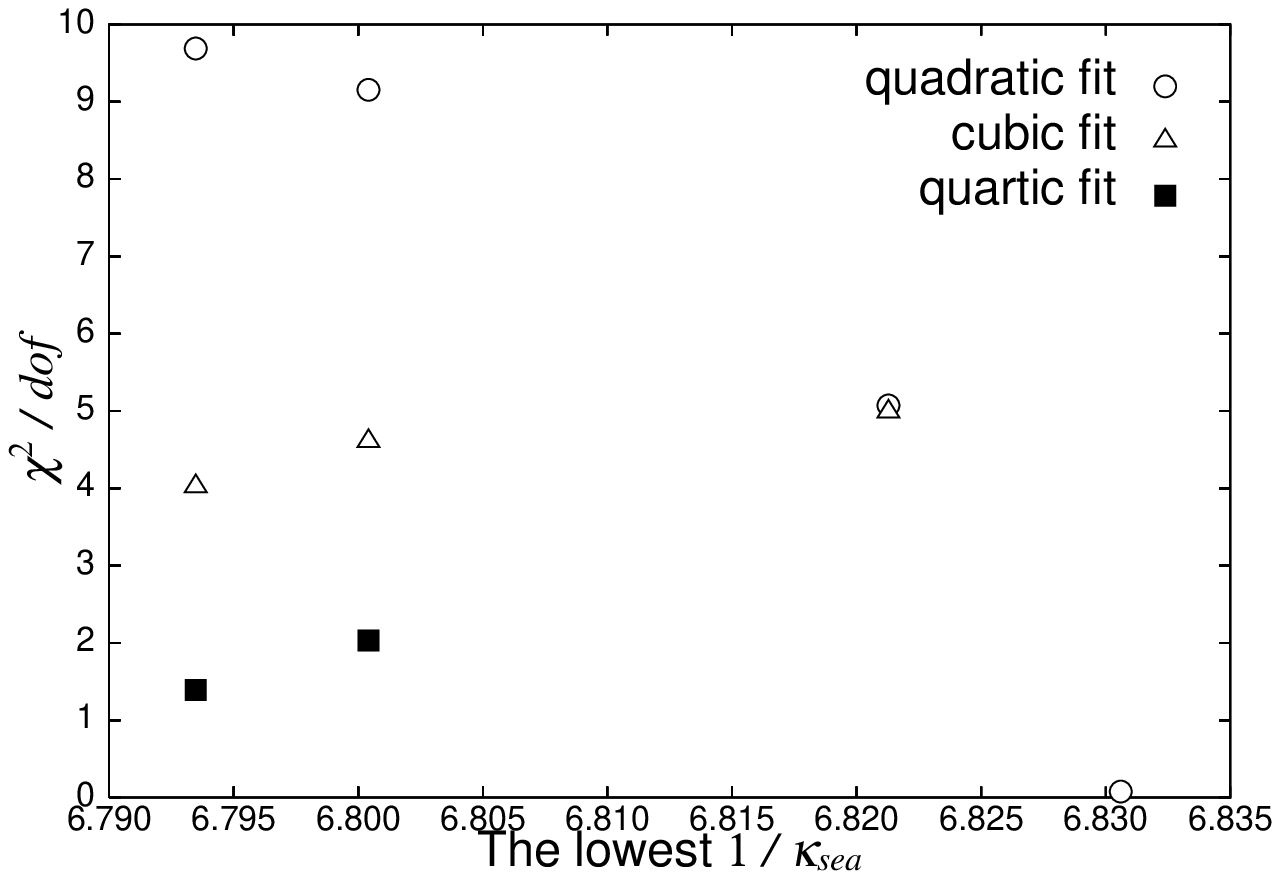}
   \includegraphics[width=75mm]
   {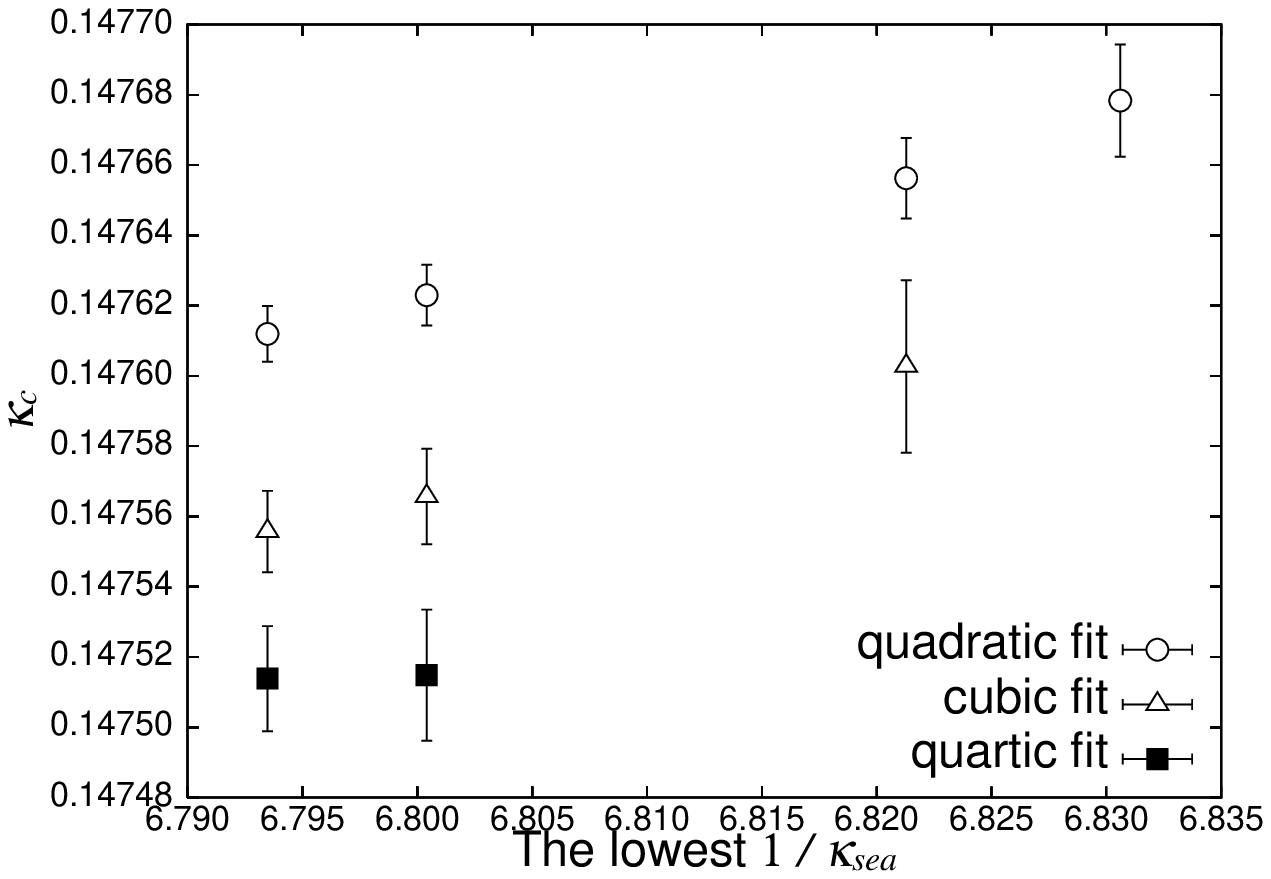}
   \caption
   {  Dependence of $\chi^2/dof$ on the fitting range and order of 
      the fitting polynomial (left panel), 
      and that of the critical hopping parameter (right panel)
      for pseudoscalar meson mass.
      Symbols are placed at the value of $1/\kappa_{sea}$ corresponding 
      to the lowest fitting range, which is changed as shown in the figure,
      while the highest is fixed to $\kappa_{sea}=0.1409$
      ($m_{PS}/m_{V} = 0.80$).
      Quadratic, cubic and quartic forms as a function of VWI quark mass
      are tested.
   }
   \label{figure:lowest_kappa_inv-m_PS2_reduced_chi2}
\vspace*{-1cm}
\end{center}
\end{figure}

A quadratic form fitted well our previous lattice data
of the pseudoscalar meson mass
with a reasonable 
$\chi^2/dof \sim 1$~\cite{Spectrum.Nf2.CP-PACS}.
As shown in Fig.~\ref{figure:kappa_inv-m_PS2}, however,
our new data at small sea quark masses
deviate significantly from the quadratic fit.
Inclusion of the small quark mass data set in the quadratic fit
rapidly increases
$\chi^2/dof$ to $\sim 10$. 
In addition, the determination of the critical hopping parameter
$\kappa_c$ becomes unstable
as shown in Fig.~\ref{figure:lowest_kappa_inv-m_PS2_reduced_chi2}.
A reasonable $\chi^2/dof$ and a stable fit are achieved only 
when we extend the polynomial to quartic, 
\begin{equation}
 m_{PS}^{2} =  B^{PS}  m_{quark}^{VWI}
             + C^{PS} (m_{quark}^{VWI})^2
             + D^{PS} (m_{quark}^{VWI})^3
             + E^{PS} (m_{quark}^{VWI})^4,
 \label{equation:m_quark_VWI_m_PS2}
\end{equation}
where
$m_{quark}^{VWI}$ is given in Eq.(\ref{equation:m_quark_VWI})
and $\kappa_c$ is taken as a fit parameter.
The quartic polynomial provides the best fit among our tests 
varying the order of polynomials.

\begin{figure}[h]
\begin{center}
   \includegraphics[width=75mm]
   {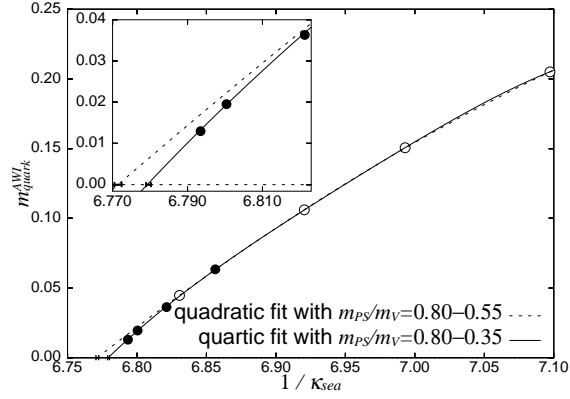}
   \caption
   {  Chiral extrapolation of AWI quark masses.
      Open symbols show the results obtained
      in the previous calculation \protect\cite{Spectrum.Nf2.CP-PACS}.
      Lines are polynomial fits as explained in the figure.
    }
   \label{figure:kappa_inv-bare_m_quark_AWI}
\end{center}
\end{figure}

\begin{figure}[h]
\begin{center}
   \includegraphics[width=75mm]
   {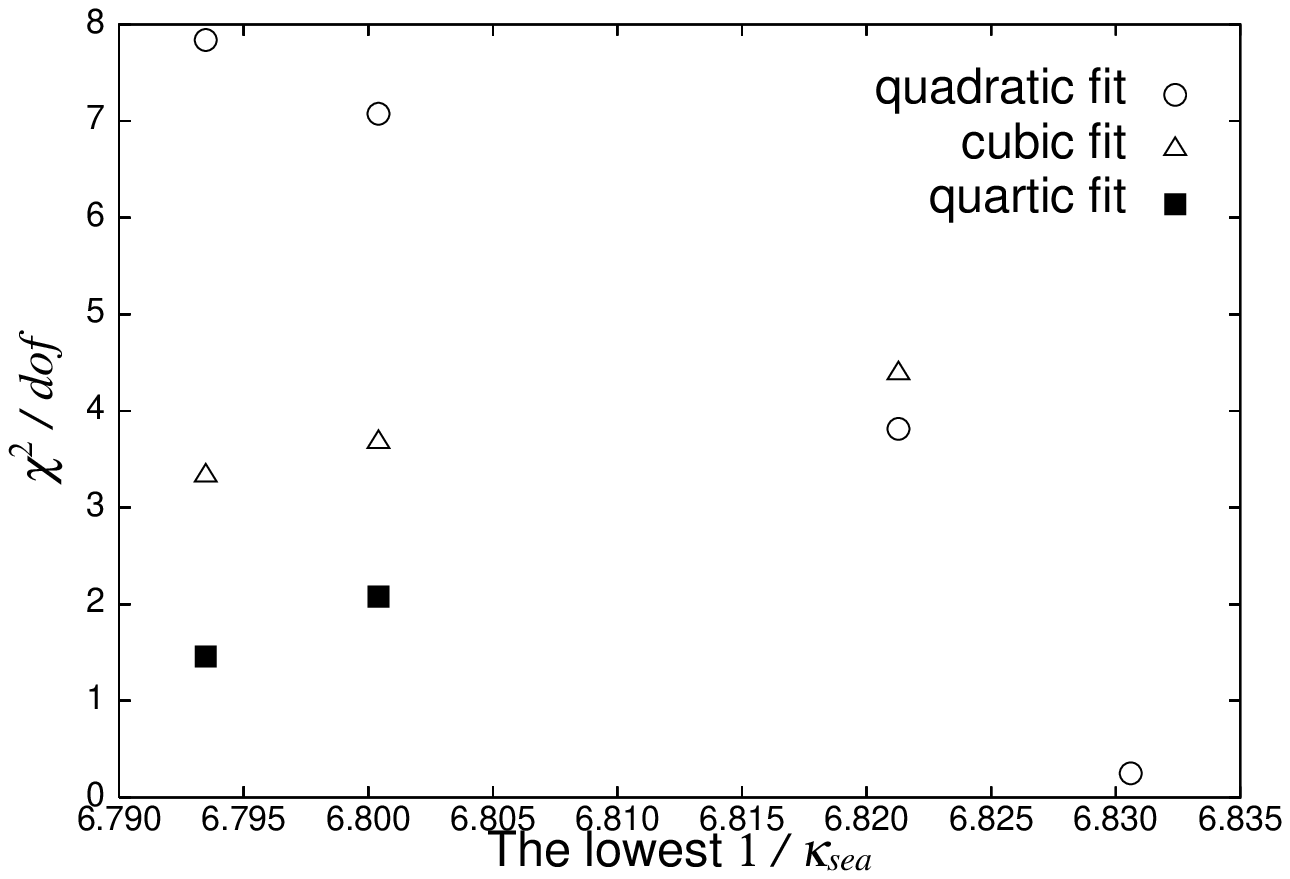}
   \includegraphics[width=75mm]
   {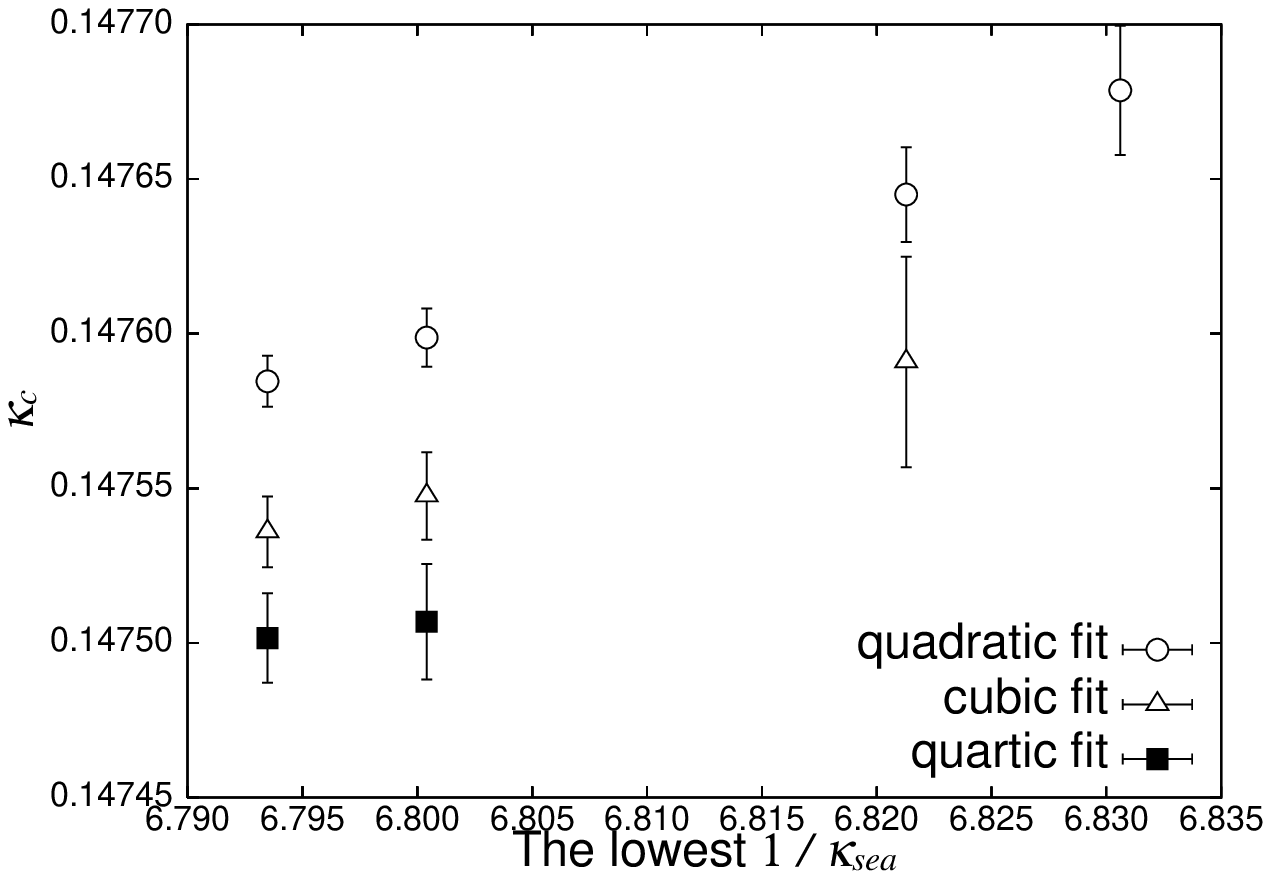}
   \caption
   {  Dependence of $\chi^2/dof$ on the fitting range and order of 
      the fitting polynomial (left panel), 
      and that of the critical hopping parameter (right panel)
      for AWI quark mass.
      Symbols are placed at the value of $1/\kappa_{sea}$ corresponding 
      to the lowest fitting range, which is changed as shown in the figure,
      while the highest is fixed to $\kappa_{sea}=0.1409$
      ($m_{PS}/m_{V} = 0.80$).
      Quadratic, cubic and quartic forms as a function of VWI quark mass
      are tested.
   }
   \label{figure:lowest_kappa_inv-bare_m_quark_AWI_kappa_c_reduced_chi2}
\end{center}
\end{figure}

%
Since $m_{PS}^2$ may be affected by the logarithmic singularity
of ChPT, we examine the convergence of extrapolations, i.e., whether it 
depends on the order of polynomials, using $m_{quark}^{AWI}$ that
has no logarithmic singularities.
Along with the case of $m_{PS}^2$, the new data at small quark masses
deviate from the quadratic fit obtained from the large quark mass data, 
as depicted in Fig.~\ref{figure:kappa_inv-bare_m_quark_AWI}.
We fit $m_{quark}^{AWI}$ by
\begin{equation}
 m_{quark}^{AWI} =  B^{AWI}  m_{quark}^{VWI}
                  + C^{AWI} (m_{quark}^{VWI})^2
                  + D^{AWI} (m_{quark}^{VWI})^3
                  + E^{AWI} (m_{quark}^{VWI})^4.
 \label{equation:m_quark_VWI_m_quark_AWI}
\end{equation}
The fit range and order dependence are given in
Fig.~\ref{figure:lowest_kappa_inv-bare_m_quark_AWI_kappa_c_reduced_chi2}.
$(m_{quark}^{VWI})^4$ terms are needed again to obtain
a reasonable $\chi^2/dof$.

%
%
We find that $\kappa_c$ determined from $m_{PS}^{2}$
agrees with that from $m_{quark}^{AWI}$
within errors.
Hence we simultaneously fit $m_{PS}^{2}$ and $m_{quark}^{AWI}$
to determine $\kappa_c$.
The resulting independent and simultaneous fits
to $m_{PS}^2$ and $m_{quark}^{AWI}$
are presented
in Tables~\ref{table:m_quark_VWI_m_quark_AWI_and_m_PS2_independent}
and \ref{table:m_quark_VWI_m_quark_AWI_and_m_PS2_simultaneous},
respectively.
The difference in mass from the fits including $(m_{quark}^{VWI})^5$
is taken as systematic errors.
These errors represent only uncertainties 
within polynomial extrapolations. 
As shown in Sec.~\ref{section:WChPT_extrapolation},
WChPT fits 
sometimes lead to values beyond these systematic errors.


\subsection{Vector meson mass}

\begin{figure}[h]
   \includegraphics[width=75mm]
   {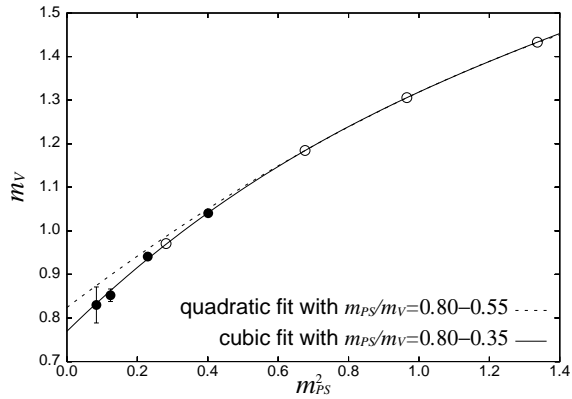}
   \caption
   {  Chiral extrapolation of vector meson mass
      in terms of pseudoscalar meson mass.
      Open symbols show the results obtained
      in the previous calculation \protect\cite{Spectrum.Nf2.CP-PACS}.
   }
   \label{figure:m_PS2_m_V}
\end{figure}

We fit vector meson mass with a 
cubic polynomial in $m_{PS}^{2}$,
\begin{equation}
 m_{V} =  A^V + B^{V} m_{PS}^{2}
        + D^{V} m_{PS}^{4}
        + F^{V} m_{PS}^{6}
 \label{equation:m_PS2_m_V}
\end{equation}
with the results shown
in Fig.~\ref{figure:m_PS2_m_V}
and Table~\ref{table:m_PS2_m_V}.
As in the case of $m_{PS}^2$ and $m_{quark}^{AWI}$,
systematic deviations from the previous fit
are observed, although the difference (7\% or 3.6$\sigma$ 
in the chiral limit) is smaller.
%
%
Inclusion of terms $m_{PS}^{4}$ and $m_{PS}^{6}$ gives a good fit
with a satisfactory $Q$.
We estimate the systematic error from higher order terms
by the difference from the fit with $m_{PS}^{8}$ term.

\begin{figure}[h]
   \includegraphics[width=75mm]
   {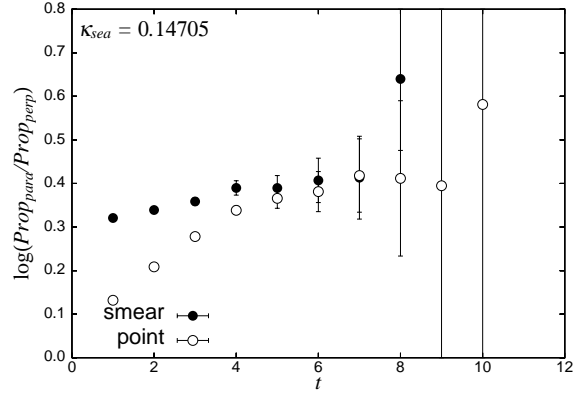}
   \caption
   {  Ratio of vector meson correlators with momentum $2 \pi / L$
      and the polarization parallel and perpendicular to it.      
   }
   \label{figure:rho_P_prop-ratio_para_perp}
\end{figure}

The effects of vector meson decays are not considered
in the fit.
If a vector meson decays into two pseudoscalar mesons,
a vector meson
with the momentum $p=2 \pi / L$ will take a different energy
depending on whether it is polarized parallel or perpendicular
to the momentum direction,
because of mixing of one vector meson state and two 
pseudoscalar meson state \cite{rho_decay.MILC,rho_decay.UKQCD}.
We find no indication of vector meson decays 
as shown in Fig.~\ref{figure:rho_P_prop-ratio_para_perp}.
Our sea quark masses and the lattice size
do not seem to be enough to allow the decay.

\subsection{Decay constants, baryon masses and Sommer scale}
\label{section:decay_constants_baryon_masses_Sommer_scale}

\begin{figure}[h]
   \includegraphics[width=75mm]
   {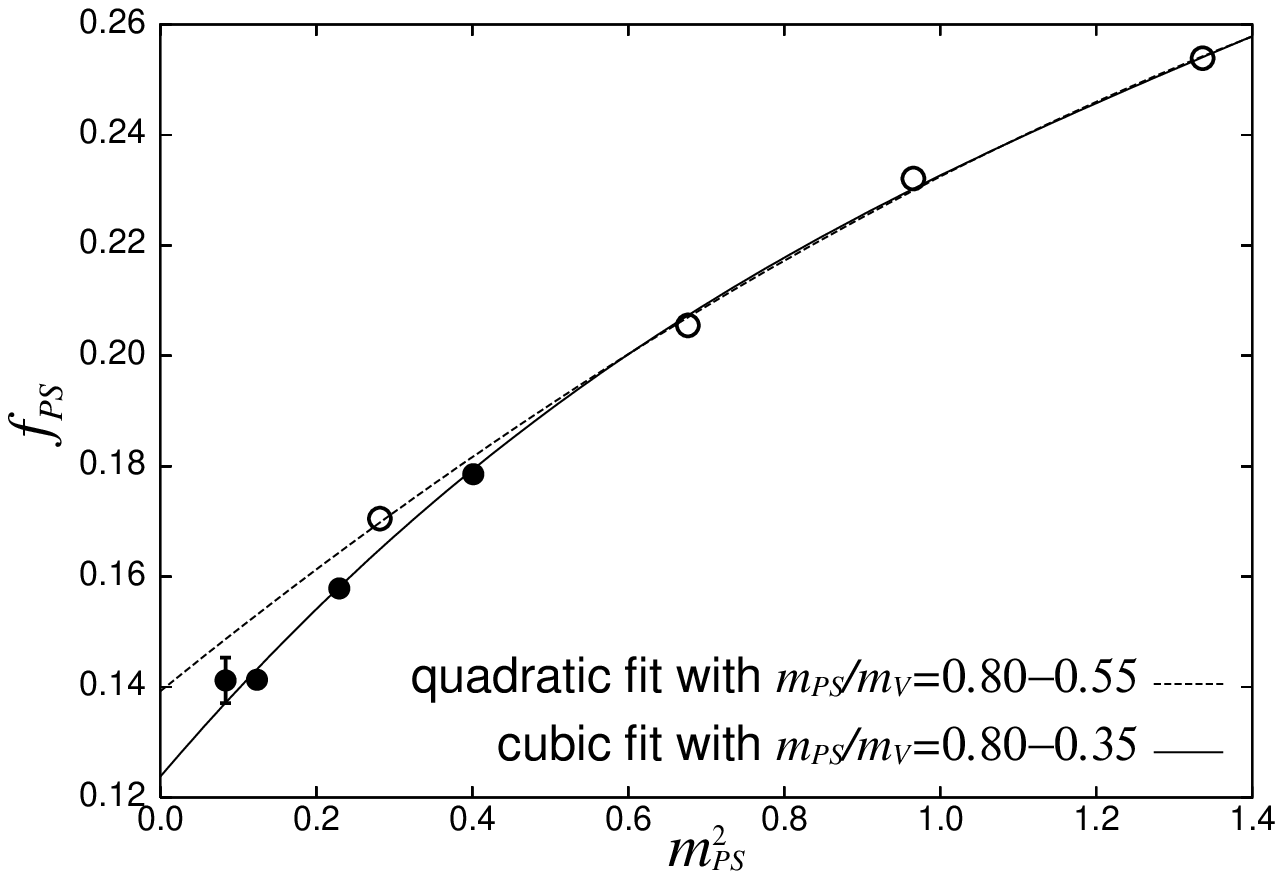}
   \includegraphics[width=75mm]
   {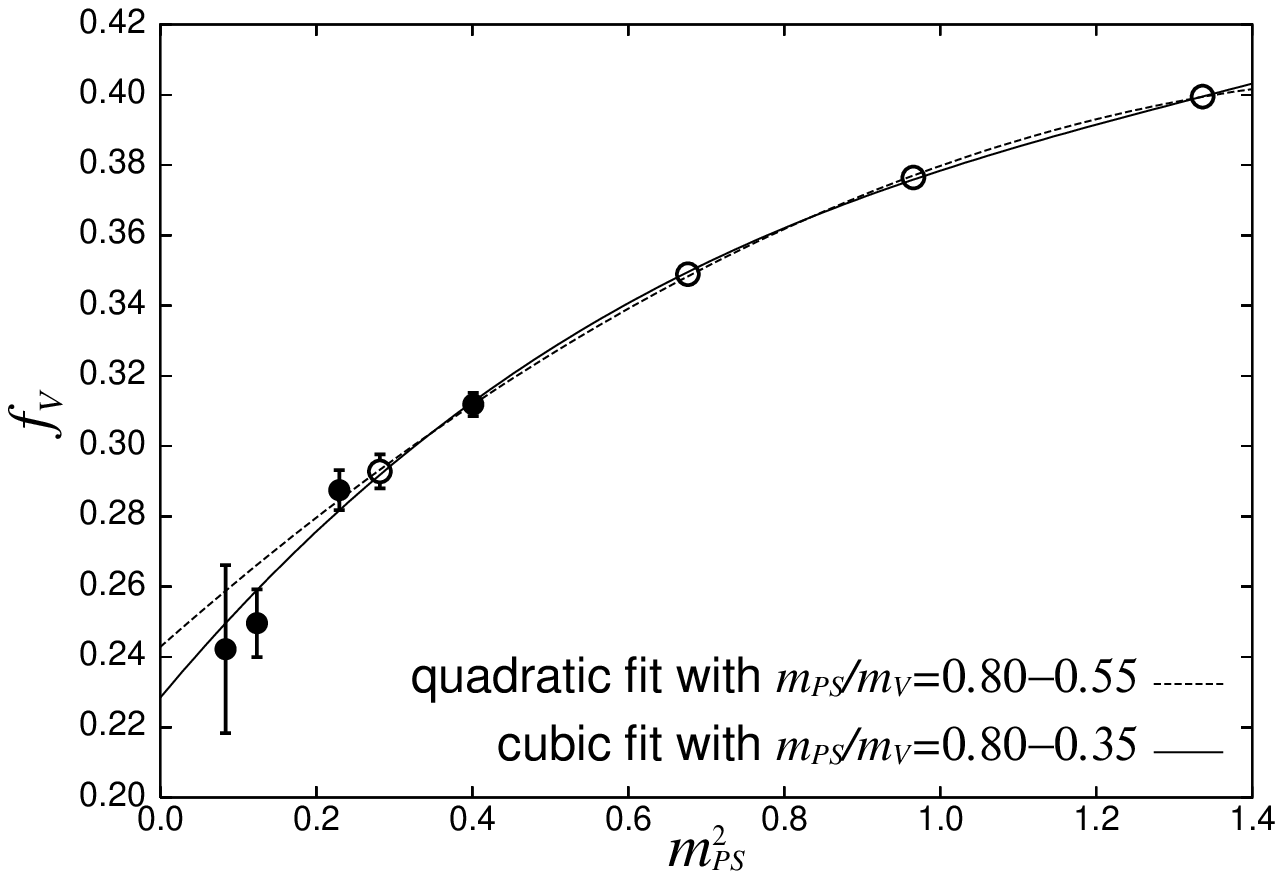}
   \caption
   {  Chiral extrapolation of pseudoscalar (left panel)
      and vector (right panel) meson decay constants.
      Open symbols show the results obtained
      in the previous calculation \protect\cite{Spectrum.Nf2.CP-PACS}.
   }
   \label{figure:m_PS2_f_PS_V}
\end{figure}

\begin{figure}[h]
 \includegraphics[width=75mm]
 {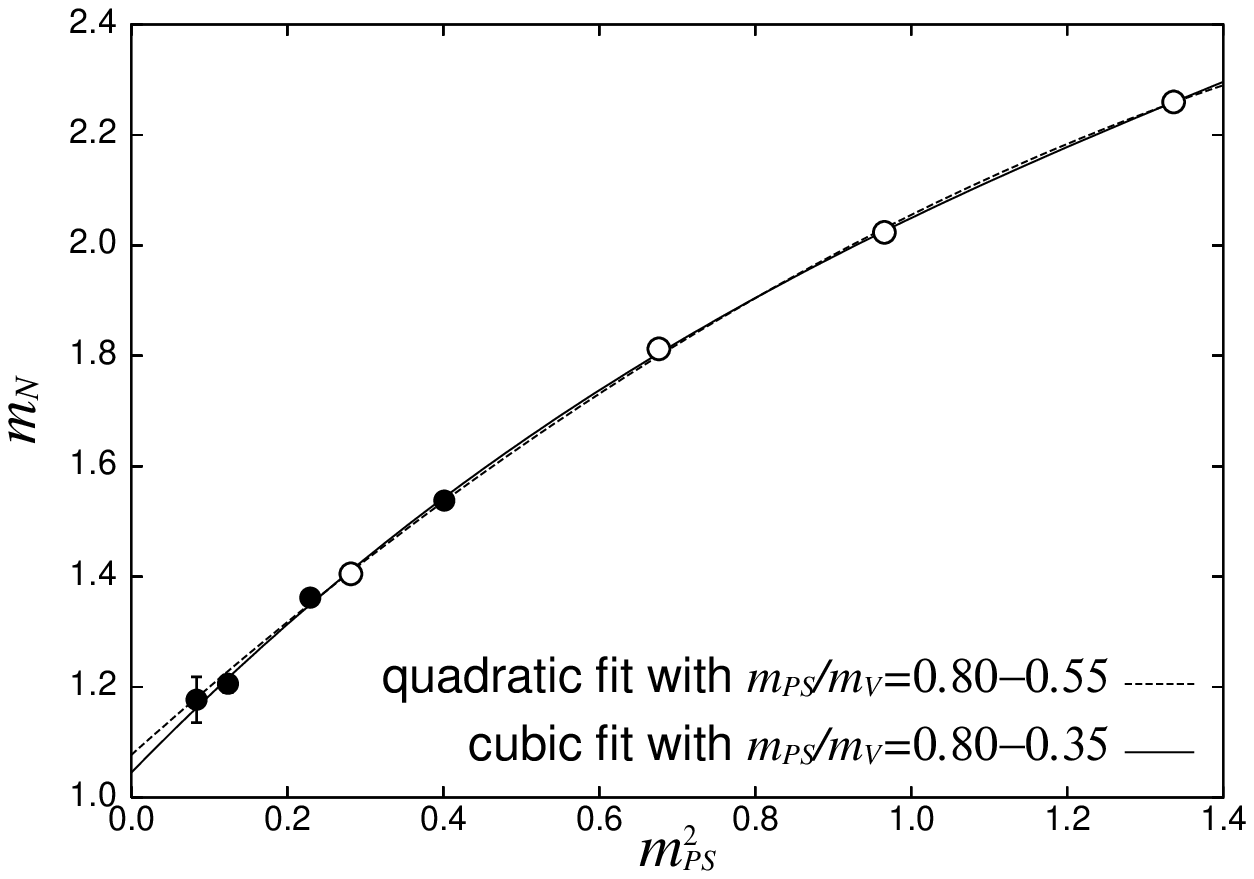}
 \includegraphics[width=75mm]
 {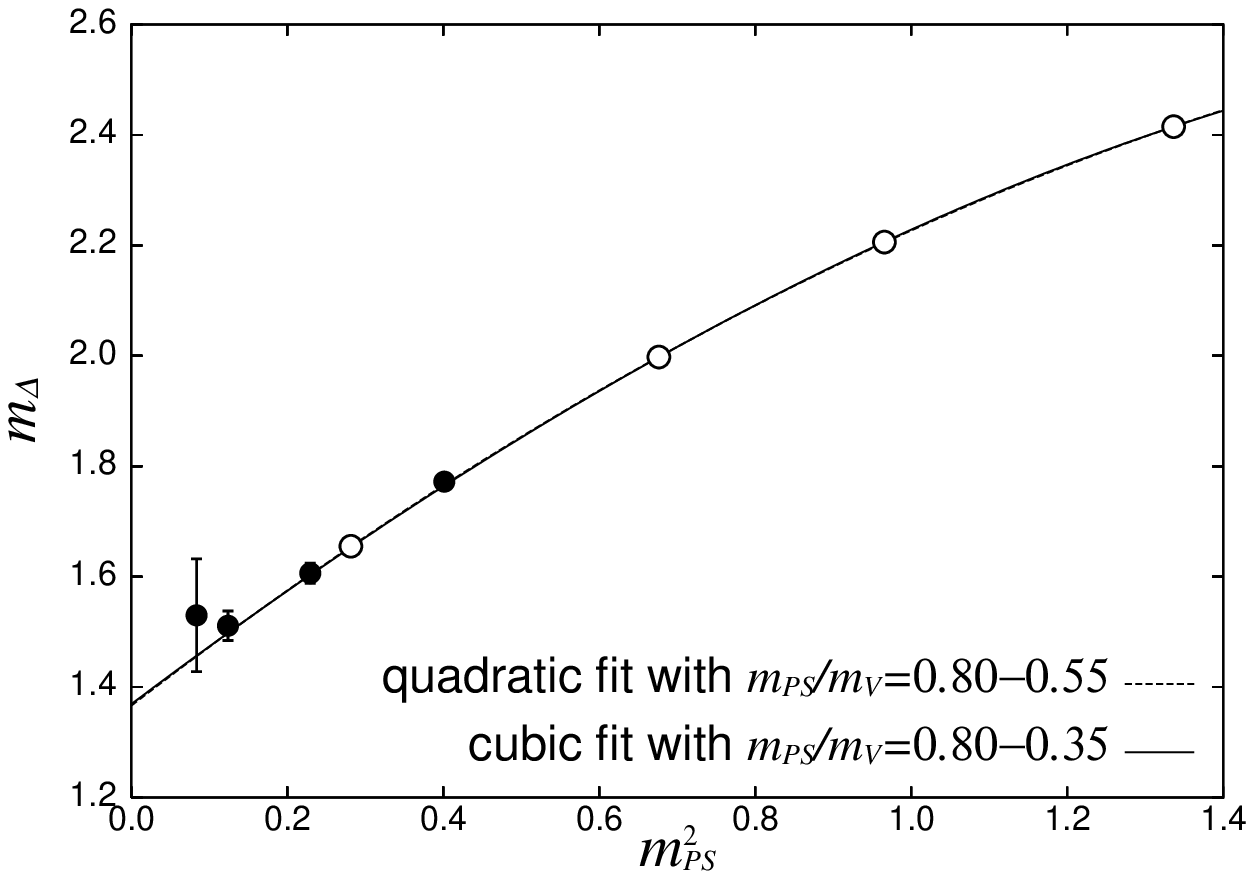}
   \caption
   { Chiral extrapolation of octet (left panel)
     and decuplet (right panel) baryon masses.
     Open symbols are the results in our previous study 
     \protect\cite{Spectrum.Nf2.CP-PACS}.
   }
   \label{figure:m_PS2_m_baryon}
\end{figure}

Chiral extrapolations are carried out for 
pseudoscalar and vector meson decay constants
and octet and decuplet baryon masses
using cubic polynomials in $m_{PS}^{2}$,
\begin{eqnarray}
 f_{PS,V} =  A^{f_{PS},f_{V}} + B^{f_{PS},f_{V}} m_{PS}^{2}
           + D^{f_{PS},f_{V}} m_{PS}^{4}
           + F^{f_{PS},f_{V}} m_{PS}^{6},\\
 \label{equation:m_PS2_decay_constants}
 m_{oct,dec} =  A^{oct,dec} + B^{oct,dec} m_{PS}^{2}
              + D^{oct,dec} m_{PS}^{4}
              + F^{oct,dec} m_{PS}^{6}.
 \label{equation:m_PS2_m_baryons}
\end{eqnarray}
The results are presented 
in Figs.~\ref{figure:m_PS2_f_PS_V} and \ref{figure:m_PS2_m_baryon}
and Tables~\ref{table:m_PS2_f_PS_V} and \ref{table:m_PS2_m_baryon}.
While the decay constants show clear deviations from 
the previous fit,
baryon masses are almost on the fit.
We gather that the latter is an accidental effect that is 
caused by a compensation of the downward shift of baryon masses
expected toward a small quark mass with an upward
finite-size shift caused by somewhat too small a lattice 
($L=2.4$~fm) for baryons (see Sec.~\ref{section:finite_size_effects}).

\begin{figure}[h]
   \includegraphics[width=75mm]
   {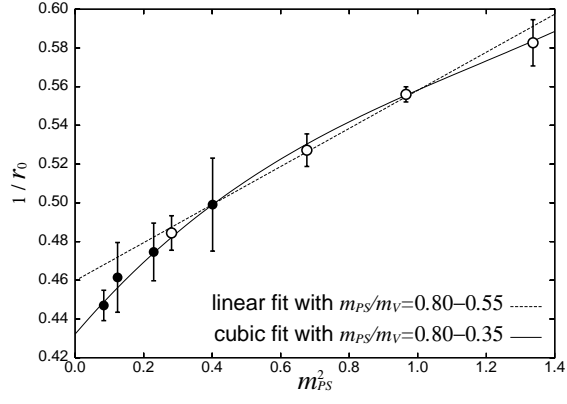}
   \caption{Chiral extrapolation of $r_0$.
            Open symbols are the results in our previous study 
		\protect\cite{Spectrum.Nf2.CP-PACS}.
           }
   \label{figure:m_PS2_r_0_inv}
\end{figure}

%
The Sommer scale $r_0$ is often extrapolated 
linearly in $m_{PS}^{2}$.
Since we find a curvature in our data, however,
we adopt the same form as that for the vector meson masses,
\begin{equation}
 \frac{1}{r_0} =  A^{r_0}
                + B^{r_0} m_{PS}^{2}
                + D^{r_0} m_{PS}^{4}
                + F^{r_0} m_{PS}^{6}.
 \label{equation:m_PS2_r_0_inv}
\end{equation}
The results are seen in Fig.~\ref{figure:m_PS2_r_0_inv}
and Table~\ref{table:m_PS2_r_0_inv}.


\subsection{Results at the physical point}

The physical point is defined by empirical
pion and $\rho$ meson masses,
$M_{\pi}=0.1350$~GeV and $M_{\rho}=0.7711$~GeV.
With our polynomial fit, 
the physical point $m_{\pi}$ for $m_{PS}$ is determined 
by solving the equation,
\begin{equation}
 \frac{ m_{\pi} }
      { A^{V} + B^{V} m_{\pi}^2 + D^{V} m_{\pi}^4 + F^{V} m_{\pi}^6 }
 =
 \frac{ M_{\pi} }
      { M_{\rho} }.
 \label{equation:chiral_lattice_spacing_rho}
\end{equation}
The $\rho$ meson mass at the physical point $m_{\rho}$ is
obtained by Eq.~(\ref{equation:m_PS2_m_V}) with $m_{PS}=m_{\pi}$,
which determines the lattice spacing $a_{\rho} = 0.2007(38)$~fm.
%
%
The lattice spacing can also be determined from $r_0$
taking its phenomenological value $R_0 = 0.49$~fm.
Using Eq.~(\ref{equation:m_PS2_r_0_inv})
instead of Eq.~(\ref{equation:m_PS2_m_V}),
we have
\begin{equation}
 \frac{ m_{\pi} }
      { A^{r_0} + B^{r_0} m_{\pi}^2 + D^{r_0} m_{\pi}^4 + F^{r_0} m_{\pi}^6 }
 = M_{\pi} R_{0}.
 \label{equation:chiral_lattice_spacing_r_0}
\end{equation}
Substitution of $m_{\pi}$ to Eq.~(\ref{equation:m_PS2_r_0_inv})
leads to $r_0$ at the physical point,
yielding an alternative lattice spacing $a_{r_0}$,
$a_{r_0} = 0.2119(61)$~fm, which
is consistent with $a_{\rho}$
within $2 \sigma$.

\begin{figure}[h]
\begin{center}
   \includegraphics[width=75mm]
   {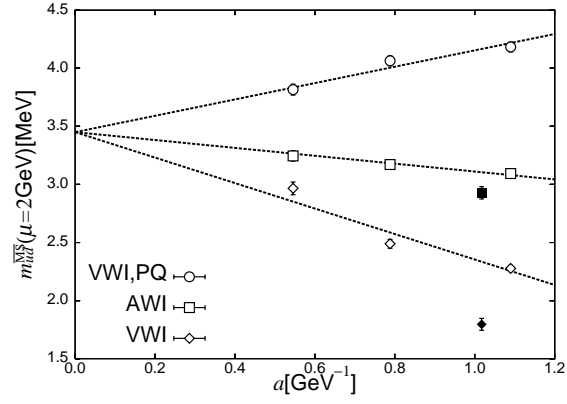}
   \caption
   { Comparison of
     degenerate up and down quark masses
     obtained by chiral extrapolations with polynomials.
     Open symbols show the results obtained
     in the previous calculation \protect\cite{Spectrum.Nf2.CP-PACS}
     and filled symbols are our new results.
     Lines are combined linear continuum extrapolations
     in the previous calculation.
   }
   \label{figure:a_inv-m_quark_VWI_AWI}
\end{center}
\end{figure}

We calculate $m_{ud}^{VWI}$ using $\kappa_{ud}$ defined by 
$m_{PS}(\kappa_{ud}) = m_{\pi}$, 
and $m_{ud}^{AWI}$ 
by Eq.~(\ref{equation:m_quark_VWI_m_quark_AWI}), and
then convert to renormalized quark masses
in the $\overline{\mbox{MS}}$ scheme at 2~GeV
(see Sec.~\ref{section:quark_mass}).
%
%
Table~\ref{table:comparisons_polynomials_previous_vs_all} presents
a summary of the parameters at the physical point,
obtained with polynomial extrapolations,  
together with comparisons with the quadratic fit in the previous work.
The difference between old and new results is generally 
4--8\% except for the VWI quark mass for which 
a difference more than 20\% is observed 
(see Fig.~\ref{figure:a_inv-m_quark_VWI_AWI}).
The latter is caused by a shift of $\kappa_c$, with which 
even a small shift leads to an amplified change in the mean
up and down quark mass.

\section{Chiral extrapolation based on ChPT}
\label{section:chiral_extrapolation_ChPT}

We first examine the one-loop formulae from continuum ChPT,
which have already been tested
in~\cite{Spectrum.Nf2.JLQCD,qq+q}.
We then attempt a fit based on WChPT including effects of 
$O(a^2)$ chiral symmetry violation due to the Wilson term.


\subsection{ChPT extrapolation}
\label{section:continuum_ChPT_extrapolation}

\begin{figure}[h]
   \includegraphics[width=75mm]
   {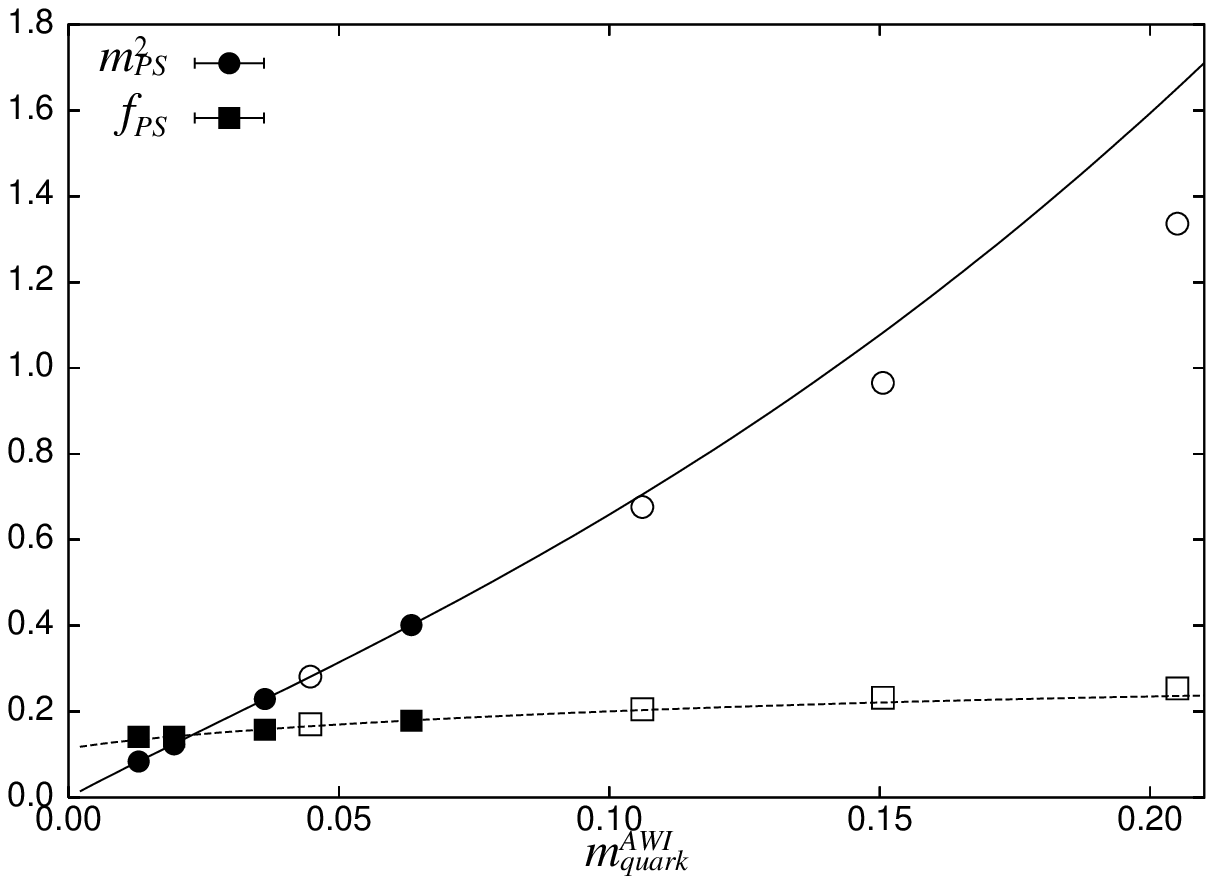}
   \includegraphics[width=75mm]
   {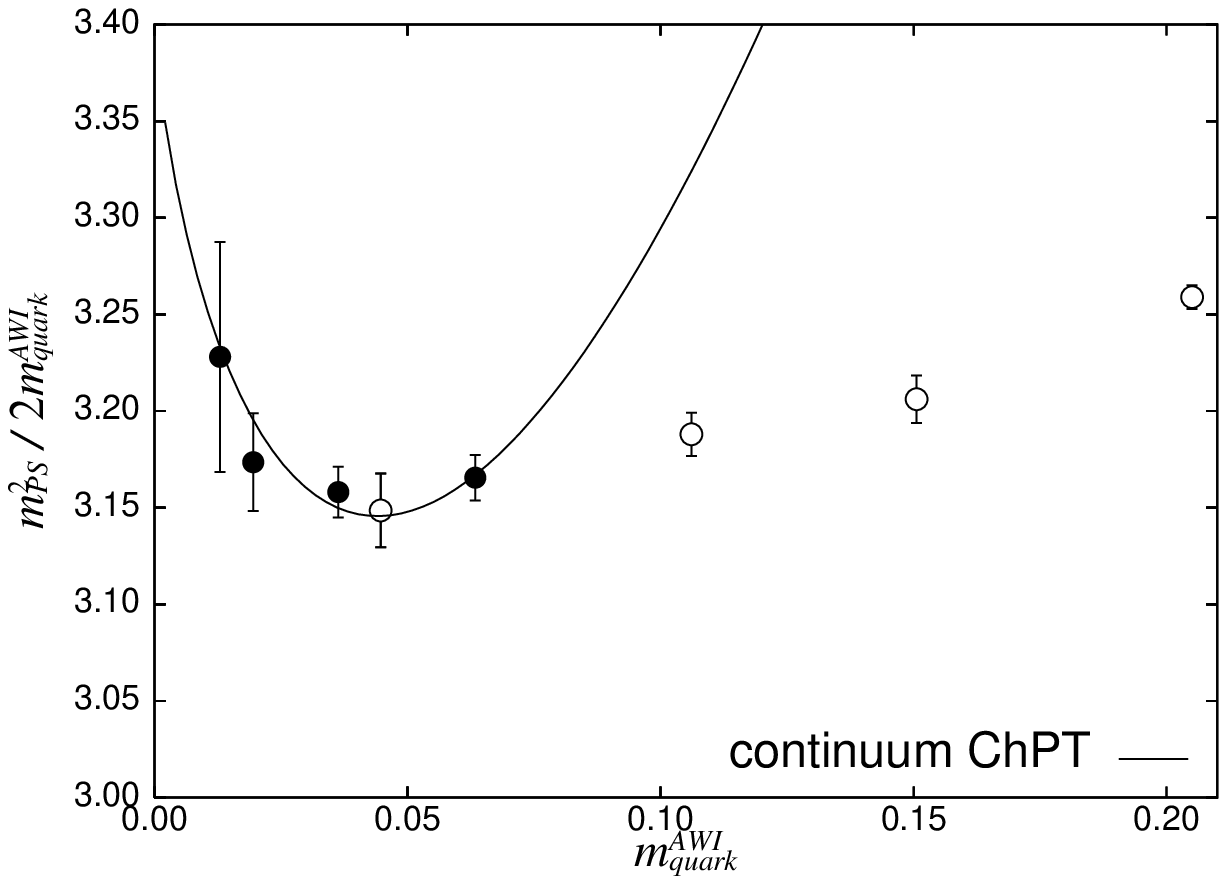}
   \caption
   {  Test of simultaneous continuum ChPT fit
      to pseudoscalar meson mass and decay constant.
      In this plot, 
      quark mass defined through 
      the axial vector Ward identity is used.
      The right panel shows the ratio $m_{PS}^2 / 2 m_{quark}^{AWI}$
      to focus on the chiral logarithm behavior.
      Open symbols are the results obtained
      in our previous study \protect\cite{Spectrum.Nf2.CP-PACS}.
   }
   \label{figure:chiral_log_m_quark_AWI_m_PS2_and_renormalized_f_PS}
\end{figure}

\begin{figure}[h]
   \includegraphics[width=75mm]
   {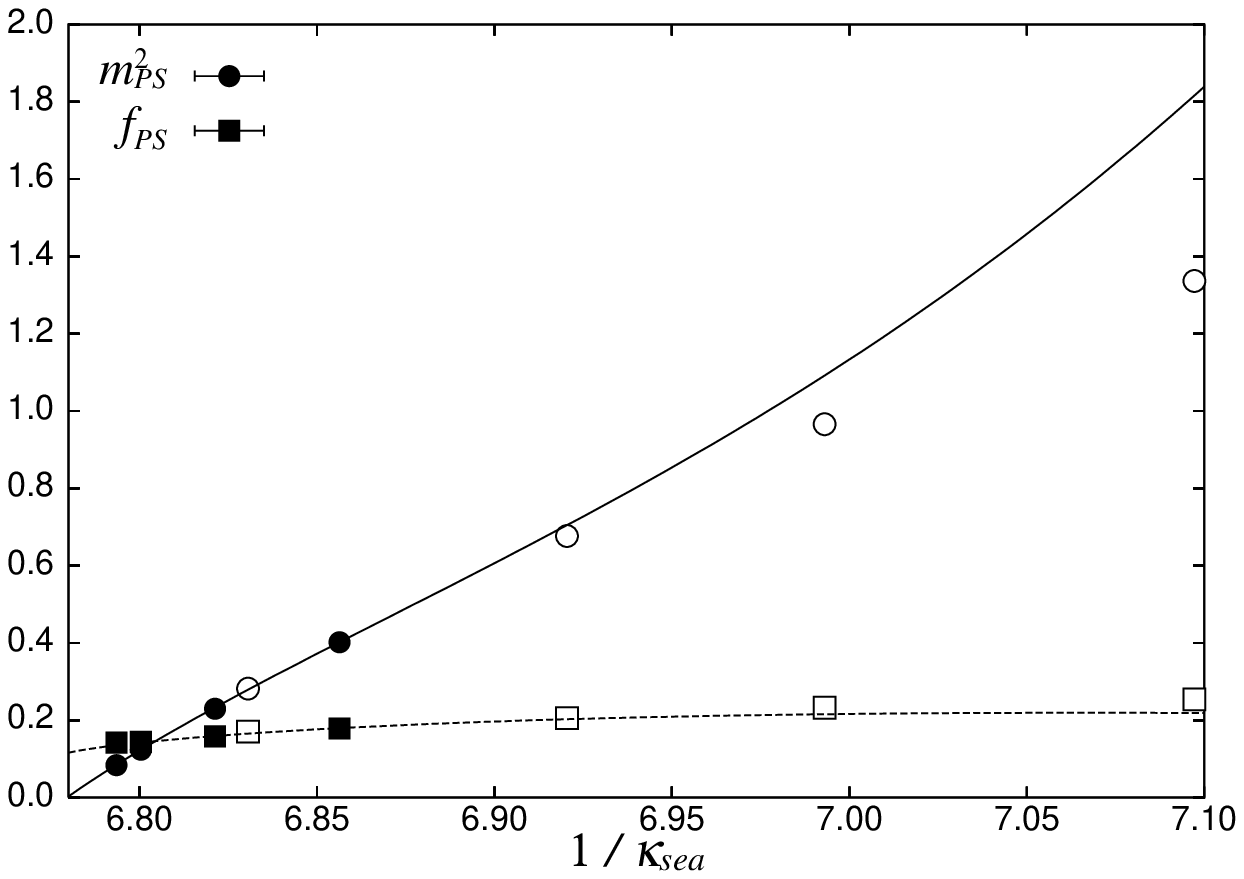}
   \includegraphics[width=75mm]
   {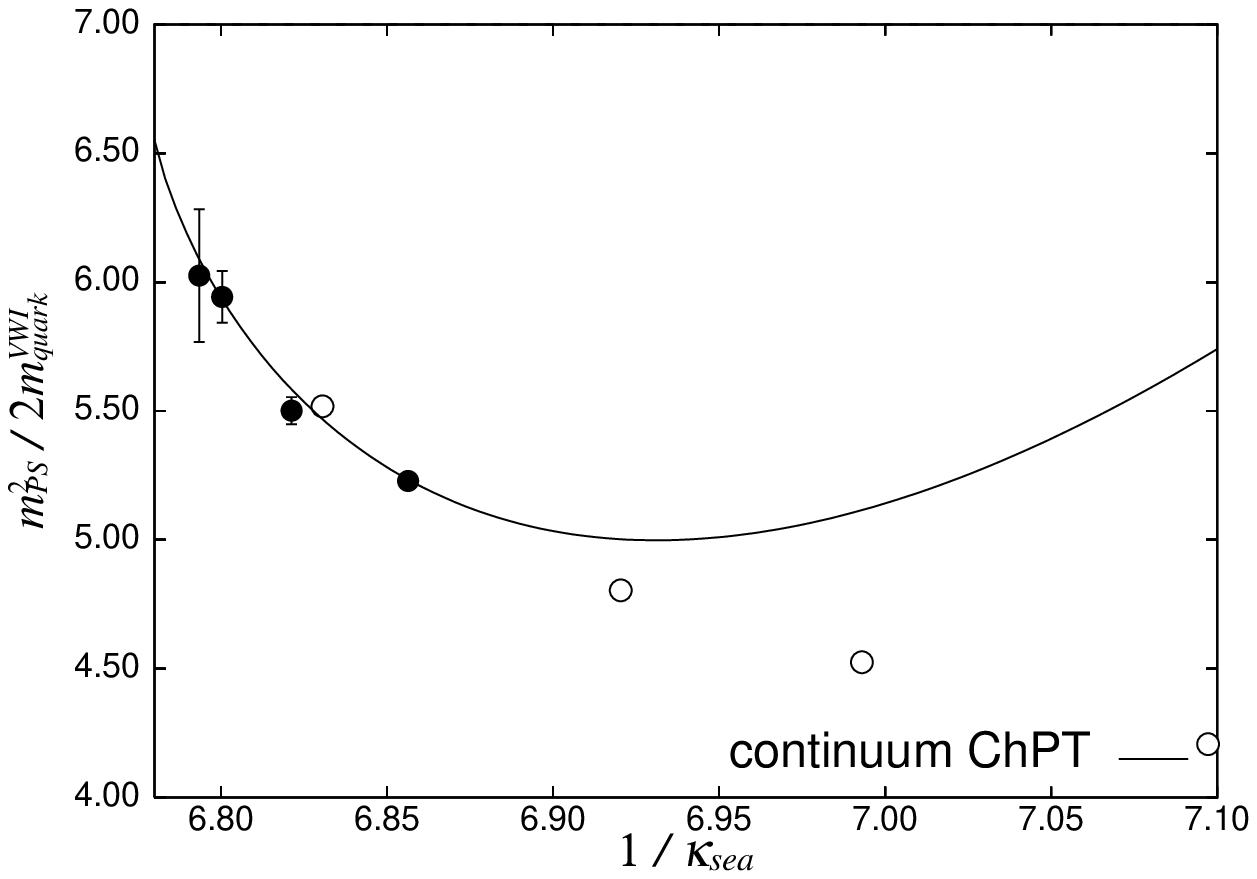}
   \caption
   { Test of simultaneous continuum ChPT fit
     with the quark mass defined through the vector Ward identity.
     Open symbols are the results obtained
     in our previous study \protect\cite{Spectrum.Nf2.CP-PACS}.
   }
   \label{figure:chiral_log_m_quark_VWI_m_PS2_and_renormalized_f_PS}
\end{figure}

The one-loop formulae~\cite{ChPT.Gasser,PQChPT.Durr} derived from
ChPT in the continuum limit are
\begin{eqnarray}
 \frac{m_{PS}^{2}}{2 B_{0} m_{quark}}
 &=& 1 + \frac{1}{2}
         \frac{2 B_0 m_{quark}}{(4 \pi f)^2}
         \log \frac{2 B_0 m_{quark}}{\Lambda_3^2}
 \label{equation:chiral_log_m_PS2} \\
 f_{PS}
 &=& f \left( 1 - 
                \frac{2 B_0 m_{quark}}{(4 \pi f)^2}
                \log \frac{2 B_0 m_{quark}}{\Lambda_4^2} \right),
 \label{equation:chiral_log_f_PS}
\end{eqnarray}
where $B_0$, $f$, $\Lambda_3$ and $\Lambda_4$ are
parameters to be obtained by fits.  
The coefficient $1/2$ in front of the logarithm is 
a distinctive prediction of ChPT. 
Since several parameters are common in the two formulae, 
we fit $m_{PS}^{2}$ and $f_{PS}$ simultaneously.
%
%
Correlations between $m_{PS}^2$ and $f_{PS}$ are neglected in the fits for
simplicity.
Thus, the $\chi^2/dof$ serves only as a guide to judge 
the relative quality of the fits.
We estimate the errors by the jackknife method.
We try both $m_{quark}^{AWI}$ and $m_{quark}^{VWI}$ 
(Cases 1 and 2 in what follows)
for $m_{quark}$ that appears in these formulae.
For $m_{quark} = m_{quark}^{VWI}$,
we use $\kappa_c$ determined
in Eq.~(\ref{equation:m_quark_VWI_m_quark_AWI})
since $m_{quark}^{AWI}$ has no logarithmic singularities
in ChPT.
From the fits summarized in 
Table~\ref{table:chiral_log_m_quark_AWI_VWI_m_PS2_and_renormalized_f_PS},
we find:
\begin{enumerate}
 \item[Case 1] ($m_{quark} = m_{quark}^{AWI}$) : 
 When we fit the data over the whole range $m_{PS}/m_{V}=0.80$--0.35,
 we are led to a large $\chi^2/dof \sim 70$. 
 By restricting the fitting interval to $m_{PS}/m_{V}=0.60$--0.35
 we obtain a reasonable fit with $\chi^2 / dof = 1.9$, which is plotted in 
 Fig.~\ref{figure:chiral_log_m_quark_AWI_m_PS2_and_renormalized_f_PS}. 
 As one observes in the second panel of this figure, 
 which shows $m_{PS}^{2}/2 m_{quark}^{AWI}$
 appearing in the left hand side
 of Eq.(\ref{equation:chiral_log_m_PS2}), the chiral logarithm
 may be visible only at $m_{PS}/m_{V} \simlt 0.40$.
%
%
 \item[Case 2] ($m_{quark} = m_{quark}^{VWI}$) : 
 In contrast to Case 1,
 $m_{PS}^{2}/2 m_{quark}^{VWI}$
 increases toward the chiral limit 
 in the whole mass range, which is seen in 
 Fig.~\ref{figure:chiral_log_m_quark_VWI_m_PS2_and_renormalized_f_PS}.
 Nevertheless, the situation is similar.
 A fit 
 over the whole range $m_{PS}/m_{V}=0.80$--0.35
 leads to $\chi^2/dof \sim 100$.
 To obtain an acceptable fit, we have to remove the  
 data at large quark masses. 
 The best fit obtained for the range $m_{PS}/m_{V}=0.60$--0.35 
 is shown in 
 Fig.~\ref{figure:chiral_log_m_quark_VWI_m_PS2_and_renormalized_f_PS}.
\end{enumerate}
In neither case do we draw the clear evidence for 
the chiral logarithm for pseudoscalar mesons.

\begin{figure}[h]
   \includegraphics[width=75mm]
   {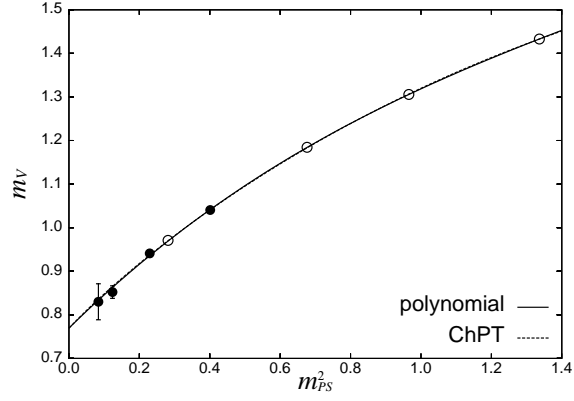}
   \caption
   { Chiral extrapolation of vector meson mass
     with a polynomial in Eq.~(\ref{equation:m_PS2_m_V})
     and a function motivated by ChPT in Eq.~(\ref{equation:m_PS2_m_V_ChPT}).
     Open symbols are the results obtained
     in our previous study \protect\cite{Spectrum.Nf2.CP-PACS}.
   }
   \label{figure:m_PS2_m_V_Poly_ChPT}
\end{figure}

For the vector meson, we adopt the formula 
based on ChPT in the static limit~\cite{ChPT.Jenkins}.
\begin{equation}
 m_{V} =  A^V + B^{V} m_{PS}^{2}
        + C^{V} m_{PS}^{3}.
 \label{equation:m_PS2_m_V_ChPT}
\end{equation}
This cubic form describes our data well as shown 
in Fig.~\ref{figure:m_PS2_m_V_Poly_ChPT}
(see Table~\ref{table:m_PS2_m_V_ChPT} for numbers).

\begin{figure}[h]
 \includegraphics[width=75mm]
 {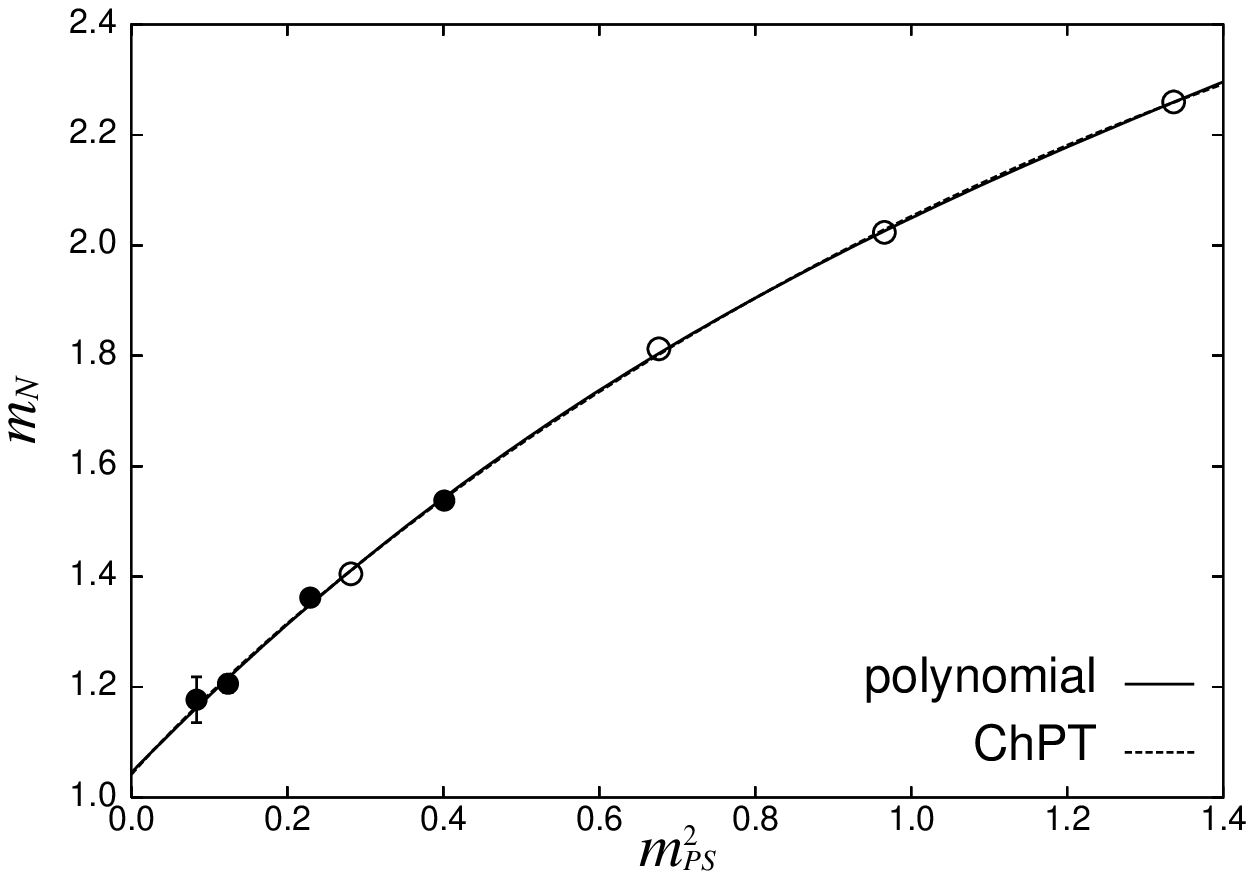}
 \includegraphics[width=75mm]
 {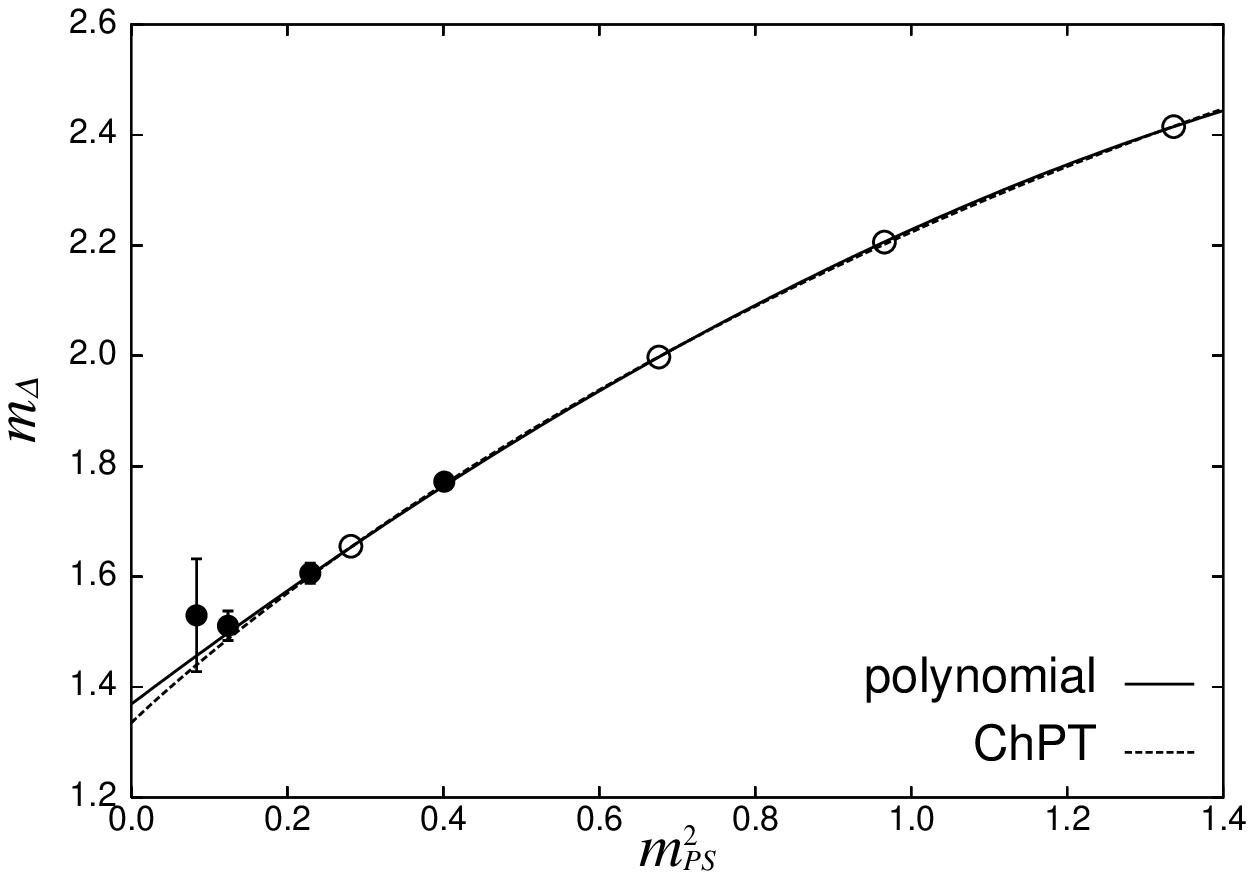}
   \caption
   { Chiral extrapolation of octet (left panel)
     and decuplet (right panel) baryon masses
     with polynomials in Eq.~(\ref{equation:m_PS2_m_baryons})
     and functions motivated by ChPT
     in Eq.~(\ref{equation:m_PS2_m_baryon_ChPT}).
     Open symbols are the results obtained
     in our previous study \protect\cite{Spectrum.Nf2.CP-PACS}.
   }
   \label{figure:m_PS2_m_baryon_ChPT}
\end{figure}

For octet and decuplet baryons 
we employ a similar cubic formula~\cite{ChPT.Bernard}
\begin{equation}
 m_{oct,dec} =  A^{oct,dec} + B^{oct,dec} m_{PS}^{2}
              + C^{oct,dec} m_{PS}^{3},
 \label{equation:m_PS2_m_baryon_ChPT}
\end{equation}
which also reproduces our data well 
(Fig.~\ref{figure:m_PS2_m_baryon_ChPT}
and Table~\ref{table:m_PS2_m_baryon_ChPT}).

In order to present predictions at the physical point, 
we carry out extrapolations using the data at $m_{PS}/m_{V}=0.60$--0.35.
From Eq.~(\ref{equation:m_PS2_m_V_ChPT}) 
the physical point $m_{\pi}$ for $m_{PS}$ is given by 
\begin{equation}
 \frac{ m_{\pi} }
      { A^{V} + B^{V} m_{\pi}^2 + C^{V} m_{\pi}^3 }
 =
 \frac{ M_{\pi} }
      { M_{\rho} }.
 \label{equation:chiral_lattice_spacing_ChPT}
\end{equation}
The lattice spacing is determined to be $a_{\rho}^{ChPT} = 0.192(10)$~fm.
For the vector meson, a fit for the whole range 
$m_{PS}/m_{V}=0.80$--0.35 is 
acceptable, as seen in Table~\ref{table:m_PS2_m_V_ChPT}.
We will use this fit in Sec.~\ref{section:WChPT_extrapolation}
with $a_{\rho}^{ChPT} = 0.2009(21)$~fm for this case.

The masses of non-strange baryons $N$ and $\Delta$
are determined by substituting $m_{\pi}$
to $m_{PS}$ in Eq.~(\ref{equation:m_PS2_m_baryon_ChPT}).
The bare quark mass at the physical point $m_{ud}$ and 
the pion decay constant $f_{\pi}$ are obtained  
from Eqs.~(\ref{equation:chiral_log_m_PS2})
and (\ref{equation:chiral_log_f_PS}).
Renormalized quark masses are calculated with $m_{ud}$
as in the case of polynomial extrapolations.
These results are compiled
in Table~\ref{table:physical_quantities_ChPT}.

%
We observe 5--10\% differences between the ChPT fits 
over $m_{PS}/m_V=0.60$--0.35 and the quadratic polynomial fits
over $m_{PS}/m_V=0.80$--0.55 obtained in the previous work. 
The numbers are tabulated in 
Table~\ref{table:comparisons_polynomials_previous_vs_all}. 
These differences are similar in magnitude to those we found with 
higher order polynomial extrapolations using the whole range  
$m_{PS}/m_V=0.80$--0.35. An exception is the VWI quark mass on which
we shall make a further comment below.

\subsection{WChPT extrapolation}
\label{section:WChPT_extrapolation}


\subsubsection{WChPT without resummation}
\label{section:orig_WChPT}

ChPT adapted to Wilson-type quark actions 
on the lattice (WChPT) has been addressed in 
Refs.~\cite{WChPT.Sharpe,WChPT.Rupak,WChPT.Oliver,WChPT.Sinya}.
An important point \cite{WChPT.Sinya} is that $O(a^2)$ chiral breaking terms
in the chiral Lagrangian are essential  
to generate the parity-flavor breaking
phase transition~\cite{WChPT.Sharpe},
which is necessary to explain the existence of massless pions for Wilson-type
quark actions~\cite{Aoki_phase.Sinya1,Aoki_phase.Sinya2,Aoki_phase.Sinya3}.
Therefore, we must include the $O(a^2)$ terms in the leading order.   
In this counting scheme, 
the one-loop formulae read~\cite{WChPT.Sinya},
\begin{eqnarray}
 m_{PS}^2 &=& A m_{quark}^{VWI}
              \left( 1 + \omega_1^{PS} m_{quark}^{VWI}
                         \log \left( \frac{A m_{quark}^{VWI}}
                                          {\Lambda_3^2} \right)
                       + \omega_0
                         \log \left( \frac{A m_{quark}^{VWI}}
                                          {\Lambda_0^2} \right)
              \right),
 \label{equation:chiral_log_m_PS2_WChPT} \\
 m_{quark}^{AWI} &=& m_{quark}^{VWI}
                     \left( 1 + \omega_1^{AWI} m_{quark}^{VWI}
                                \log \left( \frac{A m_{quark}^{VWI}}
                                                 {\Lambda_{3,AWI}^2} \right)
                              + \omega_0
                                \log \left( \frac{A m_{quark}^{VWI}}
                                                 {\Lambda_{0}^2} \right)
                     \right),
 \label{equation:chiral_log_m_quark_AWI_WChPT} \\
 f_{PS} &=& f \left( 1 - \omega_1^{f_{PS}} m_{quark}^{VWI}
                         \log \left( \frac{A m_{quark}^{VWI}}
                                          {\Lambda_4^2} \right)
              \right).
 \label{equation:chiral_log_f_PS_WChPT}
\end{eqnarray}
Here $\kappa_c$ in $m_{quark}^{VWI}$, $A$, $f$,
$\omega_0$, $\omega_1^{PS}$, $\omega_1^{AWI}$, $\omega_1^{f_{PS}}$,
$\Lambda_3$, $\Lambda_{3,AWI}$, $\Lambda_0$ and $\Lambda_4$
are free parameters, and 
the overall factor of $m_{quark}^{AWI}$
is absorbed in $\omega_0$ and $\Lambda_{0}$.
We note that $A$ consists of $O(a^0)$ and $O(a^1)$ parts,
$\omega_0 \sim O(a^2)$, $\omega_1^{AWI} \sim O(a)$ and
\begin{eqnarray}
 \omega_1^{PS}
 &=& \frac{1}{2}
     \frac{ ( A + w_1^{\pi} a) }{(4 \pi f_{0})^2},
\label{equation:WChPT_omega_1_PS}\\
 \omega_1^{f_{PS}}
 &=& \frac{ ( A + w_1^{f_{\pi}} a) }{(4 \pi f_{0})^2},
\label{equation:WChPT_omega_1_f_PS}
\end{eqnarray}
where $f_{0}$ is the pion decay constant
in the continuum and chiral limit,
which can be different from $f$ by $O(a)$. 
The constants $w_1^{\pi}$ and $w_1^{f_{\pi}}$ are $O(a^0)$.

There are two features in these formulae worth emphasizing.
First, 
the coefficients of $m_{quark}\log m_{quark}$ 
terms receive contributions of $O(a)$.  
This is in contrast to continuum ChPT, in which these 
coefficients take universal values.
Second, there are terms of the form $a^2\log m_{quark}$ 
which are more singular 
than the $m_{quark}\log m_{quark}$ terms toward the chiral limit at 
a finite lattice spacing. 
Thus WChPT formulae predict the chiral behavior at finite lattice 
spacings that is different from what is expected from 
ChPT in the continuum limit.

\begin{figure}[h]
   \includegraphics[width=75mm]
   {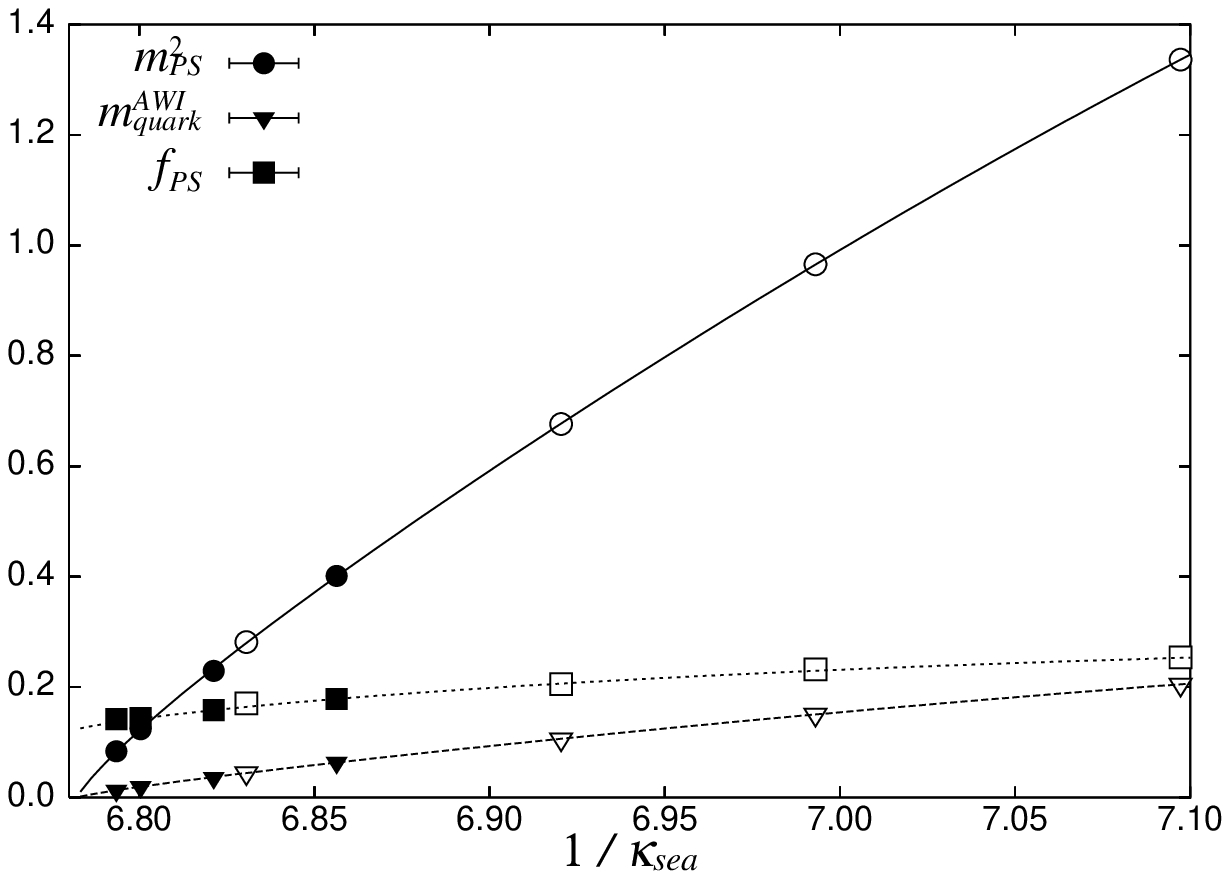}
   \includegraphics[width=75mm]
   {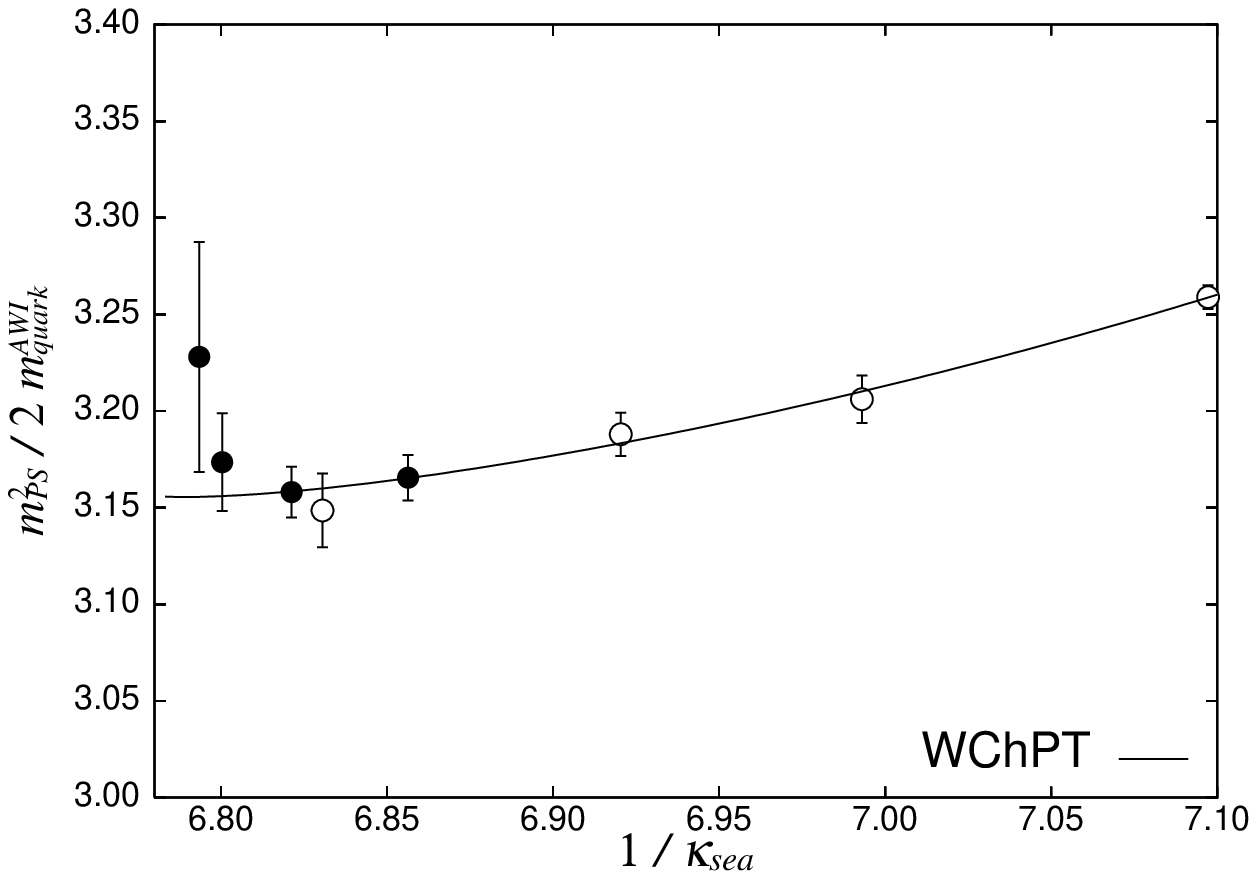}
   \caption
   { Test of the WChPT fit
     to pseudoscalar meson mass, AWI quark mass and decay constant.
     The right panel shows the ratio $m_{PS}^2 / 2 m_{quark}^{AWI}$
     to focus on the chiral logarithm behavior.
     Open symbols are the results obtained
     in our previous study \protect\cite{Spectrum.Nf2.CP-PACS}.
   }
   \label{figure:chiral_log_m_PS2_m_quark_AWI_renormalized_f_PS_WChPT}
\end{figure}

We fit $m_{PS}$ and $m_{quark}^{AWI}$ simultaneously,
neglecting correlations between them.
The errors are estimated by the jackknife method.
We then fit $f_{PS}$ with $A$ and $\kappa_c$ fixed from 
Eqs.~(\ref{equation:chiral_log_m_PS2_WChPT})
and (\ref{equation:chiral_log_m_quark_AWI_WChPT}).
%
%
We give the results in
Fig.~\ref{figure:chiral_log_m_PS2_m_quark_AWI_renormalized_f_PS_WChPT}
and
Tables~\ref{table:chiral_log_m_PS2_m_quark_AWI_WChPT}
and \ref{table:chiral_log_renormalized_f_PS_WChPT}.
Fig.~\ref{figure:chiral_log_m_PS2_m_quark_AWI_renormalized_f_PS_WChPT}
demonstrates that   
the one-loop WChPT formulae explain our data
over the whole range $m_{PS}/m_{V}=0.80$--0.35.

\subsubsection{Resummed WChPT}
\label{section:resummed_WChPT}

While fits with Eqs.~(\ref{equation:chiral_log_m_PS2_WChPT})
and (\ref{equation:chiral_log_m_quark_AWI_WChPT})
work well for the whole 
range of quark mass we measured,
extrapolation to the physical point is still
problematic because the $\omega_0 \log m_{quark}^{VWI}$ terms 
become larger than the leading terms in the chiral limit.
A way out has been proposed in Ref.~\cite{WChPT.Sinya} in which 
leading singularities around the chiral limit is resummed. 
The resulting formulae read, 
\begin{eqnarray}
 m_{PS}^2 &=& A m_{quark}^{VWI}
              \left( - \log \left( \frac{A m_{quark}^{VWI}}
                                        {\Lambda_0^2} \right)
              \right)^{\omega_0}
              \left( 1 + \omega_1^{PS} m_{quark}^{VWI}
                         \log \left( \frac{A m_{quark}^{VWI}}
                                          {\Lambda_3^2} \right)
              \right),
 \label{equation:chiral_log_m_PS2_resummed_WChPT} \\
 m_{quark}^{AWI} &=& m_{quark}^{VWI}
                     \left( - \log \left( \frac{A m_{quark}^{VWI}}
                                               {\Lambda_0^2} \right)
                     \right)^{\omega_0}
                     \left( 1 + \omega_1^{AWI} m_{quark}^{VWI}
                                \log \left( \frac{A m_{quark}^{VWI}}
                                                 {\Lambda_{3,AWI}^2} \right)
                     \right),
 \label{equation:chiral_log_m_quark_AWI_resummed_WChPT}
\end{eqnarray}
where the fitting parameters are $\kappa_c$ in $m_{quark}^{VWI}$, $A$,
$\omega_0$, $\omega_1^{PS}$, $\omega_1^{AWI}$,
$\Lambda_3$, $\Lambda_{3,AWI}$ and $\Lambda_0$.
The minus sign in the resummed part
is introduced to keep
$- \log ( A m_{quark}^{VWI} / \Lambda_0^2 )$ positive.
We note that
$f_{PS}$ is not affected by the resummation except for 
a shift of $\kappa_c$.

\begin{figure}[h]
   \includegraphics[width=75mm]
   {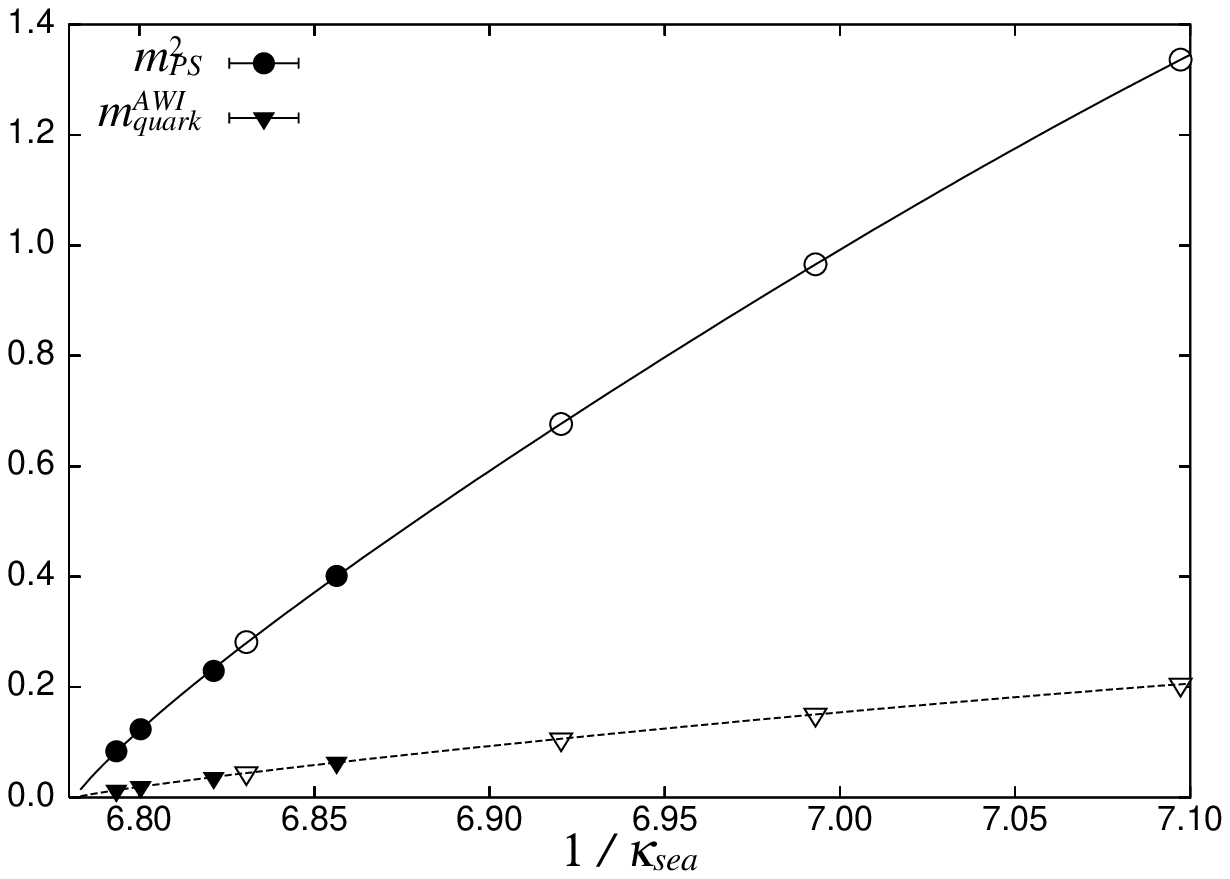}
   \includegraphics[width=75mm]
   {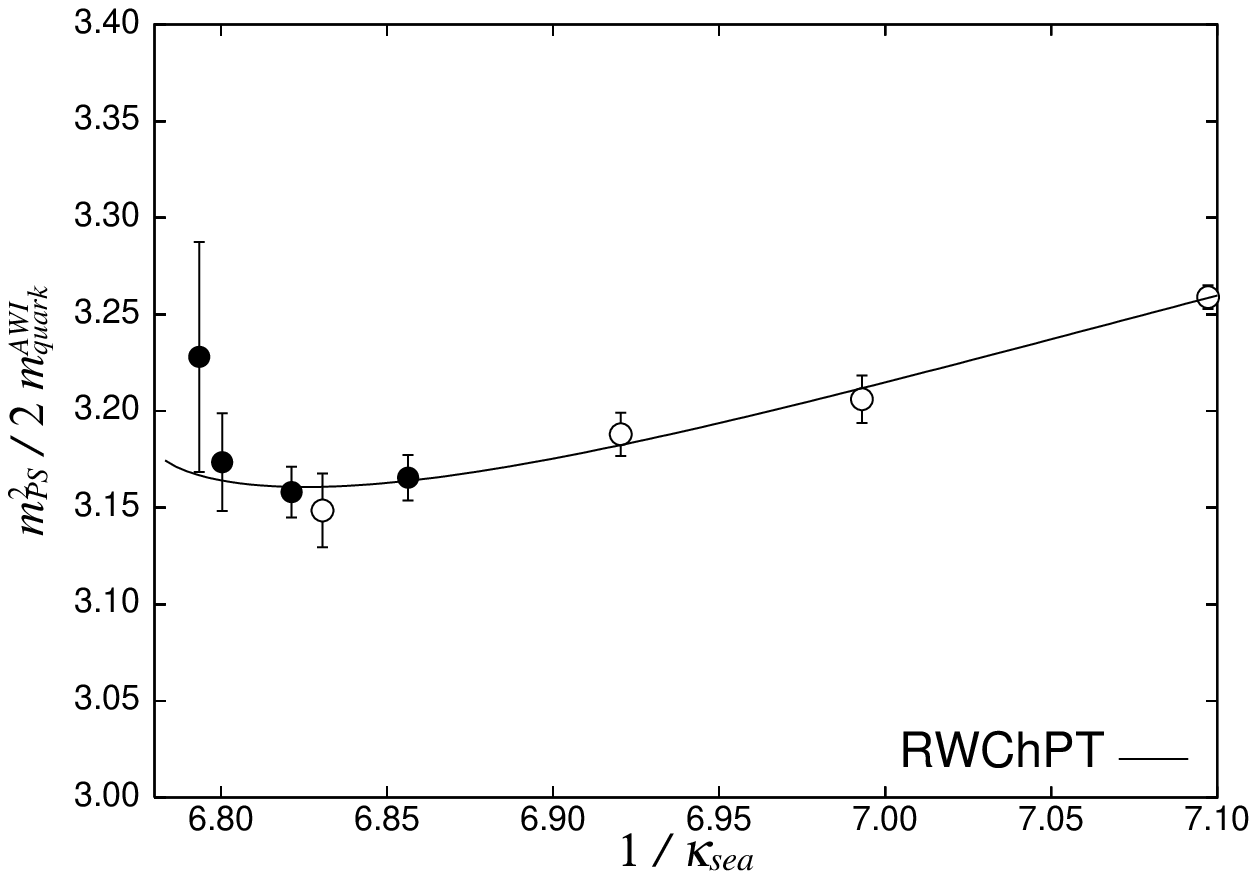}
   \caption
   { Test of the resummed WChPT fit
     to pseudoscalar meson mass and AWI quark mass.
     The right panel shows the ratio $m_{PS}^2 / 2 m_{quark}^{AWI}$
     to focus on the chiral logarithm behavior.
     Open symbols are the results obtained
     in our previous study \protect\cite{Spectrum.Nf2.CP-PACS}.
   }
   \label{figure:chiral_log_m_PS2_m_quark_AWI_renormalized_f_PS_resummed_WChPT}
\end{figure}

As with the case of WChPT without resummation, 
these resummed WChPT formulae describe our data for the whole 
range of $m_{PS}/m_{V}=0.80$--0.35.
The results are seen in 
Fig.~\ref{figure:chiral_log_m_PS2_m_quark_AWI_renormalized_f_PS_resummed_WChPT} 
and
Tables~\ref{table:chiral_log_m_PS2_m_quark_AWI_resummed_WChPT}
and \ref{table:chiral_log_renormalized_f_PS_resummed_WChPT}.

\begin{figure}[h]
   \includegraphics[width=75mm]
   {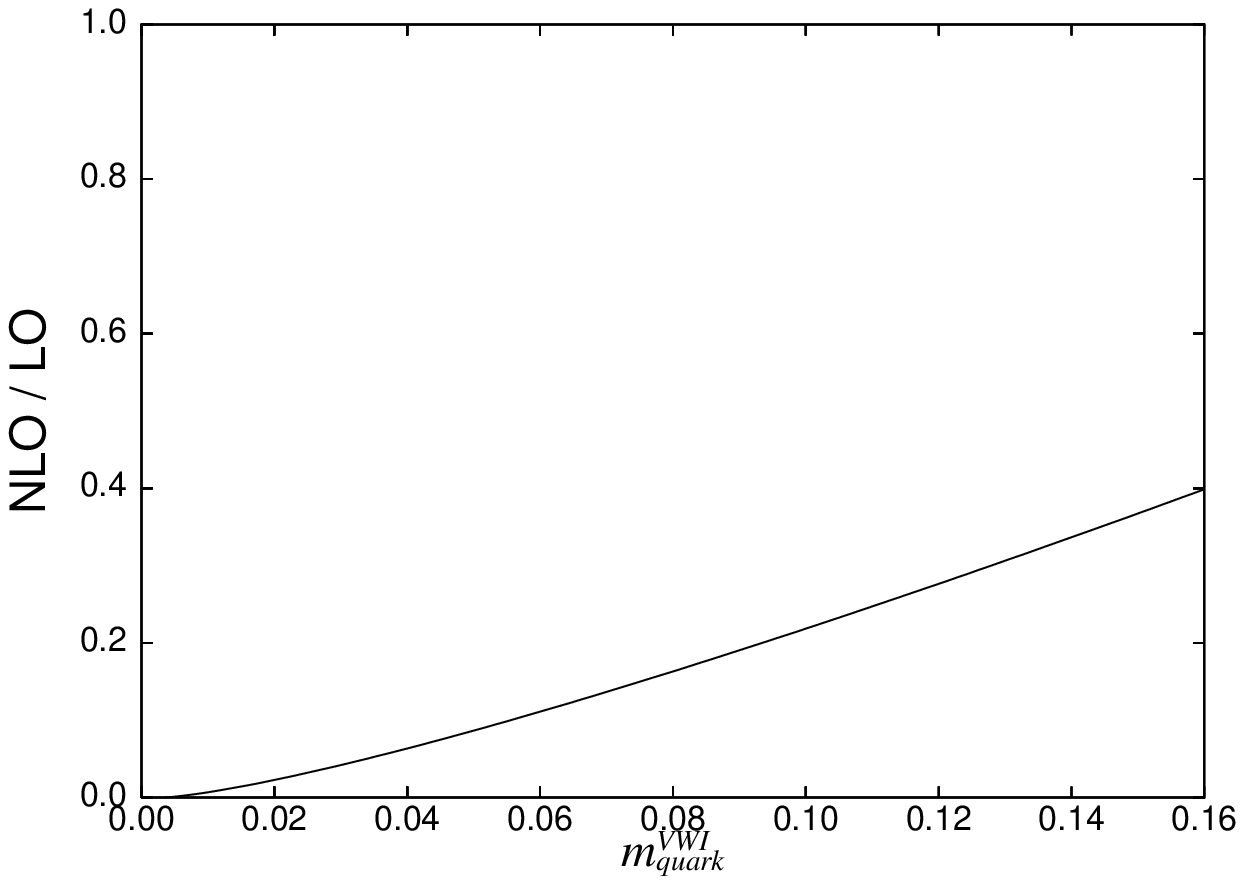}
   \includegraphics[width=75mm]
   {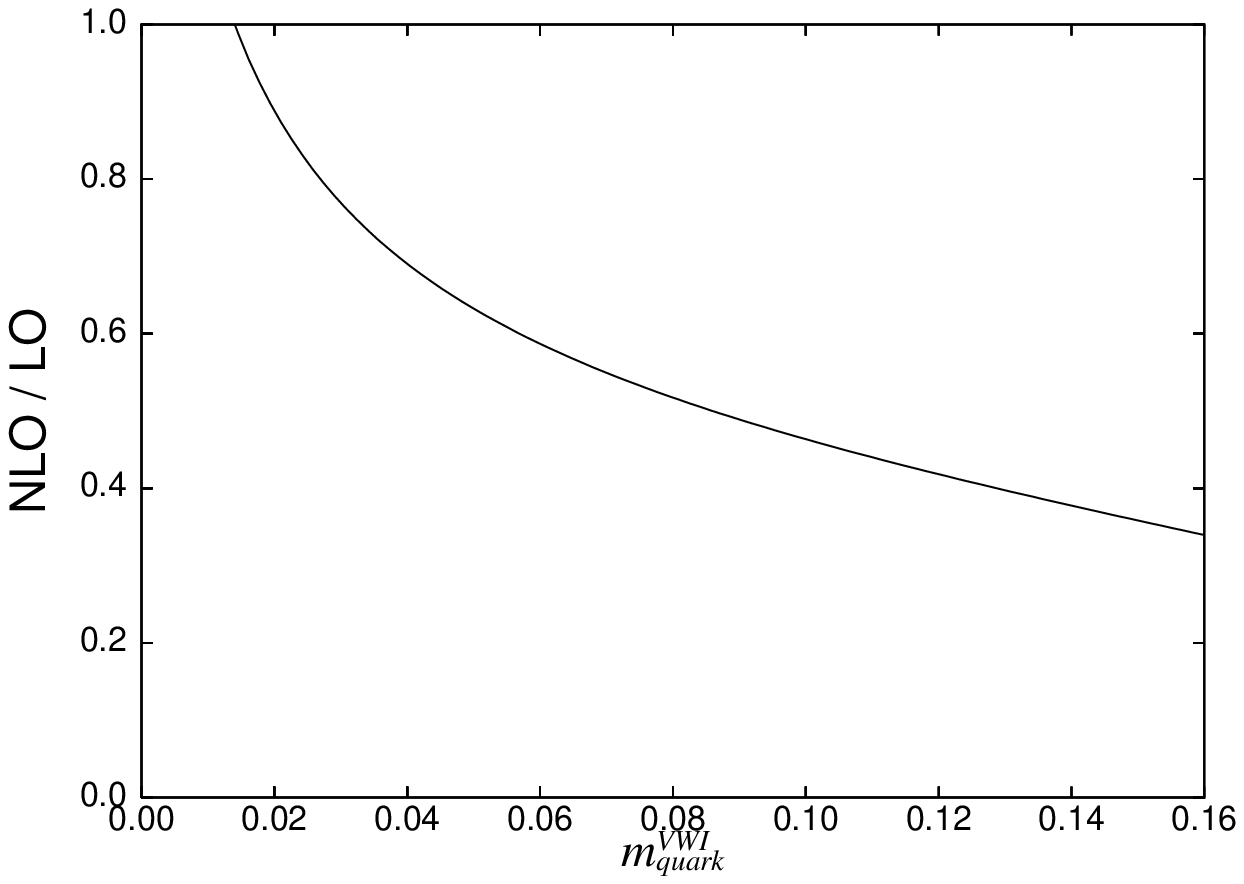}
   \caption
   {  Ratio of the next-to-leading order term to
      the leading one for $m_{PS}^2$
      with the resummed WChPT formulae (left panel)
      and with WChPT formulae without resummation (right panel) 
      as a function of $m_{quark}^{VWI}$.
   }
   \label{figure:m_quark_VWI-m_PS2_m_quark_AWI-resummed_WChPT-contribution}
\end{figure}

The magnitude of the leading and the one-loop contributions is plotted in 
Fig.~\ref{figure:m_quark_VWI-m_PS2_m_quark_AWI-resummed_WChPT-contribution} 
as a function of $m_{quark}$.
In contrast to WChPT without resummation, which is shown
in the second panel of the figure, 
the one-loop contribution of resummed WChPT fit remains small
in the the whole range of quark mass we explored, including 
the chiral limit.
This confirms the convergence of the resummed WChPT formulae.  
Furthermore, the resulting parameters are comparable with 
phenomenological estimates; we obtain $\Lambda_3 = 0.15(15)$~[GeV] and
$\Lambda_4 = 2.44(13)$~[GeV] as compared to 
$\Lambda_3 = 0.2$--2.0~[GeV] and 
$\Lambda_4 = 1.26(14)$~[GeV], respectively, 
from Ref.~\cite{ChPT.Gasser,Lambda_4.Colangelo}.
A more accurate examination requires extrapolation to the 
continuum limit, which is left for studies in the future.

%
%
In the present fit, 
the $m_{quark}\log m_{quark}$ terms are sizably 
suppressed due to $O(a)$ corrections for the pseudoscalar 
meson mass. 
In the combination $m_{PS}^2/2 m_{quark}^{AWI}$,
$(\omega_1^{PS} - \omega_1^{AWI})$ represents the strength 
of the chiral logarithm.
The resummed WChPT fit gives $(\omega_1^{PS} - \omega_1^{AWI})=0.24(13)$, 
while in continuum ChPT we expect 
$\omega_1^{PS} = A / 32 \pi^2 f_{0}^2 = 2.7$ and $\omega_1^{AWI}=0$,
with the phenomenological value of $f_{0} = 0.086$~GeV,
ignoring $O(a)$ dependence in $A$.
Namely, the coefficient of the logarithm is suppressed to 
about 10\% of the ChPT value 
by $O(a)$ contributions in $m_{PS}^2$ and $m_{quark}^{AWI}$.
It is important to repeat a similar analysis at a smaller lattice 
spacing to verify that the magnitude of the $m_{quark}\log m_{quark}$
coefficient converges toward the value predicted by ChPT.    


\subsubsection{Results at the physical point}

Since WChPT formulae are not available for the 
vector meson, 
we adopt Eq.~(\ref{equation:m_PS2_m_V_ChPT}) 
to fix the physical point for $m_\pi$. 
A fit for the whole data in the range $m_{PS}/m_V=0.80$--0.35 yields
$a_{\rho}^{ChPT} = 0.2009(21)$~fm.
Substituting $m_{\pi}$ 
to Eq.~(\ref{equation:chiral_log_m_PS2_resummed_WChPT}) and using 
$a_{\rho}^{ChPT}$, we  
obtain the VWI quark mass at the physical point $m_{ud}^{VWI}$.
Eqs.~(\ref{equation:chiral_log_m_quark_AWI_resummed_WChPT})
and (\ref{equation:chiral_log_f_PS_WChPT}) with $m_{ud}^{VWI}$ then
yield $m_{ud}^{AWI}$ and $f_{\pi}$ respectively
(Table~\ref{table:comparisons_resummed_WChPT_previous_vs_all}).

\begin{figure}[h]
   \includegraphics[width=75mm]
   {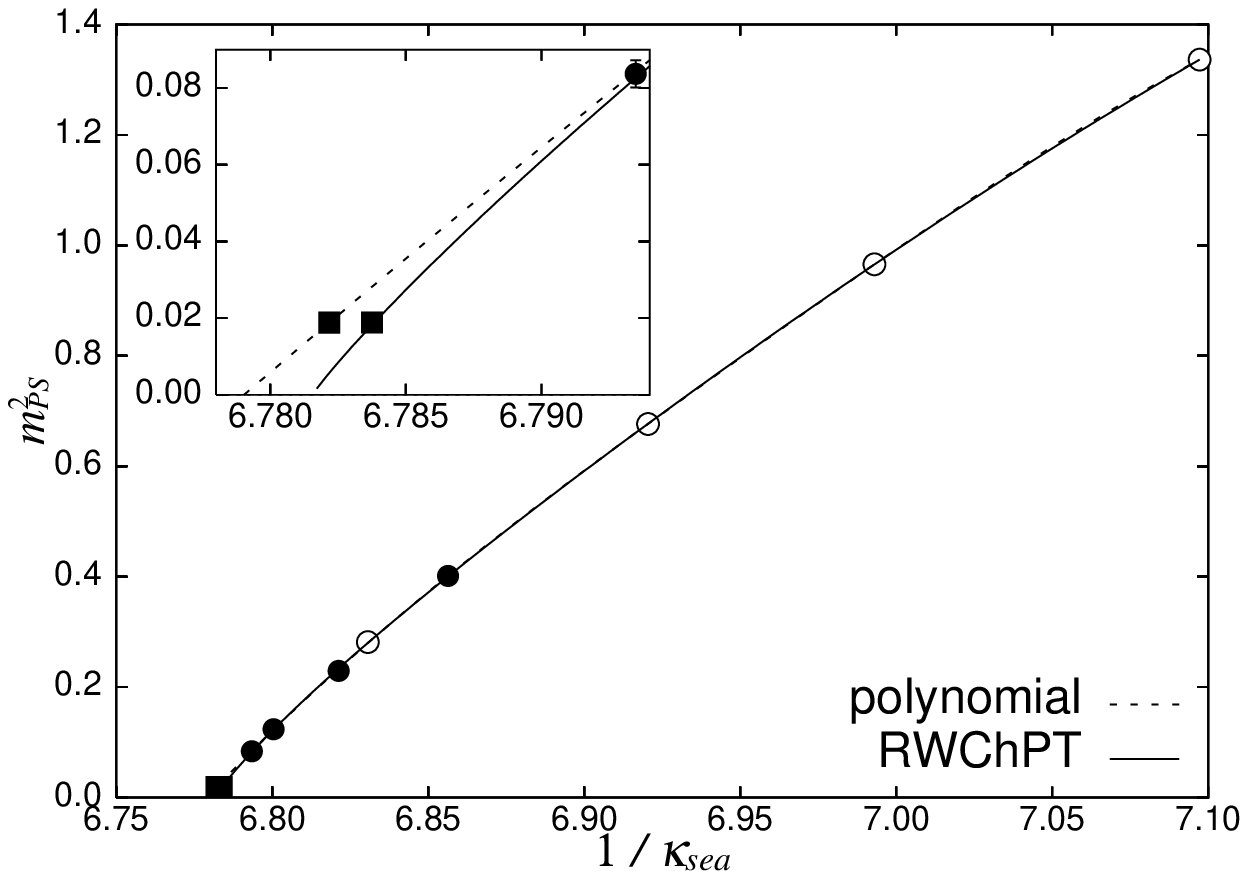}
   \includegraphics[width=75mm]
   {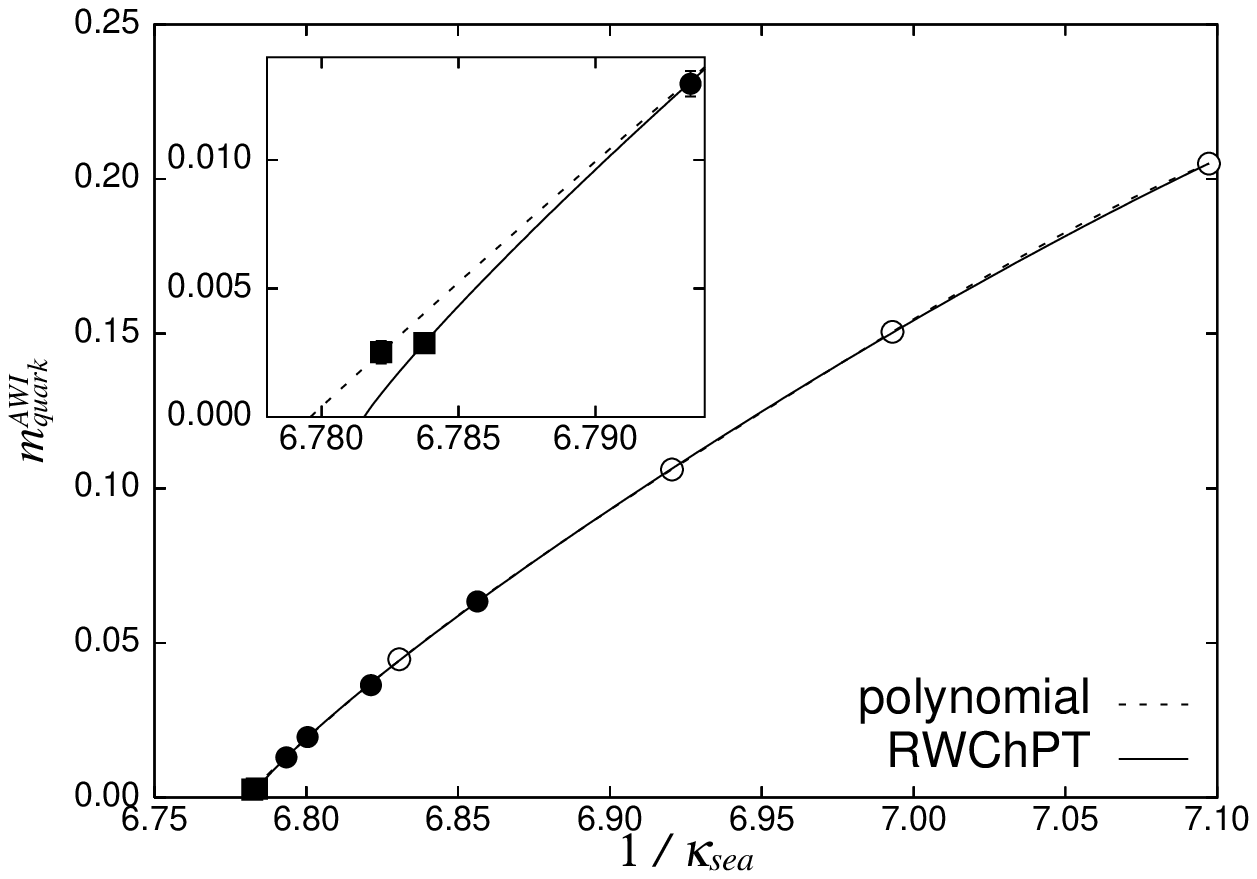}
   \caption
   { Comparisons of the polynomial and the resummed WChPT fits
     to pseudoscalar meson mass and AWI quark mass
     determined at $m_{PS}/m_V=0.80$--0.35.
     Circles show the lattice data and
     the square is the extrapolated result at the physical point.
     Open symbols are the results obtained
     in our previous study \protect\cite{Spectrum.Nf2.CP-PACS}.
   }
   \label{figure:kappa_inv_m_PS2_m_quark_AWI_polynomial_resummed_WChPT}
\end{figure}

Let us compare the resummed WChPT results with 
those of the quadratic polynomial obtained 
with the original data over the range $m_{PS}/m_V=0.80$--0.55
(Table~\ref{table:comparisons_polynomials_previous_vs_all})
and the fits using ChPT formula in the continuum limit for 
$m_{PS}/m_{V}=0.60$--0.35 (Table~\ref{table:physical_quantities_ChPT}).
The lattice spacing, the AWI quark mass, 
and the pion decay constant take similar values 
among higher order polynomials, ChPT and resummed WChPT formulae. 
An exception is the VWI quark mass which significantly
depends on the functional forms for the chiral extrapolation. 
(see Fig.~\ref{figure:kappa_inv_m_PS2_m_quark_AWI_polynomial_resummed_WChPT}).
Our final values for the light quark mass
at $a = 0.2$~fm
are:
\begin{eqnarray}
 m_{ud}^{VWI,\overline{\rm MS}}(\mu=2~\mbox{GeV})
 &=& \left\{
       \begin{array}{l}
          1.314(99)\mbox{~[MeV]} \hspace{0.5cm} \mbox{ (resummed WChPT) } \\
          1.796(51)\mbox{~[MeV]} \hspace{0.5cm} \mbox{ (polynomial) }
       \end{array}
     \right. \\
 m_{ud}^{AWI,\overline{\rm MS}}(\mu=2~\mbox{GeV})
 &=& \left\{
       \begin{array}{l}
          2.902(36)\mbox{~[MeV]}  \hspace{0.5cm} \mbox{ (resummed WChPT) } \\
          2.927(53)\mbox{~[MeV]}  \hspace{0.5cm} \mbox{ (polynomial) }
       \end{array}
     \right.
\end{eqnarray}
The sensitivity of the VWI quark mass on the functional form of chiral 
extrapolation is due to closeness of $\kappa_{ud}$ to the critical value 
$\kappa_c$.
A small variation of $\kappa_c$ is easily amplified 
in the up and down quark mass which is determined by the difference 
$1/\kappa_{ud}-1/\kappa_c$.

\subsubsection{Chiral extrapolation from large quark masses}

Finally, we test if WChPT explains the deviations of our new data 
at small quark masses from the quadratic extrapolation
of the data at $m_{PS}/m_{V}=0.80$--0.55.
A motivation of this test is
the rapid increase of the computational time
to simulate QCD toward small sea quark masses on fine lattices.
If WChPT correctly predicts the small quark mass behavior
from heavy sea quark mass simulations for
$m_{PS}/m_{V} \ge 0.5$,
it will be a great help for our studies.

\begin{figure}[h]
   \includegraphics[width=75mm]
   {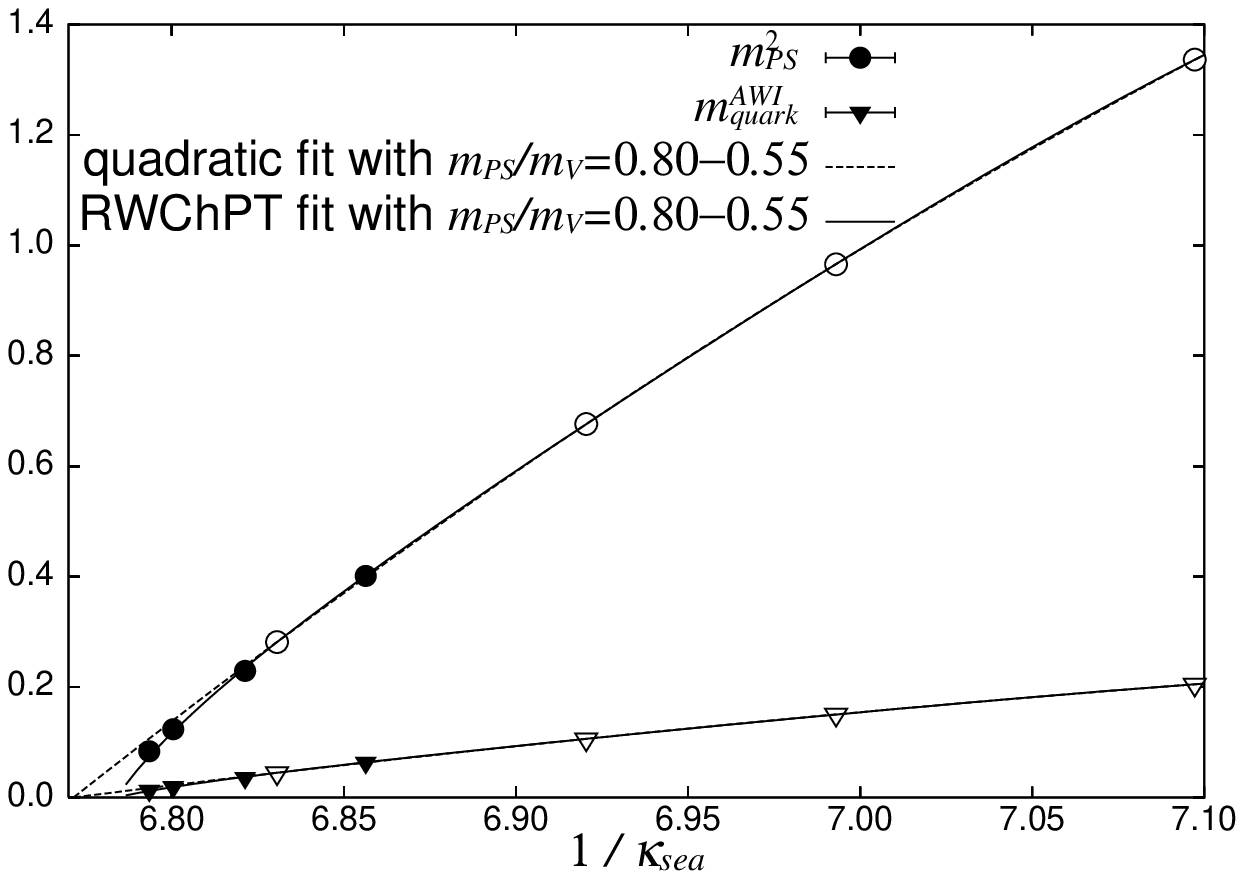}
   \includegraphics[width=75mm]
   {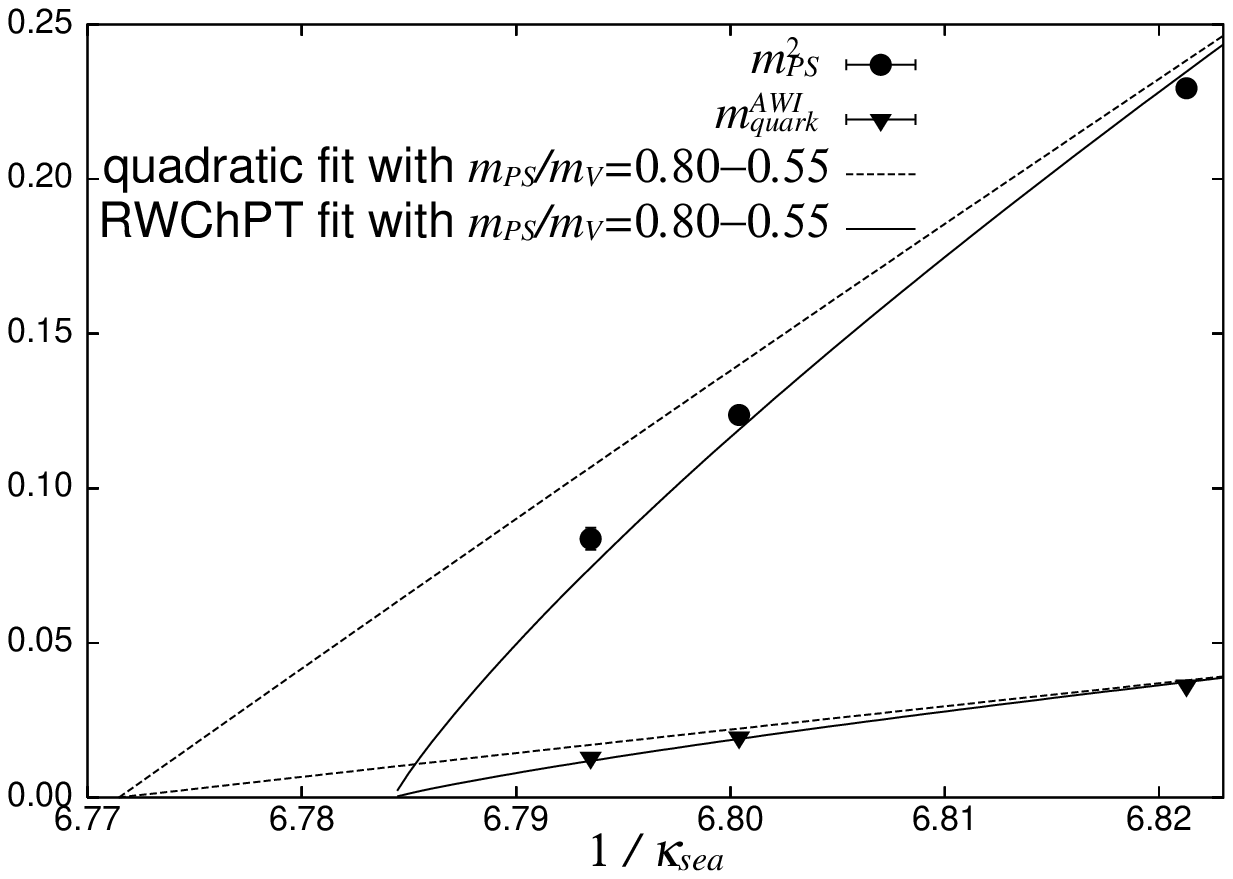}
   \caption
   {  Comparison of quadratic and resummed WChPT fits
      to pseudoscalar meson masses and AWI quark masses
      determined from the previous data
      of $m_{PS}/m_V=0.80$--0.55 
	\protect\cite{Spectrum.Nf2.CP-PACS} (open symbols)
      with the new small sea quark mass data (filled symbols).
      The right panel is an enlargement around the chiral limit.
   }
   \label{figure:kappa_inv_m_PS2_m_quark_AWI_polynomial_resummed_WChPT_previous}
\end{figure}

We apply the resummed WChPT formulae to 
the large quark mass data set at $\beta=1.80$.
Since the number of data points at $m_{PS}/m_V \geq 0.5$ is
small for a stable fitting,
we introduce a restriction:
$\Lambda_{3} = \Lambda_{3,AWI}$.
Fig.~\ref{figure:kappa_inv_m_PS2_m_quark_AWI_polynomial_resummed_WChPT_previous} (see Table~\ref{table:comparisons_resummed_WChPT_previous_vs_all}
for numerical values)
compares the fit from the large quark mass data set and
that using the data for the entire mass range. 
The resummed WChPT fit using the large quark mass data set alone
describes the small sea quark mass data very well.  
This contrasts to the polynomial extrapolation. 
Our observation suggests that WChPT may provide a valuable tool 
to carry out an accurate chiral extrapolation using simulations
with not too small quark masses.

\begin{figure}[h]
\begin{center}
   \includegraphics[width=75mm]
   {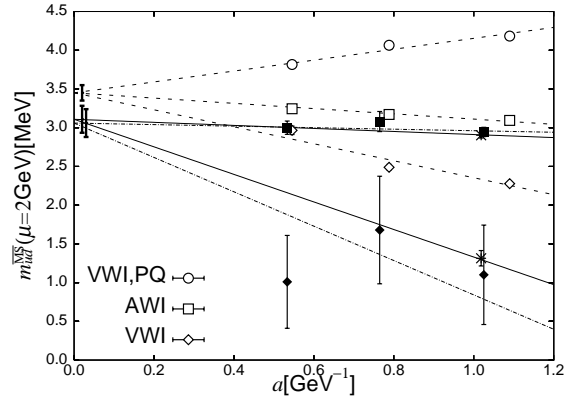}
   \caption
   { Continuum extrapolations of
     degenerate up and down quark mass
     obtained by chiral extrapolations with polynomials 
     \protect\cite{Spectrum.Nf2.CP-PACS} (open symbols)
     and the resummed WChPT formulae (filled symbols).
     The star at $a \simeq 1 \,{\rm GeV}^{-1}$ ($\beta=1.8$)
     represents the results obtained
     by the resummed WChPT formulae
     with data at $m_{PS}/m_{V}=0.80$--0.35.
     The others are the results
     with $m_{PS}/m_{V}=0.80$--0.55.
     The dashed lines are the combined linear fit
     to the quadratic chiral fit results
     and the dashed-dot lines are the ones
     to the resummed WChPT fit results,
     both with $m_{PS}/m_{V}=0.80$--0.55.
     The solid lines are the combined linear fits
     to the resummed WChPT chiral fit results
     with our whole data of
     $m_{PS}/m_{V}=0.80$--0.35 at $\beta=1.8$ and
     $m_{PS}/m_{V}=0.80$--0.55 at $\beta=1.95$ and 2.1.
   }
   \label{figure:a_inv-m_quark_VWI_AWI_poly_vs_resummed_WChPT}
\end{center}
\end{figure}

Encouraged by this, we apply the resummed WChPT to 
the two additional data sets at $m_{PS}/m_{V}=0.80$--0.55 
obtained at smaller lattice spacings at $\beta = 1.95$ and 2.1
($a = 0.16$ and 0.11~fm) in the previous work. 
A simultaneous linear continuum extrapolation
using $m_{ud}^{VWI,\overline{\rm MS}}$
and $m_{ud}^{AWI,\overline{\rm MS}}$,
combined with the results for $\beta=1.8$, 
leads to  
\begin{equation}
 m_{ud}^{\overline{\rm MS}}(\mu=2~\mbox{GeV})
  = 3.06(18)\mbox{~[MeV]}
          \hspace{0.5cm} \mbox{ (resummed WChPT fit) } 
\end{equation}
where the error is statistical only.  
When we use our whole data of $m_{PS}/m_{V}=0.80$--0.35
at $\beta=1.80$, we obtain
\begin{equation}
 m_{ud}^{\overline{\rm MS}}(\mu=2~\mbox{GeV})
 = 3.11(17)\mbox{~[MeV]}
   \hspace{0.5cm} \mbox{ (resummed WChPT fit with our whole data) }
\end{equation}
This is compared to our previous result 
using the quadratic extrapolation:  
\begin{equation}
 m_{ud}^{\overline{\rm MS}}(\mu=2~\mbox{GeV})
  = 3.45(10)\mbox{~[MeV]}
          \hspace{0.5cm} \mbox{ (quadratic fit) } 
\end{equation}
The resummed WChPT results in a 10\% decrease in the mean
up and down quark mass.
This is demonstrated in 
Fig.~\ref{figure:a_inv-m_quark_VWI_AWI_poly_vs_resummed_WChPT}.

\section{Conclusions}
\label{section:conclusion}

In this paper, we have pushed our previous study of two-flavor QCD
down to a sea quark mass as small as $m_{PS}/m_{V}=0.35$,
using the RG-improved gauge action and the clover-improved 
Wilson quark action.
We have found that our new data at $m_{PS}/m_{V}=0.60$--0.35 show clear 
deviations from the prediction of the previous chiral extrapolations
based on quadratic polynomials, which implies that
higher order terms were needed to describe
the behavior at a small sea quark mass. 
On the other hand, our current data do not show the
clear quark mass dependence expected
from ChPT in the continuum:
the chiral logarithm may appear
only below $m_{PS}/m_{V} \sim 0.4$.
This result contrasts with that of the qq+q collaboration~\cite{qq+q}
based on unimproved plaquette glue and Wilson quark actions,
but is not dissimilar to that of UKQCD~\cite{Spectrum.Nf2.UKQCD_light}.

We have provisionally ascribed the major reason for the failure 
of continuum ChPT to explicit chiral symmetry breaking 
of the Wilson term,
which is significant on our lattice of $a = 0.2$~fm.
We have then made a test of WChPT in which the effect of the Wilson 
term is accommodated, and 
found the resummed one-loop WChPT formulae
that take account of the effects up to $O(a^2)$
describe well our entire data.
Convergence tests indicate that resummed WChPT 
gives well-controlled chiral extrapolations. 
The use of WChPT generally leads to modifications of various 
physical observables at the physical point by 
about 10\%, compared with those obtained in the quadratic 
extrapolation at this lattice spacing.  
A much larger modification, however, is seen with 
the light quark mass defined through vector Ward identity: 
the WChPT extrapolation decreases it by 30\%.

We note in particular that the resummed WChPT extrapolation
from our previous data
at $m_{PS}/m_{V}=0.80$--0.55 predicts correctly the new data at 
$m_{PS}/m_{V}=0.60$--0.35.
Encouraged by this fact, we attempted a continuum extrapolation of the 
light quark mass using the resummed WChPT fits to the previous data 
at $m_{PS}/m_{V}=0.80$--0.55 but on finer lattices 
with $a = 0.16$ and 0.11~fm.
We find  in the continuum limit, 
$m_{ud}^{\overline{\rm MS}}(\mu=2~\mbox{GeV}) = 3.11(17)$~[MeV],
which is smaller than the previously reported result
by approximately 10\%.
Our work suggests that WChPT provides us with 
a valuable theoretical framework 
for chiral extrapolations.


\begin{acknowledgments}
YN thanks Oliver B\"{a}r for valuable discussions and suggestions
on the manuscript.
This work is supported in part
by Large Scale Numerical Simulation Project of 
the Science Information Processing Center, University of Tsukuba, 
and by Grants-in-Aid of the Ministry of Education 
(Nos.\ 
12304011, 
12740133, 
13135204, 
13640259, 
13640260, 
14046202, 
14740173, 
15204015, 
15540251, 
15540279, 
16028201
). 
VIL is supported by JSPS.
\end{acknowledgments}


\appendix

\section{Hadron masses}
\label{section:appendix_m_hadron}

Measured hadron masses are summarized in
Tables~\ref{table:m_PS_m_V_12x24}\,--\,\ref{table:plaq_rect_16x24}.
Our choice of the fitting range and resulting value of 
$\chi^2/{dof}$ are also given in these tables.


\clearpage


\begin{table}[h]
\caption
{  Run parameters of the present simulation.
   The step size $dt$ is given by the inverse of the
   number of the molecular dynamics steps (\#MD), and hence not listed.
   We denote the tolerance parameter in the stopping condition
   for the quark matrix inversion in calculations of the force
   by $\Delta_{force}$ and the average number of iterations by $N_{inv}$.
   Number of node (PE's) of VPP5000/80 used for the present calculation, 
   and the CPU time required per trajectory in units of hour are also given.
   The number of trajectory is denoted by $N_{traj}$.
}
\label{table:simulation_parameters}
\begin{ruledtabular}
\begin{tabular}{lllllllll}
%
%
 $12^3 \times 24$ &
 \multicolumn{2}{c}{on 4PE} & \multicolumn{2}{c}{on 4PE} &
 \multicolumn{2}{c}{on 4PE} & \multicolumn{2}{c}{on 8PE} \\
\hline
 $\kappa_{sea}$ & 
 \multicolumn{2}{c}{0.14585} & \multicolumn{2}{c}{0.14660} &
 \multicolumn{2}{c}{0.14705} & \multicolumn{2}{c}{0.14720} \\
\hline
 \#MD           &
 \multicolumn{2}{c}{200}     &   333        &    400       &
    800         &   1000     &  1250        &   1600       \\
 Accept.        &
 \multicolumn{2}{c}{0.76}    &  0.72        &   0.84       &
   0.82         &   0.90     &  0.87        &   0.91       \\
 $N_{traj}$     &
 \multicolumn{2}{c}{4000}    &  1750        &   2250       &
    680         &   3320     &   100        &   1300       \\
\hline
 $\Delta_{force}$ &
 \multicolumn{2}{c}{$10^{-10}$} & \multicolumn{2}{c}{$10^{-11}$} &
 \multicolumn{2}{c}{$10^{-11}$} & \multicolumn{2}{c}{$10^{-12}$} \\
 $N_{inv}$      &
 \multicolumn{2}{c}{  87 }   & \multicolumn{2}{c}{ 147 } &
 \multicolumn{2}{c}{ 232 }   & \multicolumn{2}{c}{ 318 } \\
 Hour/traj.     &
 \multicolumn{2}{c}{0.23}    &   0.56       &   0.69       &
    2.0         &   2.6      &   2.2        &   3.2       \\
\hline
 $m_{PS}/m_{V}$ &
 \multicolumn{2}{c}{ 0.609(2) }   & \multicolumn{2}{c}{ 0.509(5)  } &
 \multicolumn{2}{c}{ 0.413(8) }   & \multicolumn{2}{c}{ 0.349(19) } \\
\hline \hline
%
%
 $16^3 \times 24$ &
 \multicolumn{2}{c}{on 4PE} & \multicolumn{2}{c}{on 8PE} &
 \multicolumn{2}{c}{} & \multicolumn{2}{c}{} \\
\hline
$\kappa_{sea}$  &
 \multicolumn{2}{c}{0.14585} & \multicolumn{2}{c}{0.14660} &
 \multicolumn{2}{c}{0.14705} & \multicolumn{2}{c}{0.14720} \\
\hline
 \#MD           &
    200         &   250      &   333        &   500        &
 \multicolumn{2}{c}{ -- }    & \multicolumn{2}{c}{ -- }    \\
 Accept.        &
    0.61        &   0.71     &   0.79       &   0.80       &
 \multicolumn{2}{c}{ -- }    & \multicolumn{2}{c}{ -- }    \\
 $N_{traj}$     &
    800         &   1200     &   325        &   1675       &
 \multicolumn{2}{c}{ -- }    & \multicolumn{2}{c}{ -- }    \\
\hline
 $\Delta_{force}$ &
 \multicolumn{2}{c}{$10^{-10}$} & \multicolumn{2}{c}{$10^{-11}$} &
 \multicolumn{2}{c}{ -- }       & \multicolumn{2}{c}{ -- }    \\
 $N_{inv}$      &
 \multicolumn{2}{c}{ 92 }       & \multicolumn{2}{c}{ 158 }   &
 \multicolumn{2}{c}{ -- }       & \multicolumn{2}{c}{ -- }    \\
 Hour/traj.     &
 0.50           &  0.61      &   0.69       &   1.03       &
 \multicolumn{2}{c}{ -- }    & \multicolumn{2}{c}{ -- }    \\
\hline
 $m_{PS}/m_{V}$ &
 \multicolumn{2}{c}{ 0.604(3) } & \multicolumn{2}{c}{ 0.509(4) } &
 \multicolumn{2}{c}{ -- }       & \multicolumn{2}{c}{ -- }       \\
%
%
\end{tabular}
\end{ruledtabular}

\end{table}


\begin{table}[h]
\caption{
  String tension $\sigma$ and
  Sommer scale $r_0$ at simulated sea quark masses.
  The first error is statistical.
  The second and third ones are the systematic errors
  due to the choice of $t_{min}$ and $r_{min}$.
}
\label{table:potential}
\begin{ruledtabular}
\begin{tabular}{lllll}
%
%
\multicolumn{5}{l}{$12^3 \times 24$} \\
\hline
$\kappa_{sea}$ & $t_{min}$              & $r_{min}$
               & $\sigma$               & $r_0$ \\
\hline
0.14585        & 2                      & $\sqrt{2}$
               & 0.322(6)($-42$)($+91$) & 2.004(8)($+58$)($+77$) \\
0.14660        & 2                      & $\sqrt{2}$
               & 0.289(5)($-8$)($+64$)  & 2.107(8)($+37$)($+54$) \\
0.14705        & 2                      & $\sqrt{2}$
               & 0.278(5)($-38$)($+34$) & 2.167(9)($+80$)($+25$) \\
0.14720        & 2                      & $\sqrt{2}$
               & 0.255(8)($-10$)($+42$) & 2.237(17)($+10$)($+34$) \\
\hline
\hline
%
%
\multicolumn{5}{l}{$16^3 \times 24$} \\
\hline
$\kappa_{sea}$ & $t_{min}$               & $r_{min}$
               & $\sigma$                & $r_0$ \\
\hline
0.14585        & 2                       & $\sqrt{2}$
               & 0.313(11)($-10$)($+90$) & 2.011(10)($+17$)($+72$) \\
0.14660        & 2                       & $\sqrt{2}$
               & 0.270(6)($+7$)($+66$)   & 2.131(11)($+20$)($+39$) \\
\end{tabular}
\end{ruledtabular}
\end{table}

\begin{table}[h]
\caption{
  Autocorrelation time for plaquette ($\tau^{cum}_{plaq}$),
  pseudoscalar meson propagator at $N_t/4$ ($\tau^{cum}_{PS}$)
  and Wilson loop with $(r,t)=(2,2)$ ($\tau^{cum}_{W}$).
  All values are in units of HMC trajectory.
}
\label{table:autocorrelation}
\begin{ruledtabular}
\begin{tabular}{lllllll}
%
%
\multicolumn{5}{l}{$12^3 \times 24$} \\
\hline
$\kappa_{sea}$            & 0.14585     & 0.14660
                          & 0.14705     & 0.14720 \\
\hline
$\tau^{cum}_{plaq}$
                          &  7.6(1.8)   & 11.7(2.3)
                          &  9.5(2.1)   &  8.9(3.2) \\
$\tau^{cum}_{PS}$
                          &  7.9(1.6)   &  7.2(1.5)
                          &  5.3(1.2)   &  3.0(1.0) \\
$\tau^{cum}_{W}$
                          &  8.1(1.9)   & 12.6(2.9)
                          & 11.3(2.2)   & 13.0(4.4) \\
\hline
\hline
%
\multicolumn{5}{l}{$16^3 \times 24$} \\
\hline
$\kappa_{sea}$            & 0.14585     & 0.14660
                          & 0.14705     & 0.14720 \\
\hline
$\tau^{cum}_{plaq}$
                          & 14.1(3.9)   &  8.8(2.1)
                          &  --         &  -- \\
$\tau^{cum}_{PS}$
                          & 10.3(2.8)   &  4.9(1.6)
                          &  --         &  -- \\
$\tau^{cum}_{W}$
                          & 14.1(3.8)   & 10.1(4.3)
                          &  --         &  -- \\
\end{tabular}
\end{ruledtabular}
\end{table}



\begin{table}[h]
\caption
{  Parameters of independent polynomial chiral fits to AWI quark masses
   and pseudoscalar meson masses
   as a function of the VWI quark mass.
}
\label{table:m_quark_VWI_m_quark_AWI_and_m_PS2_independent}
\begin{ruledtabular}
\begin{tabular}{llllllll}
$m_{PS}/m_{V}$   & $\kappa_{c}$  & $B^{AWI}$   & $C^{AWI}$  & $D^{AWI}$
                                 & $E^{AWI}$   &
$\chi^2/dof$     & $Q$ \\ \hline
%
%
0.80--0.35       & 0.147502(14)  & 1.961(60)   & $-10.5(1.9)$ & 71(20)
                                 & $-201(67)$  &
4.38/3           & 0.22 \\ \hline
%
%
$m_{PS}/m_{V}$   & $\kappa_{c}$  & $B^{PS}$    & $C^{PS}$   & $D^{PS}$
                                 & $E^{PS}$    &
$\chi^2/dof$     & $Q$ \\ \hline
0.80--0.35       & 0.147514(15)  &  12.05(33)  &  $-55.7(9.0)$ & 359(89)
                                 & $-966(281)$   &
4.17/3           & 0.24  \\
\end{tabular}
\end{ruledtabular}
\end{table}

\begin{table}[h]
\caption
{  Parameters of simultaneous polynomial chiral fits to AWI quark masses
   and pseudoscalar meson masses
   as a function of the VWI quark mass.
   The first error is statistical and
   the second is a systematic one due to the higher order term
   for the chiral extrapolation.
}
\label{table:m_quark_VWI_m_quark_AWI_and_m_PS2_simultaneous}
\begin{ruledtabular}
\begin{tabular}{lllllll}
$m_{PS}/m_{V}$   & $\kappa_{c}$  & $B^{AWI}$   & $C^{AWI}$
                                 & $D^{AWI}$   & $E^{AWI}$   &
$\chi^2/dof$     \\ \hline
%
%
0.80--0.35       & 0.147508(14)(+7) & $1.938(54)(-60)$ & $-9.8(1.7)(+3.3)$
                                    & $65(18)(-67)$    & $-181(60)(+541)$ &
8.89/7           \\ \hline
                 &               & $B^{PS}$    & $C^{PS}$
                                 & $D^{PS}$    & $E^{PS}$    &
$Q$              \\ \hline
                 &               &  $12.18(31)(-20)$ &  $-58.9(8.6)(+7.9)$
                                 & $389(85)(-126)$   & $-1053(269)(+880)$ &
0.26             \\
\end{tabular}
\end{ruledtabular}
\end{table}

\begin{table}[h]
\caption
{ Parameters of polynomial chiral fits to vector meson mass.
  The first error is statistical and
  the second is a systematic one due to the higher order term
  for the chiral extrapolation.
}
\label{table:m_PS2_m_V}
\begin{ruledtabular}
\begin{tabular}{lllllll}
$m_{PS}/m_{V}$  & $A^{V}$     & $B^{V}$
                & $D^{V}$     & $F^{V}$     &
$\chi^2/{dof}$  & $Q$ \\ \hline
0.80--0.35      &  $0.770(15)(-6)$   & $0.790(72)(+47)$
                & $-0.304(97)(-116)$ & $0.063(39)(+113)$   &
1.10/4          & 0.89 \\
\end{tabular}
\end{ruledtabular}
\end{table}

\begin{table}[h]
\caption{Parameters of polynomial chiral fits to
         pseudoscalar and vector meson decay constants.
        }
\label{table:m_PS2_f_PS_V}
\begin{ruledtabular}
\begin{tabular}{lllllll}
%
%
$m_{PS}/m_{V}$  & $A^{f_{PS}}$  & $B^{f_{PS}}$
                & $D^{f_{PS}}$  & $F^{f_{PS}}$    &
$\chi^2/{dof}$  & $Q$ \\ \hline
0.80--0.35      & $0.1239(26)(-83)$    & $0.165(17)(+86)$
                & $-0.076(27)(-264)$   & $0.018(12)(+296)$ &
17.2/4          & 0.0018 \\ \hline
%
%
$m_{PS}/m_{V}$  & $A^{f_{V}}$   & $B^{f_{V}}$
                & $D^{f_{V}}$   & $F^{f_{V}}$     &
$\chi^2/{dof}$  & $Q$ \\ \hline
0.80--0.35      & $0.228(12)(-15)$     & $0.265(59)(+127)$
                & $-0.156(85)(-336)$   & $0.039(35)(+340)$       &
2.31/4          & 0.68  \\
\end{tabular}
\end{ruledtabular}
\end{table}

\begin{table}[h]
\caption{Parameters of polynomial chiral fits
         to octet and decuplet baryon masses.}
\label{table:m_PS2_m_baryon}
\begin{ruledtabular}
\begin{tabular}{lllllll}
%
%
$m_{PS}/m_{V}$  & $A^{oct}$       & $B^{oct}$
                & $D^{oct}$       & $F^{oct}$     &
$\chi^2/{dof}$  & $Q$ \\ \hline
0.80--0.35      & $1.051(23)(+21)$  & $1.41(12)(-19)$
                & $-0.51(18)(+56)$  & $0.097(78)(-591)$     &
4.96/4          & 0.29  \\ \hline
%
%
$m_{PS}/m_{V}$  & $A^{dec}$       & $B^{dec}$
                & $D^{dec}$       & $F^{dec}$     &
$\chi^2/{dof}$  & $Q$ \\ \hline
0.80--0.35      &  $1.381(37)(+8)$  &  $1.03(19)(-7)$
                & $-0.17(29)(+18)$  & $-0.022(126)(-186)$   &
0.75/4          & 0.95  \\
\end{tabular}
\end{ruledtabular}
\end{table}

\begin{table}[h]
\caption{Parameters of polynomial chiral extrapolation of $r_0$.}
\label{table:m_PS2_r_0_inv}
\begin{ruledtabular}
\begin{tabular}{lllllll}
 $m_{PS}/m_{V}$  &
 $A_{r_0}$       & $B_{r_0}$    &
 $D_{r_0}$       &  $F_{r_0}$   &
 $\chi^2/{dof}$  & $Q$ \\ \hline
 0.80--0.35      &
 $0.432(12)(-7)$   & $0.207(89)(+99)$  &
 $-0.11(14)(-32)$  & $0.032(64)(+364)$ &
 0.37/4          & 0.99 \\
\end{tabular}
\end{ruledtabular}
\end{table}

\begin{table}[h]
\caption
{  Results of physical quantities obtained by polynomial chiral fits 
    using data at $m_{PS}/m_{V}=0.80$--0.35.
    The results of the previous quadratic fits 
    at  $m_{PS}/m_{V}=0.80$--0.55 \protect\cite{Spectrum.Nf2.CP-PACS} 
   are also shown.
   The first error is statistical and
   the second is a systematic one due to the higher order term
   for the chiral extrapolation.
   Only statistical errors are given 
   for the previous results.
}
\label{table:comparisons_polynomials_previous_vs_all}
\begin{ruledtabular}
\begin{tabular}{c|llc}
                    & Quartic fit (this study)
             \footnote{
             For vector meson masses, decay constants
             and baryon masses,
             we employ cubic fit functions in $m_{PS}^2$ as
             Eqs.~(\ref{equation:m_PS2_m_V})--(\ref{equation:m_PS2_m_baryons}).
                      }
                    & Quadratic fit \protect\cite{Spectrum.Nf2.CP-PACS} 
                    &  Difference
\\ \hline
                         fit range in $m_{PS}/m_{V}$
                    & 0.80--0.35
                    & 0.80--0.55
                    & 
\\ \hline
 $a_{\rho}$[fm]     & $0.2007(38)(-14)$  & 0.2150(22) &  $-7\%(6.5\sigma)$
\\ \hline
 $\kappa_{ud}$      & $0.147440(13)(+7)$ & 0.147540(16) & $-0.1\%(6.3\sigma)$
\\ \hline
 $m_{ud}^{VWI,\overline{\rm MS}}(\mu=2~\mbox{GeV})$[MeV]
                    & $1.796(51)(+18)$   & 2.277(27)  & $-21\%(18\sigma)$
\\ \hline
 $m_{ud}^{AWI,\overline{\rm MS}}(\mu=2~\mbox{GeV})$[MeV]
                    & $2.927(53)(-55)$   & 3.094(35)  &  $-6\%(4.8\sigma)$
\\ \hline
 $f_{\pi}$[GeV]     & $0.1248(31)(-59)$  & 0.1288(33) &  $-3\%(1.2\sigma)$
\\ \hline
 $f_{\rho}$[GeV]    & $0.2294(74)(-111)$ & 0.2389(47) &  $-4\%(2.0\sigma)$
\\ \hline
 $m_{N}$[GeV]       & $1.060(27)(+24)$   & 1.016(16)  &  $+4\%(2.8\sigma)$
\\ \hline
 $m_{\Delta}$[GeV]  & $1.377(39)(+16)$   & 1.270(23)  &  $+8\%(4.7\sigma)$
\\
\end{tabular}
\end{ruledtabular}
\end{table}



\begin{table}[h]
\caption
{  Chiral extrapolation of pseudoscalar meson masses and decay constants
   based on the continuum ChPT formulae at one-loop
   with $m_{quark}=m_{quark}^{AWI}$ and $m_{quark}=m_{quark}^{VWI}$.
   $\kappa_c$ has been determined with $m_{quark}^{AWI}$.
   The value of $\kappa_c$ is
   shown in Table~\ref{table:m_quark_VWI_m_quark_AWI_and_m_PS2_independent}.
}
\label{table:chiral_log_m_quark_AWI_VWI_m_PS2_and_renormalized_f_PS}
\begin{ruledtabular}
\begin{tabular}{lllllll}
%
%
 $m_{PS}/m_{V}$  & $B_{0}^{AWI}$      & $f^{AWI}$
                 & $\Lambda_3^{AWI}$  & $\Lambda_4^{AWI}$ &
 $\chi^{2}/dof$  & $Q$ \\ \hline
 0.80--0.35      & 3.838(15)          & 0.12162(47)
                 & 1.553(10)          & 2.633(15)         &
 849/12          & $10^{-174}$ \\
 0.60--0.35      & 3.398(52)          & 0.1130(20)
                 & 0.902(71)          & 2.591(98)         &
 11.2/6          & 0.083 \\ \hline
%
%
 $m_{PS}/m_{V}$  & $B_{0}^{VWI}$      &  $f^{VWI}$
                 & $\Lambda_3^{VWI}$  & $\Lambda_4^{VWI}$ &
 $\chi^{2}/dof$  & $Q$ \\ \hline
 0.80--0.35      & 6.886(22)          & 0.13225(35)
                 & 2.4018(85)         & 2.463(11)         &
 1417/12         & $10^{-296}$ \\
 0.60--0.35      & 6.582(87)          & 0.1145(18)
                 & 1.645(83)          & 2.262(72)         &
 17.5/6          & 0.0076 \\
\end{tabular}
\end{ruledtabular}
\end{table}

\begin{table}[h]
\caption
{  Parameters of chiral fits to vector meson mass
   based on continuum ChPT.
}
\label{table:m_PS2_m_V_ChPT}
\begin{ruledtabular}
\begin{tabular}{llllll}
$m_{PS}/m_{V}$  & $A^{V}$     & $B^{V}$   & $C^{V}$     &
$\chi^2/{dof}$  & $Q$ \\ \hline
0.80--0.35      &  0.7692(86) & 0.897(32) & $-0.346(23)$  &
1.39/5          &  0.93 \\
0.60--0.35      &  0.731(45)  & 1.31(49)  & $-0.85(60)$   &
0.33/2          &  0.85  \\
\end{tabular}
\end{ruledtabular}
\end{table}

\begin{table}[h]
\caption{Parameters of chiral fits to octet and decuplet baryon masses
         based on continuum ChPT.}
\label{table:m_PS2_m_baryon_ChPT}
\begin{ruledtabular}
\begin{tabular}{llllll}
%
%
$m_{PS}/m_{V}$  & $A^{oct}$     & $B^{oct}$     & $C^{oct}$     &
$\chi^2/{dof}$  & $Q$ \\ \hline
0.80--0.35      & 1.043(14)     & 1.641(68)     & $-0.632(51)$    &
5.13/5          & 0.40  \\
0.60--0.35      & 1.011(52)     & 2.08(59)      & $-1.23(74)$     &
2.23/2          & 0.33  \\ \hline
%
%
$m_{PS}/m_{V}$  & $A^{dec}$     & $B^{dec}$     & $C^{dec}$     &
$\chi^2/{dof}$  & $Q$ \\ \hline
0.80--0.35      & 1.351(20)     & 1.353(88)     & $-0.481(66)$    &
1.24/5          & 0.94  \\
0.60--0.35      & 1.428(86)     & 0.52(93)      & 0.53(1.15)     &
0.23/2          & 0.89  \\
\end{tabular}
\end{ruledtabular}
\end{table}

\begin{table}[h]
\caption
{  Results of physical quantities
   obtained  by continuum one-loop ChPT chiral fits   
   using data at $m_{PS}/m_{V}=0.60$--0.35.
   For the $m_{quark}=m_{quark}^{VWI}$ case,
   $\kappa_c$ has been fixed to the value
   determined from the quartic fit to $m_{quark}^{AWI}$ shown
   in Table~\ref{table:m_quark_VWI_m_quark_AWI_and_m_PS2_independent}.
   The errors are statistical.
}
\label{table:physical_quantities_ChPT}
\begin{ruledtabular}
\begin{tabular}{r|cc}
                    & continuum ChPT~$\left( m_{quark}=m_{quark}^{AWI} \right)$
                    & continuum ChPT~$\left( m_{quark}=m_{quark}^{VWI} \right)$
\\ \hline
 $a_{\rho}$[fm]     & \multicolumn{2}{c}{0.192(10)}
\\ \hline
 $\kappa_{ud}$      & 0.147445(14) & 0.1474431(65)
\\ \hline
 $m_{ud}^{VWI,\overline{\rm MS}}(\mu=2~\mbox{GeV})$[MeV]
                    & 1.609(89)    & 1.625(81)
\\ \hline
 $m_{ud}^{AWI,\overline{\rm MS}}(\mu=2~\mbox{GeV})$[MeV]
                    &  2.66(13)    & 2.68(13)
\\ \hline
 $f_{\pi}$[GeV]     & 0.1219(64)   & 0.1231(65)
\\ \hline
%
 $m_{N}$[GeV]       & \multicolumn{2}{c}{1.074(69)}
\\ \hline
 $m_{\Delta}$[GeV]  & \multicolumn{2}{c}{1.47(11)}
\\
\end{tabular}
\end{ruledtabular}
\end{table}



\begin{table}[h]
\caption
{  Parameters of chiral fits to pseudoscalar meson and AWI quark masses
   based on WChPT.
}
\label{table:chiral_log_m_PS2_m_quark_AWI_WChPT}
\begin{ruledtabular}
\begin{tabular}{lllllllllll}
$m_{PS}/m_{V}$ & $\kappa_c$   & $A$
               & $\omega_0$   & $\omega_1^{PS}$ & $\omega_1^{AWI}$
               & $\Lambda_0$  & $\Lambda_3$     & $\Lambda_3^{AWI}$ &
$\chi^2/{dof}$ & $Q$ \\ \hline
0.80--0.35     & 0.147445(27) & 6.312(44)
               & $-0.40(13)$  & $-2.0(1.4)$     & $-2.0(1.4)$
               & 0.91(35)     & 1.95(15)        & 1.77(23)          &
11.9/8         & 0.16 \\
\end{tabular}
\end{ruledtabular}
\end{table}

\begin{table}[h]
\caption
{  Parameters of chiral fits to pseudoscalar meson decay constants
   based on WChPT.
   $\kappa_c$ and $A$ has been fixed to the values in
   Table~\ref{table:chiral_log_m_PS2_m_quark_AWI_WChPT}.
}
\label{table:chiral_log_renormalized_f_PS_WChPT}
\begin{ruledtabular}
\begin{tabular}{llllll}
$m_{PS}/m_{V}$ & $f$          & $\omega_{1}^{f_{PS}}$
               & $\Lambda_4$  &
$\chi^2/{dof}$ & $Q$ \\ \hline
0.80--0.35     & 0.1233(17)   & 3.73(30)
               & 2.44(13)     &
18.1/5         & 0.0028 \\
\end{tabular}
\end{ruledtabular}
\end{table}


\begin{table}[h]
\caption
{  Parameters of chiral fits to pseudoscalar meson and AWI quark masses
   based on the resummed WChPT.
}
\label{table:chiral_log_m_PS2_m_quark_AWI_resummed_WChPT}
\begin{ruledtabular}
\begin{tabular}{llllllllllll}
$m_{PS}/m_{V}$ & $\kappa_c$   & $A$
               & $\omega_0$   & $\omega_1^{PS}$ & $\omega_1^{AWI}$
               & $\Lambda_0$  & $\Lambda_3$     & $\Lambda_3^{AWI}$ &
$\chi^2/{dof}$ & $Q$ \\ \hline
0.80--0.35     & 0.147459(20) & 6.354(59)
               & 0.542(46)    & 0.65(51)        & 0.42(49)
               & 0.397(56)    & 0.15(15)        & 0.07(16)          &
11.0/8         & 0.20 \\
\end{tabular}
\end{ruledtabular}
\end{table}

\begin{table}[h]
\caption
{  Parameters of chiral fits to pseudoscalar meson decay constants
   based on the resummed WChPT.
   $\kappa_c$ and $A$ has been fixed to the values in
   Table~\ref{table:chiral_log_m_PS2_m_quark_AWI_resummed_WChPT}.
}
\label{table:chiral_log_renormalized_f_PS_resummed_WChPT}
\begin{ruledtabular}
\begin{tabular}{llllll}
$m_{PS}/m_{V}$ & $f$          & $\omega_{1}^{f_{PS}}$
               & $\Lambda_4$  &
$\chi^2/{dof}$ & $Q$ \\ \hline
0.80--0.35     & 0.1227(17)   & 3.78(30)
               & 2.44(13)     &
18.2/5         & 0.0028 \\
\end{tabular}
\end{ruledtabular}
\end{table}

\begin{table}[h]
\caption
{  Results of physical quantities obtained by the resummed WChPT fits 
    using data at $m_{PS}/m_{V}=0.80$--0.35. 
   The results are compared with the results of the resummed WChPT fits 
   using our previous data at $m_{PS}/m_{V}=0.80$--0.55.
}
\label{table:comparisons_resummed_WChPT_previous_vs_all}
\begin{ruledtabular}
\begin{tabular}{r|ccr}
                    & RWChPT
                    & RWChPT
                      \footnote{For $m_{PS}/m_{V}=0.80$--0.55 data,
                                we employ a restriction
                                $\Lambda_{3} = \Lambda_{3,AWI}$.
                               }
                    & Difference
\\ \hline
                         fit range in $m_{PS}/m_{V}$
                    & 0.80--0.35
                    & 0.80--0.55 &
\\ \hline
 $a_{\rho}$[fm]     & 0.2009(21) & 0.2022(38) &  $-1\%(0.3\sigma)$
\\ \hline
 $\kappa_{ud}$ & 0.147409(16) & 0.14736(22) & $+0.03\%(0.2\sigma)$
\\ \hline
 $m_{ud}^{VWI,\overline{\rm MS}}(\mu=2~\mbox{GeV})$[MeV]
                    & 1.314(99)  
			  & 1.10(64)   & $+19\%(0.3\sigma)$
\\ \hline
 $m_{ud}^{AWI,\overline{\rm MS}}(\mu=2~\mbox{GeV})$[MeV]
                    & 2.902(36)  
			  & 2.945(60)  &  $-1\%(0.7\sigma)$
\\ \hline
 $f_{\pi}$[GeV]     & 0.1238(21) 
                                      & 0.1368(43) &  $-10\%(3.0\sigma)$
\\
\end{tabular}
\end{ruledtabular}
\end{table}


\clearpage


%
%
\begin{table}[h]
\caption
{ Meson masses and bare AWI quark masses on $12^3 \times 24$ lattice.
}
\label{table:m_PS_m_V_12x24}
\begin{ruledtabular}
\begin{tabular}{c|ccc|ccc|c}
$\kappa_{sea}$ & $m_{PS}$    & [$t_{min},t_{max}$] & $\chi^2/dof$
               & $m_{V}$     & [$t_{min},t_{max}$] & $\chi^2/dof$
               & $m_{quark}^{AWI}$ \\ \hline
  0.14585      & 0.6336(14)  & [6,12]              & 0.76(84)
               & 1.0405(38)  & [6,12]              & 0.40(51)
               & 0.06340(34) \\ \hline
  0.14660      & 0.4789(23)  & [6,12]              & 1.60(1.19)
               & 0.9410(81)  & [6,12]              & 2.36(1.02)
               & 0.03632(39) \\ \hline
  0.14705      & 0.3520(29)  & [6,12]              & 0.60(77)
               & 0.8526(148) & [6,12]              & 0.67(81)
               & 0.01952(30) \\ \hline
  0.14720      & 0.2893(61)  & [6,12]              & 0.50(93)
               & 0.8300(413) & [6,12]              & 0.95(92)
               & 0.01296(49) \\
\end{tabular}
\end{ruledtabular}
\end{table}

%
%
\begin{table}[h]
\caption
{ Decay constants on $12^3 \times 24$ lattice.
  Here for the renormalization factor
  we employ $\kappa_c$ determined from a simultaneous fit
  to $m_{PS}^2$ and $m_{quark}^{AWI}$
  in Table~\ref{table:m_quark_VWI_m_quark_AWI_and_m_PS2_simultaneous}.
}
\label{table:f_PS_f_V_12x24}
\begin{ruledtabular}
\begin{tabular}{c|cc|cc}
$\kappa_{sea}$ & $f_{PS}$    & [$t_{min},t_{max}$]
               & $f_{V}$     & [$t_{min},t_{max}$]
               \\ \hline
  0.14585      & 0.1785(14)  & [6,12]
               & 0.3118(33)  & [6,12]
               \\ \hline
  0.14660      & 0.15784(87) & [6,12]
               & 0.2874(57)  & [6,12]
               \\ \hline
  0.14705      & 0.1413(14)  & [6,12]
               & 0.2496(97)  & [6,12]
               \\ \hline
  0.14720      & 0.1412(41)  & [6,12]
               & 0.2422(239) & [6,12]
               \\
\end{tabular}
\end{ruledtabular}
\end{table}

%
%
\begin{table}[h]
\caption
{ Baryon masses on $12^3 \times 24$ lattice.
}
\label{table:m_delta_m_sigma_12x24}
\begin{ruledtabular}
\begin{tabular}{c|ccc|ccc}
$\kappa_{sea}$ & $m_{N}$      & [$t_{min},t_{max}$] & $\chi^2/dof$
               & $m_{\Delta}$ & [$t_{min},t_{max}$] & $\chi^2/dof$ \\ \hline
  0.14585      & 1.5357(69)   & [5,12]              & 0.65(76)
               & 1.7722(97)   & [5,12]              & 0.74(83)
\\ \hline
  0.14660      & 1.3619(92)   & [5,12]              & 0.85(66)
               & 1.6061(183)  & [5,12]              & 1.45(97)
\\ \hline
  0.14705      & 1.2054(165)  & [5,12]              & 0.69(96)
               & 1.5110(268)  & [5,12]              & 1.28(81)
\\ \hline
  0.14720      & 1.1791(417)  & [5,12]              & 0.99(62)
               & 1.5300(1020) & [5,12]              & 0.62(1.23)
\\
\end{tabular}
\end{ruledtabular}
\end{table}

%
%
\begin{table}[h]
\caption
{ Plaquette and rectangular loops on $12^3 \times 24$ lattice.
}
\label{table:plaq_rect_12x24}
\begin{ruledtabular}
\begin{tabular}{ccc}
$\kappa_{sea}$ & $\bra W^{1 \times 1} \ket$ & $\bra W^{1 \times 2} \ket$
\\ \hline
  0.14585      & 0.504529(56)               & 0.249916(70)
\\ \hline
  0.14660      & 0.508445(69)               & 0.254866(88)
\\ \hline
  0.14705      & 0.511202(68)               & 0.258350(86)
\\ \hline
  0.14720      & 0.512632(144)              & 0.260157(186)
\end{tabular}
\end{ruledtabular}
\end{table}

\clearpage

%
%
\begin{table}[h]
\caption
{ Meson and bare AWI quark masses on $16^3 \times 24$ lattice.
}
\label{table:m_PS_m_V_16x24}
\begin{ruledtabular}
\begin{tabular}{c|ccc|ccc|c}
$\kappa_{sea}$ & $m_{PS}$    & [$t_{min},t_{max}$] & $\chi^2/dof$
               & $m_{V}$     & [$t_{min},t_{max}$] & $\chi^2/dof$
               & $m_{quark}^{AWI}$ \\ \hline
  0.14585      & 0.6333(19)  & [6,12]              & 0.72(52)
               & 1.0488(43)  & [6,12]              & 0.82(76)
               & 0.06378(47) \\ \hline
  0.14660      & 0.4781(16)  & [6,12]              & 3.55(2.04)
               & 0.9403(70)  & [6,12]              & 1.41(93)
               & 0.03642(40) \\
\end{tabular}
\end{ruledtabular}
\end{table}

%
%
\begin{table}[h]
\caption
{ Decay constants on $16^3 \times 24$ lattice.
  Here for the renormalization factor
  we employ $\kappa_c$ determined from a simultaneous fit
  to $m_{PS}^2$ and $m_{quark}^{AWI}$
  in Table~\ref{table:m_quark_VWI_m_quark_AWI_and_m_PS2_simultaneous}.
}
\label{table:f_PS_f_V_16x24}
\begin{ruledtabular}
\begin{tabular}{c|cc|cc}
$\kappa_{sea}$ & $f_{PS}$    & [$t_{min},t_{max}$]
               & $f_{V}$     & [$t_{min},t_{max}$]
               \\ \hline
  0.14585      & 0.1804(23)  & [6,12]
               & 0.3151(45)  & [6,12]
               \\ \hline
  0.14660      & 0.1592(16)  & [6,12]
               & 0.2913(48)  & [6,12]
               \\
\end{tabular}
\end{ruledtabular}
\end{table}

%
%
\begin{table}[h]
\caption
{ Baryon masses on $16^3 \times 24$ lattice.
}
\label{table:m_delta_m_sigma_16x24}
\begin{ruledtabular}
\begin{tabular}{c|ccc|ccc}
$\kappa_{sea}$ & $m_{N}$      & [$t_{min},t_{max}$] & $\chi^2/dof$
               & $m_{\Delta}$ & [$t_{min},t_{max}$] & $\chi^2/dof$ \\ \hline
  0.14585      & 1.5567(91)   & [5,12]              & 1.97(92)
               & 1.7804(113)  & [5,12]              & 0.64(54)
\\ \hline
  0.14660      & 1.3257(118)  & [5,12]              & 1.58(91)
               & 1.5899(124)  & [5,12]              & 0.96(77)
\\
\end{tabular}
\end{ruledtabular}
\end{table}

%
%
\begin{table}[h]
\caption
{ Plaquette and rectangular loops on $16^3 \times 24$ lattice.
}
\label{table:plaq_rect_16x24}
\begin{ruledtabular}
\begin{tabular}{ccc}
$\kappa_{sea}$ & $\bra W^{1 \times 1} \ket$ & $\bra W^{1 \times 2} \ket$
\\ \hline
  0.14585      & 0.504482(75)               & 0.249850(90)
\\ \hline
  0.14660      & 0.508338(61)               & 0.254739(76)
\end{tabular}
\end{ruledtabular}
\end{table}

\clearpage





\clearpage

\clearpage


\end{document}